\documentclass[aps, prd, twocolumn, floatfix, superscriptaddress, nofootinbib, amsfonts, amsmath, amssymb, floats, letterpaper, preprintnumbers, 10pt]{revtex4-2}
\usepackage{setspace, physics, soul, tcolorbox, dsfont, float, booktabs, multirow, centernot, verbatim, tikz, url}
\usetikzlibrary{positioning, decorations.markings, decorations.pathmorphing}
\tikzset{
    fermion/.style={draw=black, postaction={decorate}, decoration={markings,mark=at position 0.5 with {\arrow{>}}}},
    vector/.style={decorate, decoration={snake, amplitude=1mm, segment length=3mm, post length=0mm, pre length=0mm}}
}

\allowdisplaybreaks

\begin{document}

\preprint{MIT--CTP/5685}

\title{On the cosmology and terrestrial signals of sexaquark dark matter}

\author{Marianne Moore}
\email{mamoore@mit.edu}
\affiliation{Center for Theoretical Physics, Massachusetts Institute of Technology, Cambridge, MA 02139, USA}

\author{Tracy R. Slatyer}
\email{tslatyer@mit.edu}
\affiliation{Center for Theoretical Physics, Massachusetts Institute of Technology, Cambridge, MA 02139, USA}

\date{\today}

\begin{abstract}
    We investigate the hypothesis that sexaquarks, hypothetical stable six-quark states, could be a significant component of the dark matter. We expand on previous studies of sexaquark cosmology, accounting for the possibility that some relevant interaction cross sections might be strongly suppressed below expectations based on dimensional analysis. We update direct-detection constraints on stable sexaquarks comprising a subdominant fraction of the dark matter, as well as limits on the annihilation of an antisexaquark component from Super-Kamiokande. We argue that the scenario where sexaquarks comprise a $\mathcal{O}(1)$ fraction of the dark matter would require either a suppression of $\mathcal{O}(10^{-19})$ in sexaquark interactions with baryons, combined with a very high yield of net sexaquark number from the quark-hadron transition, or else a very strong suppression of the cross section for antisexaquark annihilation on nucleons (24+ orders of magnitude below the QCD scale). Independently, we find that a sexaquark component comprising more than $\mathcal{O}(10^{-3})$ of the dark matter can be excluded from direct-detection bounds, unless its scattering cross section is severely suppressed compared to the expected scale of strong and even electromagnetic interactions.
\end{abstract}

\maketitle

\section{Introduction}

Despite the ubiquity of dark matter in cosmological data, numerous searches with particle detectors have yet to identify it. Many searches focus on weakly interacting massive particles (WIMPs) and are placing stringent bounds on their potential mass and interaction strength. An interesting dark matter candidate loosely in the regime of WIMPs is the sexaquark (also known as the hexaquark, dihyperon, or $H$ dibaryon), which is a hypothetical bound state of two up quarks, two down quarks, and two strange quarks ($uuddss$)~\cite{Farrar1708}. We will denote this state by $S$. First proposed in 1977 as a potentially stable exotic particle~\cite{Jaffe1977}, the sexaquark is composed of Standard Model particles and does not require a dark sector or portal to mediate its interactions. As required for a dark matter candidate, the sexaquark is uncharged. If its mass falls within the appropriate range, it could have a very long lifetime, as expected for a relic particle from the early universe. However, a number of potential difficulties with this scenario have been raised in the literature, from challenges in generating the correct abundance of sexaquark dark matter in the early universe, to its non-detection in various experimental probes, such as nuclear stability bounds, double hypernuclei searches, and accelerator searches. As we will review, it has been proposed that many of these difficulties may be overcome if the sexaquark has suppressed interactions in specific channels, and that such suppression may occur naturally, e.g.~due to small wavefunction overlap factors~\cite{Farrar:2003is}.

In this paper, we investigate challenges to the sexaquark as a potential cold dark matter candidate, both revisiting some earlier arguments and identifying new constraints, in light of the possibility that one or more of the cross sections for key processes may experience a strong parametric suppression, and that sexaquarks may only make up a small fraction of the dark matter abundance. We do not re-examine the plausibility of such a suppression to the cross sections; instead, we explore the consequences should we allow such a suppression (in the absence of any such suppression it seems already well-established that the sexaquark cannot constitute any non-negligible fraction of the dark matter, e.g.~\cite{Kolb:2018bxv}). We focus on the freeze-out yield of sexaquarks and their antiparticles (``antisexaquarks''), and then discuss various possible signatures of these populations in the present day, including the electromagnetic contribution to their scattering cross section with nucleons arising from their electric polarizability, their accumulation in the Earth, and the annihilation signature of antisexaquarks.

This work primarily considers five sexaquark parameters: (1)~the sexaquark mass $m_S$, (2)~the sexaquark per-nucleon elastic scattering cross section $\sigma_{Sn}^\text{scat}$, and three thermally-averaged annihilation cross sections, (3)~the sexaquark’s self-annihilation to Standard Model final states
$\ev{\sigma v}_{S \bar{S}}^\text{ann}$, (4)~the sexaquark's breakup into two baryons via collisions with particles that do not carry baryon number, parameterized by $\ev{\sigma v}_{bb'\to SX}^\text{ann}$, and (5)~the sexaquark annihilation with antibaryons to produce baryons, $\ev{\sigma v}_{S\bar{b}\to b'X}^\text{ann}$, as well as the counterpart processes for antisexaquarks for the last two cases. For each scattering and annihilation cross section, we vary their potential strength to consider a range that spans more than ten orders of magnitude. Together, these quantities control the abundance and detectability of sexaquarks by determining the overall fraction of dark matter composed of sexaquarks and antisexaquarks, the experimental reach of dark matter direct detection experiments in searching for the sexaquark through scattering, and the capability of sensitive detectors to measure the annihilation of antisexaquarks with nuclei.

We begin by reviewing previous studies of the sexaquark, resulting constraints on its properties, and open questions in Sec.~\ref{sec:properties}. In Sec.~\ref{sec:abundance}, we examine the abundance of sexaquarks and antisexaquarks generated by equilibrium processes in the early universe, while in Sec.~\ref{sec:freezeout} we study the freeze-out of number-changing interactions to determine the relic densities of these populations, and discuss some points of confusion/disagreement in the literature. We then look at constraints from direct detection due to the sexaquark's elastic scattering with nucleons in Sec.~\ref{sec:Epol}, including estimates of its electric polarizability. We study the possible accumulation of populations of sexaquarks and antisexaquarks in the Earth in Sec.~\ref{sec:accumulation}. Finally, Sec.~\ref{sec:SuperK} investigates possible signals from annihilation of antisexaquarks in Super-Kamiokande, in the event that a relic antisexaquark population survives from the early universe. Throughout, we consider the interplay of the cross sections of interest for the sexaquark, which determine their relic abundance and whether they are detectable in terrestrial searches. Finally, we summarize our findings in Sec.~\ref{sec:summary}. We briefly review the relevant group theory in Appendix~\ref{sec:group} and explain analytically the different regimes for the sexaquark yield depending on the temperature of freeze-out in Appendix~\ref{sec:scaling}. We explore the fraction of the baryon asymmetry stored in asymmetric dark matter carrying baryon number in Appendix~\ref{app:asymmetric}, discuss and evaluate various sexaquark scattering cross sections arising from its polarizability in Appendix~\ref{sec:Pospelov}, and apply these cross sections to check that the sexaquark remains kinetically coupled to the Standard Model bath throughout freeze-out in Appendix~\ref{app:kinetic}. In Appendix~\ref{app:Earthannihilations} we explore the possibility that antisexaquark annihilations in the Earth could produce a flux of neutrinos detectable by Super-Kamiokande.

Throughout this paper, we use ${\rho_\chi = 0.3~\text{GeV/cm} ^3}$ as the local dark matter density, ${f_{S/\bar{S}} = \Omega_{S/\bar{S}} / \Omega_\text{DM}}$ as the fractions of dark matter composed of sexaquarks and antisexaquarks respectively, and denote by ${f_\chi = f_S + f_{\bar{S}} = \Omega_{\chi} / \Omega_\text{DM}}$ the total fractional dark matter abundance in (anti)sexaquarks. 

\section{\label{sec:properties}Review of previous sexaquark studies}

In this section we review previous studies on whether the sexaquark could exist as a viable bound state, and if so, whether it could contribute significantly to the abundance of dark matter. Where it is unclear if a particular limit firmly excludes the possibility of a stable sexaquark providing a substantial fraction of the dark matter abundance, we will generally assume that this possibility remains open (with any requisite conditions being satisfied), in order to see if we can exclude the resulting parameter space through independent arguments. With this approach, we do not intend to assert that arguments to the contrary (i.e. that part or all of the parameter space is already non-viable) are incorrect, only to explore complementary tests of the sexaquark dark matter hypothesis.

\subsection{Theoretical predictions for the stability and mass of the sexaquark}

Various techniques have estimated the mass of the sexaquark over a range from $1200$ to $2260$~MeV. The MIT quark bag model, for example, suggests a mass of $1760$~MeV in the SU(3)${}_f$ limit and $2150$~MeV otherwise~\cite{Jaffe1977}, while QCD sum rules predict a mass of $1878$~MeV in the SU(3)${}_f$ limit and $2190$~MeV with SU(3)${}_f$ breaking~\cite{Kodama:1994np}. A second group applied QCD sum rules and found $\sim 1200$~MeV with SU(3)${}_f$ breaking~\cite{Azizi:2019xla}. A diquark model fitted using tetraquark and pentaquark parameters finds $1200$~MeV using light tetraquark parameters and $2170$~MeV with baryon parameters~\cite{Gross:2018ivp}. Another technique, holographic models of QCD, models the sexaquark in a bulk Anti de-Sitter space for various choices of running of the anomalous dimension of the quark bilinear operator~\cite{Evans:2023zde}. They find a preferred sexaquark mass of 1750~MeV. Notably, in all of these cases, the mass is below the threshold to produce two $\Lambda$ baryons, $2230$~MeV, even in the SU(3)${}_f$ breaking limit, indicating that the sexaquark may be a bound state.

Building on previous estimates of the mass of the sexaquark, the NPLQCD and HAL QCD collaborations and the Mainz group used lattice QCD to study the binding energy $B_S$ of the sexaquark at unphysical quark masses and different values of the pion mass~\cite{Inoue:2010es, NPLQCD:2010ocs, NPLQCD:2011naw, INOUE201228, HALQCD:2019wsz, PadmanathMadanagopalan:2021exb, Green:2021qol}. The binding energy was then chiral-extrapolated to physical values of the quark masses. The resulting average value of the NPLQCD and HAL QCD groups shows that the sexaquark is unbound by ${26(11)}$~MeV, indicating a higher mass of ${2258(11)}$~MeV~\cite{Shanahan:2011su,Shanahan:2013yta}. Since under this prediction the sexaquark would not even be bound (relative to $\Lambda\Lambda$), it would certainly not be a good candidate for dark matter. 

In the remainder of this article we will thus assume the sexaquark can be more deeply bound (due to effects not captured by the lattice prediction or other calculations that yield a high mass), to explore the downstream implications of the existence of a stable sexaquark. Ref.~\cite{Farrar:2022mih} discusses mechanisms through which a deeply-bound sexaquark state might be difficult to detect in lattice calculations, distinguishing it from a more loosely bound state resembling a pair of $\Lambda$ baryons.

\subsection{Phenomenology of the sexaquark mass}

Broadly speaking, a heavier sexaquark is less stable due to the opening of viable decay channels; a lighter sexaquark would be more stable but could provide new decay channels for known hadronic states. For a viable sexaquark dark matter candidate, both its allowed decay channels and the decay channels of Standard Model particles involving the sexaquark should be either kinematically closed or sufficiently suppressed. 

The main decay channels of the sexaquark involve transforming to a pair of octet baryons, such as two nucleons, a nucleon and a $\Lambda$ hyperon, or two $\Lambda$ hyperons. All these channels are closed for a sexaquark mass ${m_S}$ less than or equal to approximately $1878$~MeV, corresponding to twice the proton mass plus twice the electron mass~\cite{Farrar1708}. Given that all other baryons possess heavier masses, their production is energetically forbidden. In a similar vein, the generation of sexaquarks from Standard Model particles is energetically unfeasible if its mass exceeds $1875$~MeV; two nucleons within a nucleus or a deuteron cannot simultaneously undergo a weak decay into the sexaquark~\cite{Farrar:2022mih}.

Still, it is possible for the sexaquark to be a cosmologically stable relic even if its mass is outside of the $1875 - 1878$~MeV range. The decay channels of the sexaquark are energetically suppressed for masses near this range~\cite{Farrar:2018hac}. As the lifetime of the sexaquark is closely related to its mass, cosmic microwave bounds dictate that should possess a minimum lifetime of $\sim 10^{24}$~s to account for all cold dark matter, which is approximately $10^7$~times longer than the age of the Universe~\cite{Slatyer:PhysRevD.95.023010}. This is a general limit, but depending on the decay channel there are almost certainly indirect-detection bounds that are stronger. This limit is less stringent if sexaquarks constitute a subdominant portion of the dark matter density. For sexaquark masses up to the mass of a nucleon plus a hyperon ($2055$~MeV), the sexaquark's decay rate to a pair of nucleons is a doubly weak process, resulting in a long enough lifetime~\cite{Kolb:2018bxv}. 

On the lower mass end, nuclear stability is ensured if ${m_S}$ is greater than or equal to $1860$~MeV, which is the mass of two nucleons minus twice the average binding energy of a nucleon in a nucleus (8~MeV) \cite{Kolb:2018bxv}, thereby preventing the nucleus from decaying to produce a sexaquark. More specifically, for masses below 1740~MeV, Super-Kamiokande would probably have observed double nucleon conversion into the sexaquark from oxygen nuclei~\cite{Farrar:2003is}. Then, at masses below 1850~MeV, using nuclear stability bounds for oxygen decay from Super-Kamiokande, \cite{Gross:2018ivp} showed that the dimensionless squared matrix element for the transition of two off-shell $\Lambda$ baryons to the sexaquark needs to be suppressed to the level of $10^{-20}-10^{-25}$ to avoid constraints. Ref.~\cite{Farrar:2023wta} discusses a similar but slightly stronger bound from deuteron decay, which extends the limit up to $1860$~MeV, and argues that the required suppression to the matrix element is much stronger than one would expect from theoretical considerations.

There are also searches for the sexaquark in the context of a broader initiative to study hypernuclei, nuclei with one or more bound hyperons such as the $\Lambda$ and $\Sigma$~\cite{Gal:2016boi}. The hunt for the sexaquark is embedded within the effort focusing on $\Lambda\Lambda$-hypernuclei. Should the sexaquark be sufficiently light compared to a pair of $\Lambda$ hyperons -- i.e. the binding energy ${B_S = 2 m_\Lambda - m_S}$ must be larger than the net binding energy of the two $\Lambda$ particles -- a double-hypernucleus could preferably decay into a sexaquark and a nucleus via the strong interaction, e.g. ${{}^6_{\Lambda\Lambda}\text{He} \to {}^4\text{He}+S}$, rather than weakly through $\Lambda \to \lbrace n,p \rbrace \pi$. If this decay has a rate typical of strong interactions, $10^{23}~\text{s}^{-1}$, then it would greatly dominate over the weak decay of hyperons within the hypernucleus, which has a rate of order $10^{10}~\text{s}^{-1}$~\cite{doi:10.1098/rspa.1989.0115}; for the decay involving the sexaquark to be subdominant to the weak decays would thus require a suppression factor of $\mathcal{O}(10^{-13})$.

As yet, these searches have found no evidence for a sexaquark state. Furthermore, the resulting limits extend up to higher sexaquark masses than can be tested with conventional nuclear decays (e.g. up to 2204~MeV for the analysis of \cite{Aoki:1991ip}). In order to evade these bounds, the strong decay involving the sexaquark would need to be suppressed by a sufficient factor to be slower than the weak decays. However, it has been argued in Ref.~\cite{Farrar:2023wta} that the strong decay of the double-hypernucleus can indeed be sufficiently inhibited by the wavefunction overlap between $\Lambda-\Lambda$ and $S$ configurations, in the event that the $S$ is deeply bound, contrasting with earlier estimates, e.g.~\cite{PhysRevD.29.433}.

Consequently, in our analysis below we consider masses from $1860$ to $1890$~MeV as a potential range for the sexaquark as a dark matter candidate, and note that our analysis could straightforwardly be extended to a wider mass range.

\subsection{Other sexaquark constraints}

In addition to the probes described above, efforts to detect or constrain the sexaquark have spanned a wide range of masses and production channels, including exotic upsilon decays~\cite{Belle:2013sba, BABAR:PhysRevLett.122.072002}, proton-proton scattering~\cite{Carroll:1978uu}, electron-proton scattering~\cite{BESIII:2023clq}, kaon-nucleus scattering~\cite{Yoon:2007aq, BNLE836:1997dwi, E224:1996njn, Aoki:1990nk, E885:2000ped, PhysRevLett.87.132504}, lead-lead collisions~\cite{ALICE:2015udw}, and bound states with helium or deuterium in gold-platinum collisions~\cite{E886:1995udz}. Some of these searches explicitly placed a cut that would render them insensitive to specific sexaquark masses; for example, ALICE~\cite{ALICE:2015udw}, Belle~\cite{Belle:2013sba}, and an experiment at BNL~\cite{Carroll:1978uu} investigated six-quark bound states with a mass above 2~GeV, so are not relevant to the mass range we consider. While other searches did not explicitly place an energy cut which would eliminate lower-mass candidates, the accumulated statistics were not sufficient to put constraints on the production and decay rate of sexaquarks in the scenario where they have highly suppressed interactions. A discussion of the impact of selected constraints on the allowed effective couplings can be found in Ref.~\cite{Farrar:2023wta}. No evidence of a sexaquark-like state has been found via these searches. 

Other studies considered the impact of sexaquarks on neutron stars. In the harsh conditions of proto-neutron stars, characterized by extreme temperatures and densities, the production of sexaquarks must be sufficiently inhibited through wavefunction overlap suppression due to their small radius. Otherwise, a large population of these particles is generated, contradicting the observed timescale and stability of proto-neutron stars~\cite{McDermott:2018ofd}. Even though the sexaquark's effective in-medium mass would be larger, it is not enough to account for all observed neutron star masses~\cite{Shahrbaf:2022upc}. The lambda hyperon suffers from a similar issue should they be present in this environment. It is possible to avoid neutron star constraints by quark deconfinement (transition to non-hadronic degrees of freedom), such that neither sexaquarks nor lambda hyperons are present in neutron stars~\cite{Shahrbaf:2022upc, Blaschke:2022knl}.

Finally, and most relevantly for this work, if sexaquarks are cosmologically long-lived, they would contribute to the dark matter abundance, although no sexaquark signal has been observed in dark matter direct detection experiments or astrophysical observations to date.

Some aspects of the freeze-out abundance of sexaquarks have been addressed in previous studies~\cite{Gross:2018ivp, Farrar:2018hac, Kolb:2018bxv}. A first analysis proposed that a 1200~MeV sexaquark could account for the observed dark matter abundance by freezing out at a temperature of approximately $30$~MeV~\cite{Gross:2018ivp}. However, a sexaquark of this mass would be expected to induce fast decays of nuclei, as discussed above. A different study examined quark equilibrium abundances in the quark gluon plasma prior to the QCD crossover and argued the correct relic abundance could be naturally obtained provided sexaquarks did not equilibrate with the thermal bath after confinement (i.e. at temperatures below $\sim 140$~MeV). This condition can be satisfied if the interaction cross sections for relevant number-changing processes are sufficiently small~\cite{Farrar:2018hac}. A third study found that sexaquarks with the proper mass and a QCD-scale interaction rate with the thermal bath would remain in equilibrium after the QCD crossover, down to temperatures around $10$~MeV, and consequently could not generate a relic density comparable to that of dark matter~\cite{Kolb:2018bxv}. In the next section, we revisit this question, considering a continuum of scenarios where the processes that produce and deplete sexaquarks can be severely suppressed, as suggested in Ref.~\cite{Farrar:2018hac}.

\section{\label{sec:abundance}Cosmological abundances in equilibrium}

To investigate whether sexaquarks can constitute a significant fraction of dark matter, we examine the expected abundance generated in the early Universe through interactions of the sexaquark and its antiparticle with the thermal bath. We begin in this section by considering the equilibrium evolution of the sexaquark and antisexaquark abundances, i.e. in the case where the rates of number-changing processes are fast compared to Hubble. We distinguish two cases, (1) where the fast number-changing processes allow the (anti)sexaquark to exchange baryon number with other hadrons, relating its chemical potential to that of the baryons, and (2) where the only rapid sexaquark-number-changing processes are those that conserve the net baryon number stored in sexaquarks and their antiparticles (e.g. $S\bar{S}$ annihilation). These two evolution pathways lead to different histories for the sexaquark chemical potential and hence the sexaquark abundance and asymmetry. The first case was studied in some detail by Ref.~\cite{Kolb:2018bxv}, but with a focus on the behavior of the equilibrium curves at relatively late times 
/ low temperatures. We will develop an analytic understanding of the behavior at higher temperatures, which will be relevant for freeze-out of suppressed interactions. We defer to the next section the detailed study of freeze-out, which depends on the strength of various interactions.

An important novel aspect of our analysis is that we study the abundance of antisexaquarks, which can be significant if the depletion cross sections are sufficiently suppressed, potentially comprising a substantial fraction of the final contribution to the dark matter density. None of the studies reviewed in Sec.~\ref{sec:properties} considered the possibility of a significant surviving fraction of antisexaquarks. This is a self-consistent assumption in Refs.~\cite{Gross:2018ivp, Kolb:2018bxv}, where sexaquarks interact quite strongly with the baryons, efficiently depleting the antisexaquark abundance. However, if we consider scenarios like that proposed in Ref.~\cite{Farrar:2018hac} where interactions with baryons that would deplete the (anti)sexaquark abundance are severely suppressed after the QCD crossover, we may also need to consider the relic abundance of stable antisexaquarks (depending on the degree to which self-annihilations are also suppressed, and the fraction of the dark matter comprised of sexaquarks). The chemical potentials in the strongly-interacting sector of the Standard Model are negligible during the QCD crossover, as noted in Ref.~\cite{Farrar:2018hac}. Consequently, the thermal bath is symmetric to a good approximation, with equal number of light quarks and antiquarks, leading us to expect the production of sexaquarks and antisexaquarks at equal rates during the crossover. If there is no subsequent depletion via interactions with the thermal bath, we would anticipate equal fractions of sexaquarks and antisexaquarks in the present day.

More generally, number-changing interactions (such as $S\bar{S}$ annihilation) may deplete both the sexaquark and antisexaquark abundance after the QCD crossover, often to differing degrees. In this case the eventual abundances depend on the evolution of the sexaquark chemical potential (which as mentioned above depends on the nature of the processes maintaining the equilibrium) and when the relevant number-changing interactions freeze out. In particular, if the depletion interaction conserves the net baryon number stored in sexaquarks and antisexaquarks as is true for $S\bar{S}$ annihilation, and is sufficiently strong, the relevant quantity determining the relic abundance is the fraction of the baryon asymmetry stored in (anti)sexaquarks, set at the time of decoupling for interactions that allow exchange of baryon number between (anti)sexaquarks and other baryons.

We assume throughout this section and the following one that (anti)sexaquarks remain at the same temperature at the rest of the Standard Model thermal bath (i.e. they remain kinetically coupled even after they are no longer in thermal equilibrium), which we justify in Appendix~\ref{app:kinetic}.

\subsection{\label{subsec:EQ_abundance}Equilibrium abundance with baryon interactions}

We concentrate our investigation on the temperature range of 155 to 1~MeV ($x=m_S/T$ from 12 to 1890), which lies between the scales of QCD confinement and nucleosynthesis~\cite{ParticleDataGroup:2020ssz}. Within this span of temperatures, strange baryons, possessing weak decay lifetimes of approximately ${10^{-10}}$~s and strong decay timescales of around ${10^{-23}}$~s, maintain thermal equilibrium amongst themselves due to their decay timescales being significantly less than the universe's age during the period in question. As such, they share a common temperature $T$. Given that their mass is much larger than the temperature, we employ the thermal equilibrium number density in the non-relativistic limit~\cite{Kolb:1990vq},
\begin{align}\label{eq:number_density_equilibrium}
    n_i^\text{eq}(T) = g_i \left( \frac{m_i T}{2\pi} \right)^{3/2} e^{-(m_i - \mu_i) / T} \ ,
\end{align}
where $g_i$, $m_i$, and $\mu_i$ denote the number of degrees of freedom, mass, and chemical potential of species $i$, respectively. We take the number of degrees of freedom to be ${g = 2}$ for baryons and ${g_S = 1}$ for the sexaquark, and we take the baryon masses from Ref.~\cite{ParticleDataGroup:2020ssz}. The relationship between the chemical potentials of baryons ($\mu_B$) and sexaquarks ($\mu_S$) is established by chemical equilibrium relations so long as chemical equilibrium is maintained, such that ${\mu_{\bar{S}} = -\mu_S}$ and ${\mu_S = 2 \mu_B}$~\cite{Gross:2018ivp, Kolb:2018bxv}. For example, either reactions of the form $S X \rightarrow b b^\prime$ (where $X$ carries no baryon number and $b,b^\prime$ denote baryons) or $S \bar{b}\rightarrow b' X$ are sufficient to enforce the second relation.

We treat the alternate case where there are no sexaquark-number changing reactions to relate the chemical potential of (anti)sexaquarks to the one of baryons, and hence $\mu_S$ may diverge from $2\mu_B$, in Sec.~\ref{subsec:equilibrium}. This corresponds to the reaction where sexaquarks' only fast number-changing process is annihilation with their antiparticle into lighter Standard Model states.

The entropy is given by ${s(T)=2 \pi^2 g_{\ast S}(T) T^3/45}$, with $g_{\ast S}(T)$ being the effective number of degrees of freedom for particles whose mass is much smaller than the temperature, which we take from \cite{ParticleDataGroup:2020ssz}. The baryon asymmetry is defined as the ratio ${Y_B = n_B/s}$, where ${n_B = \sum_b n_b - n_{\bar{b}}}$ is the sum of the number densities of baryon species $b$. We will generally use the term ``yield'' to describe relic number densities normalized to entropy density. 

\begin{figure}[ttt]
    \centering
    \includegraphics[width=\columnwidth]{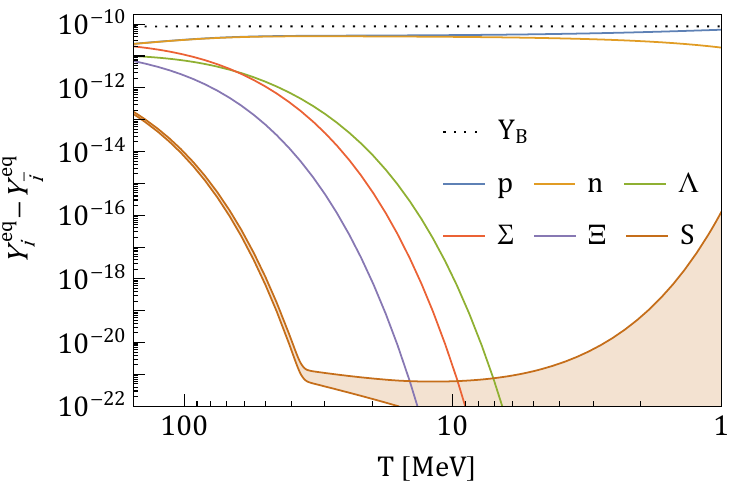}
    \caption{Evolution of the baryon asymmetry with temperature for the thermal equilibrium number densities. The label $\Sigma$ indicates the asymmetry in $\Sigma^+,\Sigma^0,\Sigma^-$, the label $\Xi$, the asymmetry in $\Xi^0,\Xi^-$, and the label $S$, the asymmetry in sexaquarks including the factor coming from baryon number 2. The shaded brown region indicates the range of late-time baryon asymmetry in the sexaquark for a mass in the range ${1860-1890}$~MeV, with the larger asymmetry obtained for lighter sexaquarks. The black dotted line represents $Y_B$.}
    \label{fig:asymmetry}
\end{figure}

Refs.~\cite{Gross:2018ivp,Kolb:2018bxv} define the baryon asymmetry including the contribution from sexaquarks as ${Y_{BS} = Y_B + 2 Y_S }$, where ${Y_S = Y_B m_p \Omega_\text{DM}/(m_S \Omega_B)}$. However, there are a few limitations of this definition that mean it is not directly applicable to our generalized calculation. First, it implies that sexaquarks are all of dark matter and that the abundance of antisexaquarks can be neglected. The formula can be corrected by using $\Omega_S = (f_S - f_{\bar{S}}) \Omega_\text{DM}$ rather than $\Omega_\text{DM}$. If the fractions of sexaquarks and antisexaquarks are equal, the dark matter sector does not contribute to the baryon asymmetry; we will see this situation approximately holds if sexaquarks chemically decouple from the baryons before the chemical potential gets large. Second, we find that in chemical equilibrium, most of the total baryon asymmetry is carried by the protons and neutrons, while sexaquarks carry at most $\mathcal{O}(2 \times 10^{-3})$ of the baryon asymmetry, as shown in Fig.~\ref{fig:asymmetry}. Thus, using the definition of $Y_{BS}$ given above inflates the overall baryon asymmetry, and results in a larger-than-observed asymmetry in the proton and neutron yields. In the following we will generally use $Y_B = Y_{BS}$, tacitly assuming $Y_S \ll Y_B$, in contrast with previous works which sought to explain the full dark matter abundance with sexaquarks only (not antisexaquarks). In Appendix~\ref{app:asymmetric} we discuss prospects for obtaining the full dark matter abundance solely with asymmetric baryonic dark matter. We assume there are no processes that violate baryon number between the QCD phase transition and today such that we can use today's measured value of $Y_B = 8.52 \times 10^{-11}$.

Ensuring that the total baryon asymmetry equals
\begin{align}\label{eq:baryon_asymmetry}
    \sum_b \left[ n_b^\text{eq} - n_{\bar{b}}^\text{eq} \right] + 2 \left[ n_S^\text{eq} - n_{\bar{S}}^\text{eq} \right] = s(T) Y_{B} \ ,
\end{align}
where the sum covers all octet baryons ${b = \lbrace p, n, \Lambda, \Sigma^0, \Sigma^\pm, \Xi^0, \Xi^- \rbrace}$, we can establish the thermal equilibrium abundance of (anti)sexaquarks. This equation represents the baryon asymmetry and would also hold true out of equilibrium, in the absence of baryon-number-violating processes, if we replaced the thermal equilibrium number densities with the correct number densities. Once the sexaquarks freeze-out, we will use $n_S$ and $n_{\bar{S}}$ instead of their equilibrium number densities. Solving Eq.~\eqref{eq:baryon_asymmetry} for the chemical potential as a function of temperature allows us to determine the equilibrium number density of baryons and sexaquarks in a comoving volume.

\begin{figure}[ttt]
    \centering
    \includegraphics[width=\columnwidth]{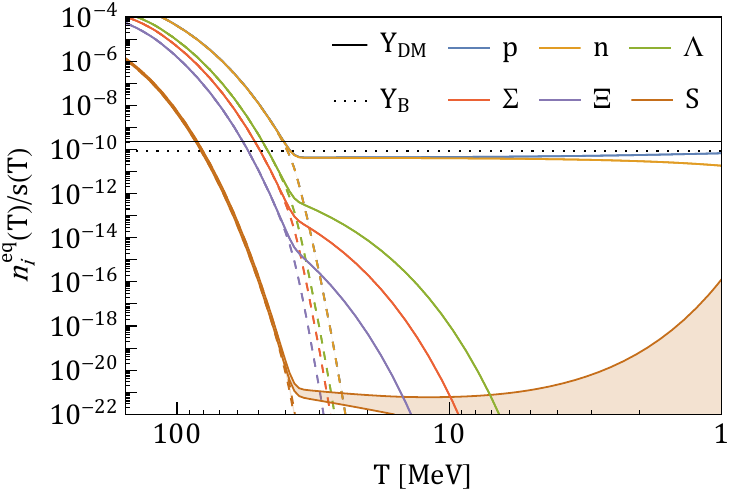}
    \caption{Equilibrium yield of octet baryons and sexaquarks for temperatures below the QCD confinement temperature. Time runs from left to right. The shaded brown area indicates the range of late-time number density of sexaquarks for a mass in the range of ${1860 - 1890}$~MeV. The larger number density at lower temperatures is obtained for lighter sexaquarks. The colored dashed lines represent the number densities of the corresponding antiparticles. The black dotted (full) line shows the observed baryonic (dark matter) abundance today.}
    \label{fig:equilibrium_abundance}
\end{figure}

Fig.~\ref{fig:equilibrium_abundance} illustrates the equilibrium yield of baryons, sexaquarks, and their antiparticles, obtained by numerically solving Eq.~\eqref{eq:baryon_asymmetry}. The horizontal lines denote the expected baryon and dark matter yields derived from the baryon-to-photon ratio, ${\eta^0\sim 6 \times 10^{-10}}$~\cite{ParticleDataGroup:2020ssz}, through ${Y_B = \eta^0 (1.8 g_{\ast S}^0)^{-1}}$ and ${Y_\text{DM} = (\Omega_\text{DM} m_p)/(\Omega_\text{B} m_S) Y_B}$. The superscript $0$ indicates the value measured today and we used the relationship between the photon number density and the entropy density, $s(T) = 1.80 g_{\ast S}(T) n_\gamma(T)$.

Even though the sexaquark and baryon yields respect the baryon asymmetry of Eq.~\eqref{eq:baryon_asymmetry}, the curves for the sexaquarks and all baryons but protons do not flatten out at late times, as is typical in the standard asymmetric dark matter scenario (see e.g.~\cite{Gelmini:2013awa}). This is because the asymmetry of each species, ${Y_i^\text{eq} - Y_{\bar{i}}^\text{eq}}$, is not constant. The asymmetry is shared between sexaquarks and each baryon species and evolves with temperature, as depicted in Fig.~\ref{fig:asymmetry}.

The insights from Fig.~\ref{fig:equilibrium_abundance} can be interpreted by analytically solving Eq.~\eqref{eq:baryon_asymmetry} using only the baryon equilibrium number densities, since the contribution from the sexaquarks is exponentially suppressed by the Boltzmann factor due to the sexaquark's higher mass. In principle, the evolution of the chemical potential could lift this suppression, but we have checked that this effect is small in the relevant epoch. This yields the chemical potential
\begin{align}\label{eq:mu_b}
    \mu_B(T) &\approx T \ \text{arcsinh} \left( \frac{Y_{B} s(T) (2\pi)^{3/2}}{4\sum_b (m_b T)^{3/2} e^{-m_b/T}} \right) \ .
\end{align}
This chemical potential exhibits two distinct regimes. At high temperatures, ${\mu_B \sim \mathcal{O}(10^{-3}~\text{MeV}) \ll m_b}$ and it can be neglected in comparison with the particle mass, while at low temperatures, it becomes comparable to the mass of a single baryon. The transition between these regimes occurs at around $40$~MeV for sexaquarks and $30$~MeV for baryons. For even smaller temperatures, less than $10$~MeV, the sum over baryons in Eq.~\eqref{eq:mu_b} can be replaced by its dominant term due to the proton. By inserting the chemical potential of Eq.~\eqref{eq:mu_b} in the equation for the (anti)sexaquark equilibrium number density (Eq.~\eqref{eq:number_density_equilibrium}), we find the following expressions for the two regimes:
\begin{align}
    n_{S,\bar{S}}^\text{eq, high T} (T) &\approx \left( \frac{m_S T}{2\pi} \right)^{3/2} e^{-m_S/T} \ , \label{eq:number_density_highT} \\
    n_S^\text{eq, low T} (T) &\approx \left( \frac{2\pi m_S}{T} \right)^{3/2} \nonumber \\
    &\times \frac{Y_{B}^2 s^2(T) e^{-(m_S-2m_p)/T}}{4\left[ m_p^{3/2} + \sum_{b\not = p} m_b^{3/2} e^{-(m_b-m_p)/T} \right]^{2} } \ , \label{eq:number_density_lowT} \\
    n_{\bar{S}}^\text{eq, low T} (T) &\approx \frac{4 e^{-m_S/T}}{Y_B^2 s^2(T)} \left( \frac{m_S T}{2\pi} \right)^{9/2} \nonumber \\
    &\times \bigg[ \sum_{b} \bigg(\frac{m_b}{m_S}\bigg)^{3/2} e^{-m_b/T} \bigg]^{2} \ .
\end{align}
At high temperatures, chemical potentials are negligible and interactions are rapid. Both the sexaquark and antisexaquark equilibrium yields decrease as $e^{-m_S/T}$. As the temperature decreases and the chemical potential turns on, the behavior of the sexaquark yield is influenced by the exponential term involving the difference ${m_S - 2m_p}$, with ${2m_p \sim 1877}$~MeV. This results in two possibilities, depicted in the shaded brown region of Fig.~\ref{fig:equilibrium_abundance}: either an increase in the yield at late times for small sexaquark masses or a second phase of depletion for larger masses. This behavior resembles the phenomenon observed in bouncing dark matter scenarios~\cite{Puetter:2022ucx}. However, in our case, the bounce is driven by the mass difference rather than a departure from chemical equilibrium. In contrast, the antisexaquark population continues to deplete exponentially with a factor of $e^{-(m_S + 2m_p)/T}$ and becomes negligible quickly past ${T \sim 40}$~MeV.

Similarly, the baryon number density decreases exponentially as $e^{-m_b/T}$ at high temperatures. Due to their lighter mass relative to the sexaquark, the baryon abundances are enhanced with respect to sexaquarks. Once the antibaryons decouple, the yield of baryon $i$ follows a milder slope given by $[\sum_b e^{-(m_b-m_i)/T}]^{-1}$, resulting in an almost constant yield for the lightest baryons, namely the proton and neutron. The behavior of other baryons is determined by the exponential involving the difference of their mass with that of the proton, leading to the fastest depletion of the heaviest baryons. Antibaryons, on the other hand, continue to deplete exponentially fast and their yield quickly becomes negligible.

As shown in Fig.~\ref{fig:asymmetry}, in the equilibrium state most of the baryon asymmetry is carried by protons and neutrons, with a negligible fraction carried by sexaquarks. The highest values for the asymmetry carried by sexaquarks are obtained early on, immediately after the QCD crossover, but are still suppressed by at least two orders of magnitude compared to protons/neutrons. There is a potential upturn at late times if sexaquarks have $m_S < 2 m_p$, in which case their equilibrium asymmetry grows to reach $\mathcal{O}(10^{-16})$ at 1~MeV. Otherwise, their asymmetry continues to decrease exponentially. In any case, this remains much less than the asymmetry carried by protons and neutrons, reflecting that the sexaquark equilibrium abundance is always either very small or very similar to the antisexaquark equilibrium abundance.

\subsection{\label{subsec:equilibrium}Equilibrium abundance with fixed sexaquark asymmetry}

Another possibility is that sexaquarks do not undergo reactions involving baryons once formed in the QCD phase transition, but continue to interact with the thermal bath of photons and other relativistic species. The chemical potential of sexaquarks is related to the antisexaquark through $\mu_S = -\mu_{\bar{S}}$ and evolves separately from the baryon chemical potential. This can be viewed as a standard ``asymmetric dark matter'' scenario, with the asymmetry being fixed by an initial condition. It is in contrast with the previous case (Sec.~\ref{subsec:EQ_abundance}), which assumes that at least one of the strong interaction rates involving (anti)baryons remains fast relative to Hubble. There, the chemical potential of sexaquarks obeyed both $\mu_{\bar{S}} = -\mu_S$ and $\mu_S = 2\mu_B$ until sexaquarks chemically decouple from the bath at freeze-out.

We will generally assume that the evolution follows the path described in Sec.~\ref{subsec:EQ_abundance} in the hadronic phase, down to some temperature $T_*$ where the reactions enforcing $\mu_S=2\mu_B$ decouple; after that point, the sexaquark chemical potential would evolve to conserve the net baryon number stored in sexaquarks and their antiparticles. Alternatively, if the interactions enforcing this relation are never fast after the quark-hadron transition, the sexaquark chemical potential would presumably be set during the quark-hadron transition; we  expect the natural scale of this chemical potential to be comparable to $2 \mu_B$, but discuss this point in more depth in Sec.~\ref{sec:earlier}.

We can calculate how the equilibrium yield of (anti)sexaquarks shown in Fig.~\ref{fig:equilibrium_abundance} would be affected in this scenario. We enforced ${\mu_S = 2\mu_B}$ at early times until $T_\ast$, the temperature at which baryon-number-exchanging reactions are turned off, and determined the baryon asymmetry in baryons ($Y_B'$) as well as the remainder in sexaquarks (${Y_B - Y_B'}$). For ${T < T_\ast}$, we enforced separately
\begin{align}
    \sum_b \left[ n_b^\text{eq} - n_{\bar{b}}^\text{eq} \right] &= s(T) Y_{B}' \ , \\
    2 \left[ n_S^\text{eq} - n_{\bar{S}}^\text{eq} \right] &= s(T) (Y_B - Y_B') \ . \label{eq:YB_sexaquarks}
\end{align}
Thus, if we substitute in the equilibrium number densities for these species, we obtain the ratio
\begin{align}\label{eq:ratioYB}
    \frac{2 m_S^{3/2} e^{-m_S/T_\ast}}{\sum_b m_b^{3/2} e^{-m_b/T_\ast}} &= \frac{Y_B - Y_B'}{Y_B'} \ .
\end{align}

In this section we presume that the sexaquark annihilation cross section is large, and so maintains equilibrium with the relativistic bath, such that the final relic abundance is set by the asymmetry when interactions that exchange baryon number with the bath become slow relative to the expansion rate.

If the decoupling of these baryonic interactions occurs before the chemical potential starts influencing the shape of the sexaquark and baryon curves (which is certainly true if the decoupling occurs during the QCD phase transition, but also at somewhat later times), the sexaquark yield plateaus as is typical of asymmetric dark matter scenarios~\cite{Gelmini:2013awa}. If the baryonic interactions remain in equilibrium to ${T\lesssim 40}$~MeV, as discussed in the Sec.~\ref{subsec:EQ_abundance}, the final sexaquark abundance is almost independent of the subsequent evolution (and the antisexaquark abundance is negligible).

\begin{figure}[ttt]
    \centering
    \includegraphics[width=\columnwidth]{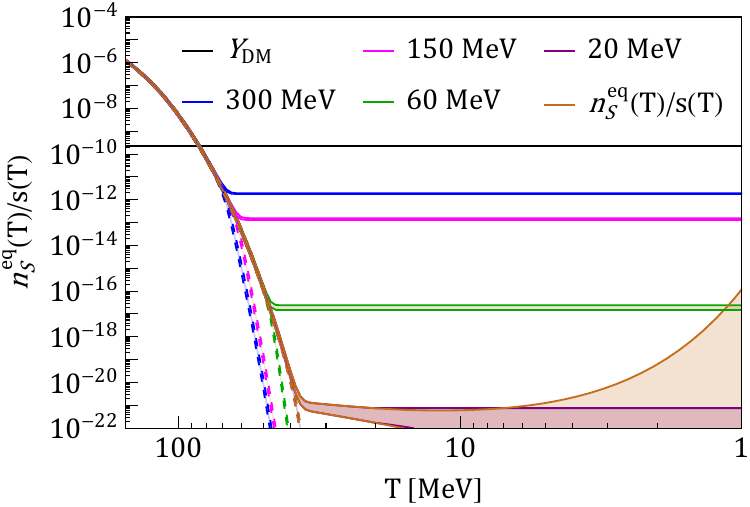}
    \caption{(Anti)sexaquark equilibrium yield in scenarios where the (anti)sexaquarks remain coupled to the Standard Model thermal bath, but depart from chemical equilibrium with the baryons after a temperature $T_*$. We show results for four choices of $T_*$ (300~MeV, 150~MeV, 60~MeV, and 20~MeV) indicated by the colored lines, as well as the case where $T_*$ is very low and we recover the previous equilibrium result. The black horizontal line shows the full dark matter yield.}
    \label{fig:alternate_eq}
\end{figure}

From Eq.~\eqref{eq:YB_sexaquarks} we obtain the sexaquark yield after solving for the chemical potential and reinserting in the number density, giving ${n_S^\text{eq, low T} = (Y_B - Y_B')/(2s(T))}$. We show the evolution of the (anti)sexaquark equilibrium abundance in Fig.~\ref{fig:alternate_eq}, for different choices of the temperature $T_\ast$ at which (anti)sexaquarks chemically decouple from baryons and no longer efficiently exchange baryon number. Note that the equilibrium yield of octet baryons does not change appreciably compared to Fig.~\ref{fig:equilibrium_abundance}.

We observe that even for values of $T_\ast$ that are very high (including unphysically high values before the QCD crossover, where the hadrons are not the relevant degrees of freedom), the sexaquark yield is always at least 2-3 orders of magnitude below that required to obtain the full dark matter abundance. This reflects the fact, shown in Fig.~\ref{fig:asymmetry}, that while the relation ${\mu_S=2\mu_B}$ is maintained, the (anti)sexaquarks always carry only a small fraction of the overall cosmic baryon asymmetry, due to the presence of lighter degrees of freedom that also carry baryon number. In Appendix~\ref{app:asymmetric}, we present the required dark matter chemical potential as a function of its mass to obtain all of dark matter. The sexaquark would require a baryon number $\mathcal{O}(10)$ to make up the full dark matter abundance.

\section{\label{sec:freezeout}Freeze-out of the (anti)sexaquark abundance}

Let us now consider how the (anti)sexaquark relic abundance is set by freeze-out from the equilibrium histories discussed in the previous section. We consider several illustrative interaction processes which should be the dominant channels and/or give rise to detectable signals, and show how they span the range of possible outcomes. We largely follow the methods of Refs.~\cite{Gross:2018ivp,Kolb:2018bxv}, but extend these studies by considering a continuum of scenarios where the processes that produce and deplete sexaquarks can be severely suppressed, as suggested in Ref.~\cite{Farrar:2018hac}. This leads us to account for a wider range of processes than these previous works, since if one process is highly suppressed, others may be relevant. 

In this calculation, we focus on the predictive scenario where (anti)sexaquarks are in chemical equilibrium with the Standard Model bath, with interactions that couple their chemical potential to that of the baryons (as discussed in Sec.~\ref{subsec:EQ_abundance}), for at least a short time at the end of or after the QCD crossover (this also covers the case where ${\mu_S\approx 2\mu_B}$ at the end of the crossover, even if this relation is not enforced by equilibrium). We will calculate the limits this assumption (of some form of post-crossover equilibrium) imposes on the cross sections for various processes. In this case we lose sensitivity to any difficult-to-model details of the crossover. At later times the interactions may freeze out completely, or alternatively interactions that change the sexaquark number density but not the net baryon number in (anti)sexaquarks may remain fast, leading to the equilibrium evolution discussed in Sec.~\ref{subsec:equilibrium}.

We do not study further the cosmology of the case where there is no depletion of (anti)sexaquarks at all after the QCD crossover. In that case, for the purpose of setting constraints in later sections, we only assume that the sexaquark and antisexaquark abundances are close to equal (based on the smallness of the chemical potential during the quark-hadron transition) and that the Universe is not overclosed. We likewise only briefly discuss (in Sec.~\ref{sec:earlier}) the case where $\mu_S$ is very different from $2\mu_B$ due to (unknown) processes occurring during the quark-hadron crossover, and then ${\mu_S=2\mu_B}$ is not restored after the crossover, but self-annihilation remains fast and efficiently depletes the antisexaquark abundance. However, this case only differs from cases we do study (where ${\mu_S=2\mu_B}$ is fixed at the quark-hadron transition but not thereafter, and self-annihilation is rapid) by a rescaling of the initial condition, and so many of our results can be extrapolated.

\subsection{\label{subsec:FO_formalism}Freeze-out formalism}

Sexaquarks and antisexaquarks eventually depart from their equilibrium distribution once their interaction rate with the baryon-photon bath falls below the Hubble rate. In this section, we analyze this freeze-out behavior in detail. However, some key results can already be concluded from Fig.~\ref{fig:equilibrium_abundance} and the discussion in Sec.~\ref{sec:abundance}, in the event where baryon-number-exchanging processes dominate and hence ${\mu_S=2\mu_B}$ until freeze-out, as assumed in previous studies:

\begin{itemize}
    \item The sharp flattening in the sexaquark equilibrium abundance at ${T \lesssim 40}$~MeV, due to the onset of the baryon chemical potential, means that the final relic abundance will be quite insensitive to the time of freeze-out if it occurs after this point. Thus we expect a fairly narrow band of predictions for the relic abundance for any scenario where there is at least one process fast enough to maintain sexaquarks in chemical equilibrium with the bath (including ${\mu_S=2\mu_B}$) down to ${T \sim 40}$~MeV. This sharp prediction for the abundance is roughly 10 orders of magnitude below the relic abundance, with the possibility of slightly higher values for light sexaquarks that remain in equilibrium to late times (${T \sim 1}$~MeV). In this scenario, sexaquarks decouple earlier than antisexaquarks from the baryon-photon bath.

    \item In contrast, at ${T \gtrsim 40}$~MeV, the abundance depends sensitively on the freeze-out temperature for baryonic number-changing processes, and hence on the cross section for these processes. In this regime, we also expect very similar initial yields of sexaquarks and antisexaquarks and minimal dependence on the exact sexaquark mass. A sufficiently strong self-annihilation cross section may maintain a partial equilibrium after baryon-number-exchanging processes freeze out, resulting in a relic abundance determined by the annihilation cross section (in the symmetric regime where annihilations are too weak to efficiently deplete all the antisexaquarks) or by the freeze-out temperature for baryonic interactions (in the asymmetric regime with strong annihilation). To achieve the observed dark matter density, the freeze-out of sexaquarks should take place at a relatively high temperature, approximately ${T \sim 90}$~MeV, as can be concluded from Fig.~\ref{fig:equilibrium_abundance}, and there should be no subsequent efficient self-annihilation.
\end{itemize}

In the remainder of this section, we quantitatively map out the behavior in these two regimes in terms of the processes responsible for maintaining chemical equilibrium. We assume that all strongly-interacting states other than the sexaquark maintain chemical equilibrium throughout the sexaquark freeze-out and follow their equilibrium distributions with appropriate chemical potentials.

The reactions we consider in the freeze-out comprise the sexaquark self-annihilation ${S \bar{S} \to Y\bar{Y}}$ where $Y,\bar{Y}$ represent Standard Model final states, the breakup reaction ${S X \to b b'}$, and the annihilation reaction ${S \bar{b} \to b'X}$, as well as the counterpart reactions for antisexaquarks. As previously, $X$ represents (possibly multiple) particles which do not carry baryon number, such as mesons and photons, and $b,b'$ are octet baryons. The second reaction is expected to dominate over the third, since the comoving number density of light states $X$ is enhanced compared to baryons, and was the primary process studied in previous works \cite{Gross:2018ivp, Kolb:2018bxv}. However, the third interaction has the potential for interesting experimental signatures at late times, which will be addressed in Sec.~\ref{sec:SuperK}. If the second and third reactions are very small but the first remains fast, we find ourselves in the asymmetric dark matter scenario discussed in the previous subsection. For the self-annihilation, we assume that the sexaquark mass is greater than the $Y$ mass, making the reaction exothermic. For the breakup reaction, we consider strangeness- and isospin-conserving possible final states $bb' = \lbrace \Lambda\Lambda, \Lambda\Sigma^0, \Sigma^0\Sigma^0, \Sigma^+ \Sigma^- , n \Xi^0, p \Xi^- \rbrace$ and sum over all these channels. The exothermic reactions involve both two- and three-body final states (e.g. ${\Lambda\Lambda \to S\pi^0\pi^0}$, ${\Lambda\Sigma^0 \to S\pi^0}$), but mesons stay abundant and in thermal equilibrium with the baryon-photon bath throughout the process, and thus we do not need to track their abundance explicitly in these equations. For the annihilation cross section, we also only consider the strangeness- and isospin-conserving reactions. For ease of reading, we label these cross sections by their initial states including the (anti)sexaquark. 

The three reactions we consider have initial/final states with baryon number 2 (${S X \to b b'}$), 1 (${S \bar{b} \to b' X}$), or 0 (${S \bar{S} \to Y\bar{Y}}$). Sexaquarks are likely engaged in further interactions with larger total baryon number, but we expect these interactions to be suppressed.

The Boltzmann equations governing the freeze-out of sexaquarks and antisexaquarks then read as
\begin{align}\label{eq:Boltzmann}
    \dv{n_S}{t} + 3 H n_S &= - \ev{\sigma v}_{S \bar{S} }^\text{ann} (n_S n_{\bar{S}} - n_S^\text{eq} n_{\bar{S}}^\text{eq})\\
    & - \sum_{b,b'} \big[ \Gamma_{S X }^\text{fo} + \Gamma_{S \bar{b} }^\text{fo} \big] \big[ n_S - n_S^\text{eq} \big] \ ,\nonumber \\
    \dv{n_{\bar{S}}}{t} + 3 H n_{\bar{S}} &= - \ev{\sigma v}_{S \bar{S} }^\text{ann} (n_S n_{\bar{S}} - n_S^\text{eq} n_{\bar{S}}^\text{eq})\\
    & - \sum_{b,b'} \big[ \Gamma_{{\bar{S}} X }^\text{fo} + \Gamma_{{\bar{S}} {b} }^\text{fo} \big] \big[ n_{\bar{S}} - n_{\bar{S}}^\text{eq} \big] \ ,\nonumber
\end{align}
where ${H(T)=1.66 g_\ast^{1/2}(T) T^2/m_\text{Pl}}$ is the Hubble parameter, $\Gamma_{S X }^\text{fo}$ is the thermally-averaged breakup rate to two baryons, and $\Gamma_{\bar{S}b}^\text{fo}$ is the thermally-averaged annihilation rate. All thermally-averaged cross sections $\ev{\sigma v}^\text{ann}$ are assumed to be energy-independent $s$-wave annihilation. The rates are given in the non-relativistic limit by
\begin{align}\label{eq:<sigma_v>}
    \Gamma_{S X}^\text{fo} &= \ev{\sigma v}_{b b' \to SX}^\text{ann} \frac{n_b^\text{eq} n_{b'}^\text{eq}}{n_S^\text{eq}} \ ,\\
    \Gamma_{\bar{S}b}^\text{fo} &= \ev{\sigma v}_{\bar{S}b \to \bar{b}' X}^\text{ann} n_{b}^\text{eq} \ .\label{eq:<sigma_v>_Bbar}
\end{align}
The first reaction receives contribution from various pairs $bb'$ of baryons, except for temperatures below $10$~MeV where $\Lambda\Lambda$ becomes largely dominant. The second reaction is dominated throughout the temperature range of interest by the proton and neutron contributions.

\subsection{\label{subsec:parameterization}Parameterization of cross sections}

We parameterize the cross sections as ${\ev{\sigma v}_{b b' \to SX}^\text{ann} =A m_\pi^{-2}}$ in alignment with \cite{Kolb:2018bxv}, and ${\ev{\sigma v}_{\bar{S}b \to \bar{b}' X}^\text{ann}=B m_\pi^{-2}}$. Here, $A$ and $B$ are dimensionless number that we varied between unity, corresponding to a strong interaction, and $10^{-19}$ for $A$ ($10^{-17}$ for $B$), which is where $\Gamma/H$ falls below unity at $T = 155$~MeV. Lower values would imply that sexaquarks and antisexaquarks never reach equilibrium with the baryon-photon bath after the QCD crossover.

While $m_\pi^{-2}$ is a typical scale for strong interaction processes, some studies suggest that the sexaquark might exhibit a significantly suppressed interaction with baryons which could arise from a small radius and minimal wavefunction overlap during an interaction~\cite{Farrar:2022mih, Farrar:2023wta}. For instance, using an estimate of $0.2$~fm ($0.15$~fm) for the sexaquark' spatial extent, Ref.~\cite{McDermott:2018ofd} finds that the thermally-averaged cross section $S\gamma \to \Lambda\Lambda$ could be suppressed by a factor of $10^{-10}$ ($10^{-12}$). Ref.~\cite{Farrar:2023wta} proposes a theory band for the suppression factor of sexaquark-baryon-baryon cross sections extending down to $\sim 10^{-20}$. We expect $A$ to be parametrically similar to this suppression factor, up to some other numerical factors.

For the self-annihilation cross section, $\ev{\sigma v}_{S \bar{S}}^\text{ann}$, the geometric cross section would be $\mathcal{O}(\pi r_S^2) \sim 10^{-17}$ cm$^3$/s, taking the sexaquark radius to be 0.15~fm \cite{McDermott:2018ofd,Farrar:2018hac}. The actual cross section will depend on the effective sexaquark couplings, and we treat it as a free parameter.

The (anti)sexaquark breakup rate parameterized by $A$ scales with temperature as ${\Gamma_{S X}^\text{fo} = \Gamma_{\bar{S} X}^\text{fo} \sim A \ T^{3/2} e^{-B_S/T}}$, while the reaction parameterized by $B$ shows different scalings due to the chemical potential; for ${T > 50}$~MeV, it is ${\Gamma_{S \bar{b}}^\text{fo} \sim \Gamma_{\bar{S} b}^\text{fo} \sim B T^{3/2} e^{-m_p/T}}$, while for ${T < 50}$~MeV, we have
\begin{align}
    \Gamma_{S \bar{b}}^\text{fo} &\sim B \ T^{3/2} e^{-2m_p/T} \ , \\
    \Gamma_{\bar{S} b}^\text{fo} &\sim B \ T^{3/2}  \ . \label{eq:rate_Sbarb_FO}
\end{align}
For temperatures less than 10~MeV, the yield of baryons other than protons and neutrons becomes exponentially suppressed. Freeze-out occurs quickly unless the cross sections parameterized by $A$ and $B$ are exponentially large, except for antisexaquarks freezing out through $B$, which have a smoother temperature dependence (Eq.~\eqref{eq:rate_Sbarb_FO}) and can freeze out later. In this case the antisexaquark abundance is always negligible.

One might ask whether this parameterization could miss important temperature dependence in the cross sections. The variation in the temperature during freeze-out is modest compared to the parameter ranges we consider for $A$ and $B$, but for the $B$ parameter we will later extrapolate to annihilation in the present day, where the temperature difference could be relevant. We will assume that the relevant matrix elements are not momentum-dependent and so any temperature dependence should arise only from phase-space effects; we expect $\ev{\sigma v}^\text{ann}$ for two-body exothermic processes to be roughly constant unless the mass energy liberated in the interaction is small enough to be comparable to the kinetic energy of the incoming particles. This is never true for the process controlled by $B$, but may be true for the process controlled by $A$; however, at most this should be a $\mathcal{O}(1)$ effect.

\subsection{\label{subsec:FO_results}Freeze-out results}

Let us begin by considering the scenario where only the process controlled by $A$ is important and we effectively set $B$ and $\ev{\sigma v}_{S\bar{S}}^\text{ann}$ to zero; this approach will allow us to check consistency with previous studies in the large-$A$ limit~\cite{Kolb:2018bxv}.

\begin{figure}[ttt]
    \centering
    \includegraphics[width=\columnwidth]{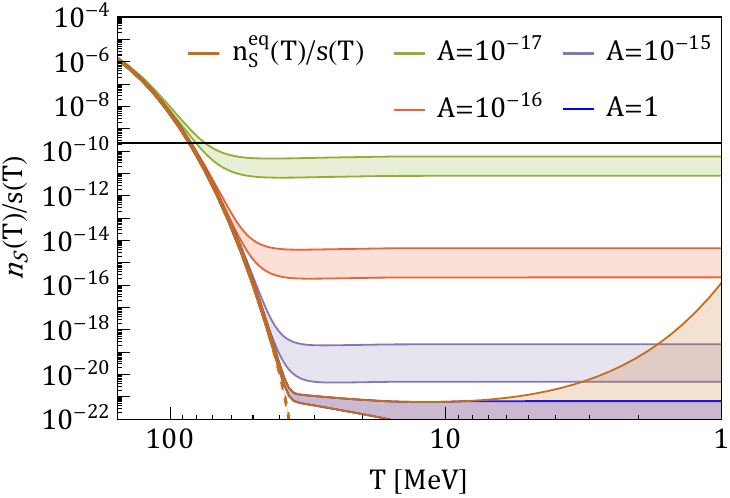}
    \caption{Freeze-out yield of sexaquarks as a function of their mass and cross section. The colors correspond to the choice of thermally averaged cross section, parameterized by ${\ev{\sigma v}_{SX}^\text{ann} = A m_{\pi}^{-2}}$ with ${A=\lbrace 10^{-17}, 10^{-16}, 10^{-15}, 1 \rbrace}$, and the width of each color band indicates the sexaquark mass range $1860-1890$~MeV. Smaller cross sections result in earlier freeze-out and a larger abundance closer to the expected relic dark matter abundance. The black line represents the observed dark matter yield.}
    \label{fig:freeze_out_ni}
\end{figure}

For this case, we show in Fig.~\ref{fig:freeze_out_ni} the evolution of the sexaquark abundance for a number of different example cross sections, illustrating the stable yield across a large range of values of $A$, from ${A=1}$ down to ${A = 10^{-15}}$. We see that in these cases the freeze-out occurs at different temperatures, but the final relic abundance is quite similar due to the flatness of the equilibrium abundance with respect to temperature. In contrast, decreasing $A$ from $10^{-15}$ down to $10^{-17}$ induces a sharp change in the relic abundance, due to the steep scaling of the equilibrium abundance with temperature above ${T \sim 40}$~MeV.

\begin{figure} 
\begin{minipage}{\columnwidth} 
\includegraphics[width=\columnwidth]{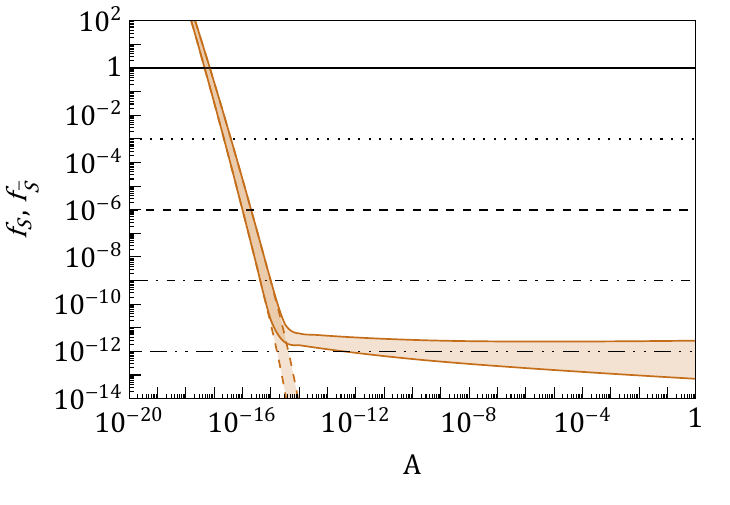} 
\end{minipage} 

\vspace{0.05cm}
\begin{minipage}{\columnwidth} 
\includegraphics[width=\columnwidth]{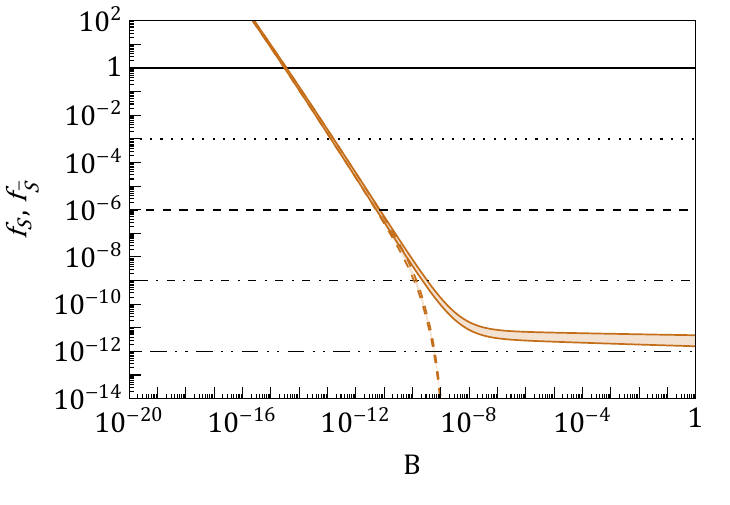} 
\end{minipage} 
\begin{minipage}{\columnwidth} 
\includegraphics[width=\columnwidth]{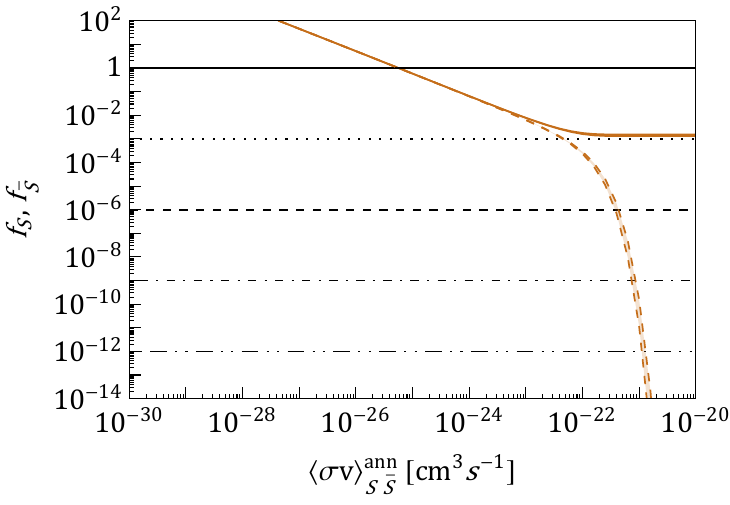} 
\end{minipage} 

\caption{Fraction of the total dark matter density composed of sexaquarks $f_{S/\bar{S}} = \Omega_{S/\bar{S}} / \Omega_\text{DM}$ as a function of the thermally averaged cross sections (top) $\ev{\sigma v}_{SX\to bb'}^\text{ann} = A m_\pi^{-2}$, (middle) $\ev{\sigma v}_{\bar{S}b \to \bar{b}' X}^\text{ann} = B m_\pi^{-2}$, and (bottom) $\ev{\sigma v}_{S \bar{S}}^\text{ann}$ for ${T_\ast = 150}$~MeV. See Fig.~\ref{fig:alternate_eq} and text for details. The horizontal black lines represent the values of $f_\chi$ for ${f_\chi=1}$~(full), $10^{-3}$~(dotted), $10^{-6}$~(dashed), $10^{-9}$~(dot-dashed), and $10^{-12}$~(dot-dot-dashed). The width of the band corresponds to the range of sexaquark mass ${1860 - 1890}$~MeV. The dashed orange lines correspond to the antisexaquark freeze-out abundance.}
\label{fig:freeze_out_abundance_fchi}
\end{figure}

In the top panel of Fig.~\ref{fig:freeze_out_abundance_fchi}, we plot the behavior of $f_S$ and $f_{\bar{S}}$, the final relic sexaquark abundance as a fraction of the dark matter density, as a continuous function of $A$. We observe two distinct regimes, associated with freeze-out at temperature above and below ${T \sim 40}$~MeV respectively; we obtain analytic descriptions of the $A$-dependence of $f_S$ in these regimes in Appendix~\ref{sec:scaling}. The steep decline of $f_S$ at small $A$ follows the scaling ${f_S \propto A^{-m_S/B_S}}$, with the binding energy $B_S$ being ${B_S = 2m_\Lambda - m_S}$; we observe that because ${B_S \ll m_S}$, this is a much steeper scaling than the inverse dependence of the relic abundance on the annihilation cross section in the standard thermal freeze-out calculation. At larger $A$, corresponding to decoupling at ${T \lesssim 40}$~MeV, we find that $f_S$ is nearly independent of $A$, because the equilibrium abundance varies only slowly with temperature in this regime, as discussed in Sec.~\ref{subsec:FO_formalism}. However, the value of $f_S$ does depend somewhat on the sexaquark mass, shown by the broad region at the bottom of the top panel of Fig.~\ref{fig:freeze_out_abundance_fchi}, as anticipated from Fig.~\ref{fig:equilibrium_abundance}.

To compare our results with previous literature, we note that both previous studies of the sexaquark freeze-out considered a relatively large cross section (${10^{-29} - 10^{-27}}$~cm${}^2$ in \cite{Gross:2018ivp} and ${10^{-36} - 10^{-24}}$~cm${}^2$ in \cite{Kolb:2018bxv}), i.e. ${A \in \lbrace 10^{-11},10^2 \rbrace}$. This range does not allow for sexaquarks in the preferred mass range to reach the full observed dark matter abundance.

No previous work on the sexaquark has explicitly considered the annihilation interaction $S\bar{S}$ and the cross section $S\bar{b}$ ($\bar{S}b$). We now consider setting $A$ to zero and activating these other freeze-out interactions. We take $T_\ast = 150$~MeV for the case where $\ev{\sigma v}_{S\bar{S}}^\text{ann}$ dominates, which maximizes the sexaquark abundance. The lower panels of Fig.~\ref{fig:freeze_out_abundance_fchi} show these cases. We found that with all these interactions we can obtain the full dark matter abundance in sexaquarks, which would yield a symmetric population of sexaquarks and antisexaquarks, since in all these cases the freeze-out occurs early while the Universe is still symmetric. The scenario where $\ev{\sigma v}_{S\bar{S}}^\text{ann}$ controls the freeze-out corresponds to the standard thermal relic calculation. Table~\ref{tab:freeze-out_summary} presents a summary of the approximate freeze-out parameter values (1) which lead to the full dark matter abundance when the other parameters are set to zero, (2) corresponding to the minimum value to equilibrate after the QCD crossover, and (3) corresponding to the maximum value for which the population after freeze-out is symmetric

\begin{table}[ht]
    \centering
    \caption{\label{tab:freeze-out_summary}Approximate values of the freeze-out parameters that, if the other parameters are set to zero, correspond to (1) the minimum value to equilibrate after the QCD crossover, (2) the value that gives the full dark matter abundance ${f_\chi = f_S + f_{\bar{S}} = 1}$, and (3) the maximum value where the sexaquark and antisexaquark relic abundances are comparable ($f_S = f_{\bar{S}}$).}
    \begin{tabular}{l c c c}
         \hline
         & $A$ ($bb' \to SX$) & $B$ ($\bar{S}b \to \bar{b}'X$) & $\ev{\sigma v}_{S\bar{S}}^\text{ann}$ [$\text{cm}^3/\text{s}$] \\ \hline
         min & $10^{-19}$ & $10^{-17}$ & $10^{-28}$ \\
         $f_\chi = 1$ & $10^{-17}$ & $10^{-15}$ & $10^{-26}$ \\
         $f_S = f_{\bar{S}}$ & $10^{-15}$ & $10^{-10}$ & $10^{-22}$ \\
         \hline
    \end{tabular}
\end{table}

More broadly, we explored the effect of varying $A$ and $B$, and $\ev{\sigma v}_{S\bar{S}}^\text{ann}$ on the freeze-out abundance over wide range of parameter values with one or more of these interactions turned on. When all of these cross sections are smaller than the minimum values presented in Table~\ref{tab:freeze-out_summary}, the (anti)sexaquarks do not equilibrate with the baryon-photon bath. Conversely, when any one of these cross sections is larger than the maximum value where the sexaquark and antisexaquark relic abundances are comparable ($f_S = f_{\bar{S}}$), the chemical potential becomes effective at depleting the antisexaquark yield. For higher cross sections, the abundance of antisexaquarks quickly becomes negligible, while the abundance of sexaquarks becomes almost insensitive to the variation of the parameters $A$ and $B$, and $\ev{\sigma v}_{S\bar{S}}^\text{ann}$. Varying these parameters over 10 orders of magnitude results in a change of the abundance of less than a few percent. For cases where the processes controlled by $A$ or $B$ remain fast (compared to Hubble) to this point, the result is a relic abundance that is significantly too small to explain all the dark matter ($f_S \sim 10^{-11}$), as noted in Ref.~\cite{Kolb:2018bxv}. If instead $A$ and $B$ are both small and only $\ev{\sigma v}_{S\bar{S}}^\text{ann}$ drives the antisexaquark abundance, the abundance can plateau as high as ${f_S \sim 10^{-3}}$ of the total dark matter abundance, depending on the temperature when the processes controlled by $A$ and/or $B$ decoupled, due to following the alternative equilibrium curve described in Sec.~\ref{subsec:equilibrium}. In this scenario, sexaquarks annihilate with antisexaquarks rather than the baryon and photon bath, and as soon as the antisexaquarks abundance depletes due to the chemical potential, no number-changing reaction can occur for the
sexaquarks.

The opposite limit, when all of $A$, $B$, $\ev{\sigma v}_{S\bar{S}}^\text{ann}$ are very small, is where the full dark matter relic abundance can be obtained. In this scenario, the freeze-out process occurs at higher temperatures, leading to a larger relic abundance, which is less sensitive to the dark matter mass. As expected, in this scenario the dark matter abundance is composed of approximately equal fractions of sexaquarks and antisexaquarks since the antisexaquarks have not been depleted by the time freeze-out occurs in the early Universe. 

\begin{figure*}[ttt]
\centering
\begin{minipage}[b]{0.43\textwidth}
        \centering
        \includegraphics[width=\textwidth]{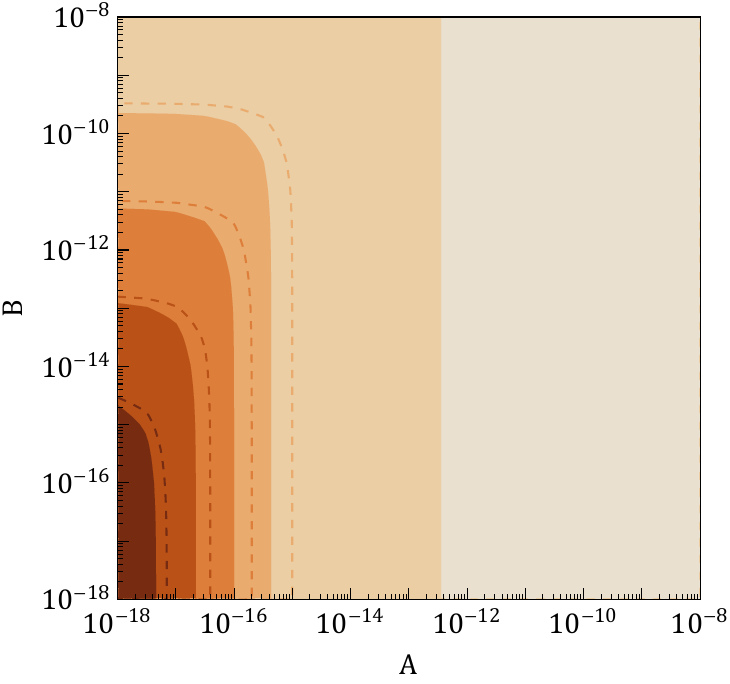} 
  \end{minipage}~~
  \begin{minipage}[b]{0.43\textwidth}
    \centering
    \includegraphics[width=\textwidth]{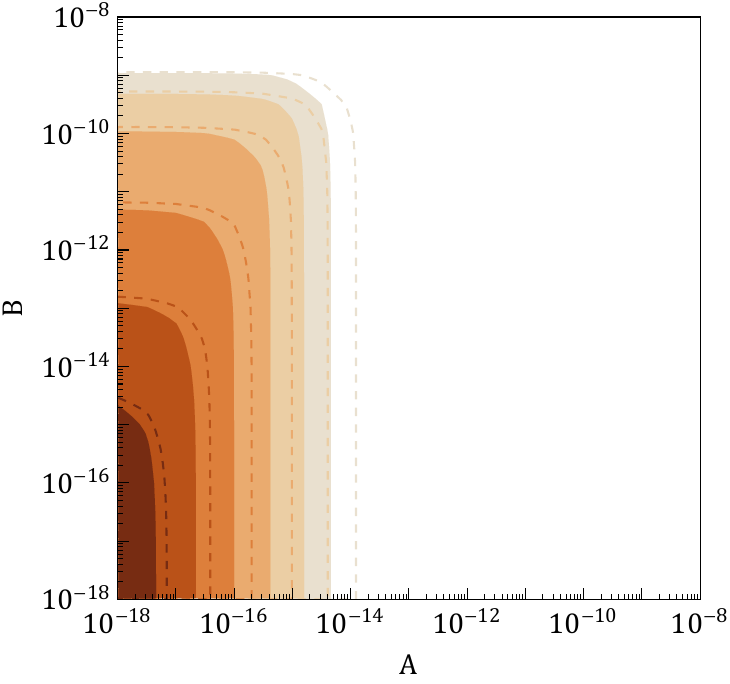}
  \end{minipage}~~
\begin{minipage}[t]{0.14\textwidth}
    \centering
    \includegraphics[width=.6\textwidth]{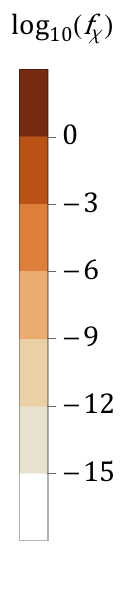}
\end{minipage}
  \caption{Fraction of the total dark matter density composed of sexaquarks $f_{S/\bar{S}} = \Omega_{S/\bar{S}}/\Omega_\text{DM}$ as a function of the thermally averaged annihilation cross sections parameterized by $\ev{\sigma v}_{S X \to bb'}^\text{ann} =A m_\pi^{-2}$ and $\ev{\sigma v}_{\bar{S} b \to \bar{b}'X}^\text{ann} = Bm_\pi^{-2}$. The {\it left} panel shows the abundance of sexaquarks, and the {\it right} panel the abundance of antisexaquarks. The contours (dashed lines) are for 1890~MeV (1860~MeV) (anti)sexaquarks.}
  \label{fig:freeze_out_AB}
\end{figure*}

We show the effect of simultaneously turning on $A$ and $B$ in Fig.~\ref{fig:freeze_out_AB} for 1860 and 1890~MeV sexaquarks. Generally, either one of them is important. The sexaquark population never gets depleted below $f_S \sim 10^{-12}$, such that no white space is visible.

Our findings indicate that sexaquarks can only produce a substantial fraction of the dark matter abundance when $A,B$ are much smaller than the typical strong interaction scale. The suppression by up to 12 orders of magnitude claimed to be due to the small radius and minimal wavefunction overlap is not sufficient to reach the required values on the order of $10^{-17}$ needed to account for all of dark matter. There may be an additional tunneling factor coming from the six-quark configuration change between the sexaquark and a pair of baryons. This factor could be as small as $10^{-4}$ which can bring the cross section further from the strong interaction scale~\cite{Farrar:2023wta}.

Additionally, while previous studies of the sexaquark have treated it as a pure flavor singlet \cite{Farrar1708, Farrar:2022mih, Farrar:2023wvm}, as discussed in Appendix~\ref{sec:group} we would expect flavor symmetry breaking to induce mixing with other representations\footnote{We thank Bob Jaffe for this point.}. This mixing could mitigate the strong suppression found by these earlier references.

In the first panel of Fig.~\ref{fig:freeze_out_abundance_fchi}, where the freeze-out is controlled by $A$, the contribution from the other thermally averaged annihilation cross sections can be neglected for $B \leq 10^{-16}$, $\ev{\sigma v}_{S \bar{S}}^\text{ann} \leq 10^{-26}~\text{cm}^3/\text{s}$. For the second panel, the results are unaffected (i.e. the assumption that $B$ controls the freeze-out is valid) for $A \leq 10^{-18}$, $\ev{\sigma v}_{S \bar{S}}^\text{ann} \leq 10^{-27}~\text{cm}^3/\text{s}$. Finally, the results in the third panel are valid for for $A \leq 10^{-19}$ and $B \leq 10^{-17}$.

Some other works on the sexaquark have considered a wider mass range, extending to 2054~MeV. A heavier mass would give a more suppressed equilibrium abundance, as well as a lower freeze-out abundance for the same annihilation rates, and would not qualitatively change our conclusions.

\subsection{\label{sec:earlier}Comparison with earlier studies}

Comparing our results with previous studies, we note some discrepancies related to the requirements for sexaquark chemical equilibration at the end of the QCD phase transition and wish to address some of these differences. We also discuss in this section the expected values of our cross section parameters in the framework of Ref.~\cite{Farrar:2023wta}.

Notably, Ref.~\cite{Kolb:2018bxv} finds that a freeze-out solely parameterized by $A$ requires ${A \leq 10^{-20}}$ to avoid (anti)sexaquarks equilibrating with the baryon-photon bath after the QCD phase transition. In comparison, we find a slightly milder limit of ${A \leq 10^{-19}}$, based on requiring $\Gamma/H > 1$ at $T=155$~MeV. Part of this discrepancy may be due to assuming a slightly different evolution for $g_*$ during and after the QCD phase transition (we find closer agreement with the equilibrium evolution of $\Gamma/H$ at lower temperatures). Since $A$ in any case only parameterizes our ignorance of the cross sections and even in the best case would have factor-of-few uncertainties, we do not consider this difference critical.

More significantly, Ref.~\cite{Farrar:2023wta} states that the dominant $S$-number changing process for freeze-out is given by ${K^+ S \to p \Lambda}$, comparing the rate of this process with ${X S \to b b'}$ where $X$ is one or two pions or a photon. In contrast, we agree with Ref.~\cite{Kolb:2018bxv} that the two-pion process dominates. We suspect that Ref.~\cite{Farrar:2023wta} may have extracted the scaling for the rate from the arXiv version v1 of Ref.~\cite{Kolb:2018bxv}, which contains a typo in the estimated rate for $\Gamma_{\Lambda\Lambda \to S \pi\pi}$, with the exponential factor scaling as $e^{-(2m_\Lambda + m_S)/T}$. We note that the published version of this work is correct, with an exponential factor of $e^{-(2 m_\Lambda - m_S)/T}$. This means that Ref.~\cite{Farrar:2023wta} artificially suppresses the rate of this process by a factor of $e^{-2m_S/T} \ll 1$ (the suppression exceeds 10 orders of magnitude at 155~MeV). Using the corrected rate, we find that the process ${X S \to b b'}$ requires a much smaller coupling than ${K^+ S \to p \Lambda}$ to avoid equilibrating with baryons in the bath.

Ref.~\cite{Farrar:2023wta} parameterizes the interactions of the sexaquark with hadrons in terms of an effective Lagrangian, 
\begin{equation}
    \mathcal{L}_\text{eff} \supset \frac{\tilde{g}}{\sqrt{40}} \bar{\psi}_B \gamma^5 \psi_{B^{\prime}}^c \phi_S + g_{SSV} \phi^\dagger_S \partial_\mu \phi_S V^\mu + h.c. \ , \label{eq:hadlagrangian}
\end{equation}
where $\psi_B$ describes baryons, $V^\mu$ describes vector mesons, and $\phi_S$ is the sexaquark field.

In this framework, the rate of ${\pi \pi S \to \Lambda \Lambda}$ is controlled by $\tilde{g}$; Ref.~\cite{Farrar:2023wta} takes our rate parameter $A$ to be equal to $\tilde{g}^2/40$ for this process. By reproducing their calculation with the corrected rate, we found that the required value of the coupling to avoid equilibration after the QCD phase transition is ${\tilde{g} \lesssim 4 \times 10^{-9}}$, in broad agreement with our result that $A\lesssim 10^{-19}$ is needed to avoid equilibration when including all processes, and contrasting with the original claim of Ref.~\cite{Farrar:2023wta} that only $\tilde{g} \lesssim 2\times 10^{-6}$ is required. The new criterion excludes most (but not quite all) of the theory prediction band for $\tilde{g}$ given in Ref.~\cite{Farrar:2023wta} for the mass range we consider, implying that a strong tunneling suppression to $\tilde{g}$ would be needed in addition to a suppression from the overlap of baryon and sexaquark wavefunctions. This prediction also assumes the sexaquark is a pure flavor singlet, which would be unexpected in the presence of flavor symmetry breaking, as we discuss in Appendix~\ref{sec:group}.

One reason there has been considerable interest in the possibility that sexaquarks never couple to the baryons after the quark-hadron transition is the argument from Ref.~\cite{Farrar:2018hac} that in this case the correct abundance of dark matter can be naturally obtained. In reviewing \cite{Farrar:2018hac}, we identified a numerical error in the latest arXiv version (v3) (which also persists in \cite{Farrar:2020zeo}), which we have discussed with the author and confirmed as a bug. Correcting this issue modestly decreases the inferred sexaquark abundance generated during the crossover, giving $\Omega_\text{DM}/\Omega_B \sim 2.5$ instead of $5.3$. This result still uses the expression given in Ref.~\cite{Farrar:2018hac} for the efficiency with which strange quarks are confined in sexaquarks, 
\begin{equation}
    \kappa_s(m_S, T) = \frac{1}{1 + (r_{\Lambda,\Lambda} + r_{\Lambda,\Sigma} + 2 r_{\Sigma,\Sigma} + 2 r_{N,\Xi})} \ , \label{eq:efficiency}
\end{equation}
where $r_{1,2}\equiv \text{exp}\left[-(m_1 + m_2 - m_S)/T\right]$. The author has indicated to us that there should be an additional factor of 2 or 4 preceding the bracketed factor in the denominator for the spin degrees of freedom of the baryons~\cite{Farrar:private_comm}, which would further decrease the sexaquark yield. More generally, since the goal of this equation is to determine the partition of strange quarks among the possible hadronic states, it is not clear to us why the denominator should omit color-singlet states containing strange quarks that do not share the full quark content and quantum numbers of the sexaquark (we also note that weak decays that do not preserve strangeness are already fast relative to Hubble at this temperature).

In the analysis of Sec.~\ref{subsec:equilibrium}, where we assume ${\mu_S=2\mu_B}$ down to some cutoff temperature $T_*$, we obtain Eq.~\eqref{eq:ratioYB} describing the fraction of the baryon asymmetry stored in (anti)sexaquarks, as an analogue to Eq.~\eqref{eq:efficiency}. We have confirmed that taking $T_*\sim 150$~MeV and substituting into this equation yields ${\Omega_\chi/ \Omega_\text{DM} \approx 10^{-3}}$, consistent with our numerical calculation (as presented in Fig.~\ref{fig:alternate_eq}); the strong suppression relative to the original result of Eq.~\eqref{eq:efficiency} is due to the large number of degrees of freedom lighter than the sexaquark that can carry baryon number, and analogously we would expect a suppression to $\kappa_s$ due to degrees of freedom lighter than the sexaquark that contain strange quarks. Ref.~\cite{Andronic:2017pug} cites experimental results showing good agreement between hadron yields from the quark-hadron transition in data from ALICE, and a simple statistical model where the yield per degree of freedom is proportional to $m^{3/2} e^{-m/T}$, i.e. to the non-relativistic number densities. If we evaluate the expected fraction of strange quarks contained in sexaquarks versus baryons according to these yields, we obtain:
\begin{equation}
    \frac{2 n_S}{\sum_b s_b n_b} \approx 0.003 \ . 
\end{equation}
If we also included strangeness in kaons, this ratio would drop to $\sim 7 \times 10^{-4}$.

We also emphasize that the key quantity may not be per se the efficiency with which strange quarks are confined in sexaquarks, but the efficiency with which strange quark {\it asymmetry} is confined in (anti)sexaquarks, as we will show in Sec.~\ref{sec:SuperK} that if antisexaquarks are not efficiently depleted in the early universe, they are expected to have very striking signals in the present day. Our calculation indicates that the chemical potential for sexaquarks would need to be enhanced by three orders of magnitude at the end of the QCD phase transition, relative to the equilibrium expectation $\mu_S=2\mu_B$, in order for this asymmetry to generate the full dark matter abundance (see also Appendix~\ref{app:asymmetric}).

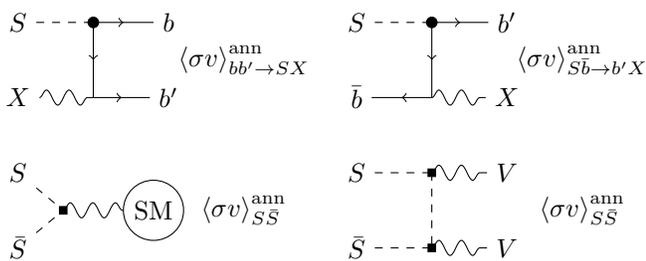
\begin{figure}[htt]
\centering
\begin{tikzpicture}

\begin{scope}[shift={(0,0)}]
\node (s) at (0,1) {$S$};
\coordinate (sb) at (1,1);
\node (b) at (2,1) {$b$};
\node (v) at (0,0) {$X$};
\coordinate (vbp) at (1,0);
\node (bp) at (2,0) {$b'$};
\draw[dashed] (s) -- (sb);
\draw[fermion] (sb) -- (b);
\draw[vector] (v) -- (vbp);
\draw[fermion] (vbp) -- (bp);
\draw[fermion] (sb) -- (vbp);
\filldraw [black] (1,1) circle (2pt);
\node at (3,0.5) {$\ev{\sigma v}_{bb' \to SX}^\text{ann}$};
\end{scope}

\begin{scope}[shift={(4.5,0)}]
\node (s2) at (0,1) {$S$};
\coordinate (sb2) at (1,1);
\node (b2) at (2,1) {$b'$};
\node (v2) at (2,0) {$X$};
\coordinate (vbp2) at (1,0);
\node (bp2) at (0,0) {$\bar{b}$};
\draw[dashed] (s2) -- (sb2);
\draw[fermion] (sb2) -- (b2);
\draw[fermion] (sb2) -- (vbp2);
\draw[fermion] (vbp2) -- (bp2);
\draw[vector] (vbp2) -- (v2);
\filldraw [black] (1,1) circle (2pt);
\node at (3,0.5) {$\ev{\sigma v}_{S \bar{b} \to b'X}^\text{ann}$};
\end{scope}

\begin{scope}[shift={(0,-2)}]
\node (s5) at (0,1) {$S$};
\node (sbar3) at (0,0) {$\bar{S}$};
\coordinate (ssbar) at (0.6,0.5);
\node (sm) at (1.8,0.5) {SM};
\coordinate (smsmbar) at (1.4,0.5);
\draw[vector] (ssbar) -- (smsmbar);
\draw[dashed] (s5) -- (ssbar);
\draw[dashed] (sbar3) -- (ssbar);
\node at (3,0.5) {$\ev{\sigma v}_{S\bar{S}}^\text{ann}$};
\draw[fill=black] (0.55,0.45) rectangle ++(0.1,0.1);
\draw (sm) circle (0.4cm);
\end{scope}

\begin{scope}[shift={(4.5,-2)}]
\node (s4) at (0,1) {$S$};
\coordinate (sv4) at (1,1);
\node (v4) at (2,1) {$V$};
\node (sbar2) at (0,0) {$\bar{S}$};
\coordinate (svbar) at (1,0);
\node (vbar) at (2,0) {$V$};
\draw[dashed] (s4) -- (sv4);
\draw[dashed] (sbar2) -- (svbar);
\draw[vector] (sv4) -- (v4);
\draw[vector] (svbar) -- (vbar);
\draw[dashed] (sv4) -- (svbar);
\draw[fill=black] (0.95,-0.05) rectangle ++(0.1,0.1);
\draw[fill=black] (0.95,0.95) rectangle ++(0.1,0.1);
\node at (3,0.5) {$\ev{\sigma v}_{S\bar{S}}^\text{ann}$};
\end{scope}

\end{tikzpicture}
\caption{\label{fig:feynman}Feynman diagrams corresponding to the three annihilation processes governing sexaquark freeze-out, using the effective hadronic Lagrangian of Ref.~\cite{Farrar:2023wta}, presented in Eq.~\eqref{eq:hadlagrangian}. Arrows indicate flow of baryon number. We identify by a black circle the vertex $Sbb'$ and by a black square the vertex $Vbb'$, with $V$ a vector meson. SM represents Standard Model final states.}
\end{figure}

Finally, while we have treated the three types of cross sections we consider as independent, we could in principle relate them using the Lagrangian of Eq.~\eqref{eq:hadlagrangian}. Some caution may be needed as the $S\bar{S}$ and $\bar{S}b$ annihilation processes have center-of-mass energies of $3-4$~GeV and produce relativistic particles in the final state, such that it may not be fully correct to treat them with a low-energy hadronic effective Lagrangian. Nonetheless, if we proceed with this approach, representative Feynman diagrams for the three processes are given in Fig.~\ref{fig:feynman}, with the vertices controlled by $\tilde{g}$ and $g_{SSV}$ marked by black circles and squares respectively. We observe we would expect both $A$ and $B$ to scale with $\tilde{g}^2$, while contributions to $\langle \sigma v\rangle^{\text{ann}}_{S\bar{S}}$ will be controlled primarily by $g_{SSV}^2$, similarly to the elastic scattering cross section.

\section{\label{sec:Epol}Elastic scattering and direct detection}

Direct dark matter searches may be applied to constrain the (anti)sexaquark fraction of dark matter through its elastic scattering on Standard Model particles. For any relic antisexaquarks, there will also be stringent constraints on their annihilation with nuclei (i.e. the process parameterized by the $B$ coefficient described in Sec.~\ref{sec:freezeout}), which we will discuss in Sec.~\ref{sec:SuperK}. However, these constraints can also depend on the elastic scattering cross section via their accumulation in the Earth (Sec.~\ref{sec:accumulation}). For this reason, and because they can also constrain the sexaquark population in scenarios where antisexaquarks are depleted early on, we introduce elastic scattering searches first.

As composite particles, sexaquarks can interact strongly with nuclei through a few channels, including a repulsive single vector meson exchange -- with the leading strong force contribution potentially arising from ${\omega-\phi}$ mixing~\cite{Farrar:2020zeo} -- and an attractive two-pion exchange. Being strong-interaction processes, an estimate for the strength of this cross section would be given by ${\sigma_0 \sim m_\pi^{-2} = 2 \times 10^{-26}~\text{cm}^2}$. These elastic scattering cross sections are not anticipated to be suppressed by large factors, as may be present for number-changing interactions. Previous studies of sexaquark direct detection signals have relied on these cross sections being large enough that sexaquarks do not reach underground detectors, or do so with such tiny kinetic energies that they are not observable, in order to evade stringent limits from those experiments~\cite{Xu:2020qjk, Farrar:2022mih}.

In the event that these cross sections {\it are} suppressed, another source of elastic scattering would come from the sexaquark's electromagnetic interaction through two-photon exchange. This would yield an independent contribution to the total cross section, which would not depend on the coupling to mesons, and which can be estimated from lattice studies. The two-photon exchange process corresponds to an interaction due to the sexaquark's static electric or magnetic polarizability. We outline an estimate for this cross section in Sec.~\ref{subsec:EM_estimate}. We then discuss current direct detection constraints on the parameter space in Sec.~\ref{subsec:current_DD_limits}, which can be used to constrain the elastic scattering cross section arising from meson exchange and/or electromagnetic effects.

\subsection{\label{subsec:EM_estimate}Estimate of scattering via electromagnetic interactions}

The behavior of an electrically-neutral composite particle in response to an external electromagnetic field is determined by the distribution of quarks and gluons within the bound state. The dipole, quadrupole, and anapole moment, the charge radius, and the polarizabilities of such composite particle dictate its electromagnetic interaction with external fields or particles~\cite{Pospelov:2000bq}.

The sexaquark, being neutral with a total spin of zero, does not possess dipole, quadrupole, or anapole moments. Instead, the particle's leading structural deformation under the influence of an external electric or magnetic field is described by its polarizabilities and its charge radius. The electric and magnetic polarizabilities correspond to a shift in the energy quadratic in the electromagnetic field, given by the effective low-energy Hamiltonian ${H=- (\alpha_S/2) \mathbf{E}^2} {- (\beta_S/2) \mathbf{H}^2}$. The factor of $1/2$ in our calculations is a conventional choice that aligns with the Hamiltonian employed by \cite{Pospelov:2000bq}. The typical choice in lattice QCD and experiments is different from ours and we can use $\alpha_S = 4\pi \alpha_S^\text{LQCD/exp}$ to convert from the lattice QCD and experimental convention to our choice. 

Classically, the polarizability of a charged sphere scales with its volume. Thus if the neutron's radius is $\mathcal{O}(0.8~\text{fm})$ and the corresponding value for the sexaquark could be as small as 0.15~fm~\cite{McDermott:2018ofd,Farrar:2018hac}, we may expect the sexaquark polarizability to be within a few orders of magnitude of that of the neutron. The measured neutron polarizability is $\alpha_n^\text{LQCD/exp} = 11.8 \times 10^{-4}~\text{fm}^3$~\cite{ParticleDataGroup:2020ssz}, giving in our convention ${\alpha_n = 148 \times 10^{-4}}~\text{fm}^3$.
We use the neutron as a benchmark since it is a neutral quark bound state with a well-measured polarizability value. This inferred value serves as a rough order-of-magnitude estimate; we consider sexaquark polarizabilities from ${(0.01-1)\times}$~this value. We are also aware of a preliminary lattice QCD calculation that finds a value for the sexaquark polarizability similar to the neutron's~\cite{Detmold:private_comm}.

To model the cross section resulting from the effective Hamiltonian, we considered several models for the spherically symmetric nuclear charge density, including the homogeneously charged sphere~\cite{Pospelov:2000bq}, the Gaussian distribution, and the exponentially decreasing distribution. The first model is suitable for heavy nuclei, the second for lithium-like nuclei, and the third for an individual proton~\cite{Borkowski:1975ume}. We determined the corresponding electric field and evaluated ${\ev*{-\frac{1}{2} \alpha_S \mathbf{E}^2}}$ in the center of mass frame, from which we extracted the nucleus cross section $\sigma_{S N}^\text{scat}$, and subsequently converted it into an effective per-nucleon cross section $\sigma_{S n}^\text{scat}$ through the Born approximation, ${\sigma_{S n}^\text{scat} = \sigma_{S N}^\text{scat} (\mu_{S n} / A \mu_{S N})^2 }$. Here $\mu_{S N}$ is the reduced mass of the sexaquark-nucleus system, as opposed to $\mu_{S n}$ for the sexaquark-nucleon system. Throughout this section we use $n$ to denote nucleon (in previous sections, we have used $n$ to denote neutrons only). At the energy transfer of interest, the Helm form factor can be approximated by unity.

For all three models, the cross section follows the same scaling, differing only in the dimensionless prefactor of the leading order term, which we label $C$. We discuss the derivation of the cross section in detail in Appendix~\ref{sec:Pospelov}, and only present the final result here. In the limit where the momentum transfer $q$ times the length scale $R$ is small ($q R \ll 1$), or at small angles for any energy, we obtain the effective per-nucleon cross section
\begin{align}\label{eq:sigma_chi_n}
    \sigma_{S n}^\text{scat} = C \mu_{S n}^2 \alpha_S^2 \alpha^2 \frac{ Z^4}{R^2 A^2} \ .
\end{align}
The parameter $\mu_{S n}=m_S m_n / (m_S + m_n)$ represents the reduced mass of the nucleon and dark matter species, $\alpha$ refers to the fine structure constant, while $(Z,A)$ denote the number of protons and nucleons, respectively. The cross section is isotropic. In contrast to single photon-mediated interactions, such as Rutherford scattering, the leading term in the cross section remains independent of the momentum transfer, as is also the case for scattering due to the charge radius and quadrupole moment of the particle~\cite{Pospelov:2000bq, Appelquist:2015}.

Typically, the length scale of the atom, denoted as $R$, increases with the number of nucleons according to the relation $R \propto A^{1/3}$. As indicated by the last factor in Eq.~\eqref{eq:sigma_chi_n}, the scaling of the nucleon cross section is given by $\sigma_{S n}^\text{scat} \propto Z^4 A^{-8/3}$, which results in an enhancement of the cross section for heavier atoms. This is to be contrasted with the familiar scaling where the nucleus cross section scales as $A^4$ (for dark matter much heavier than nuclei) and the nucleon cross section does not change. The values of the dimensionless constant $C$ obtained in this work and the atom's scale $R$, are provided in Table~\ref{tab:C_and_R}. The value of $C$ for the homogeneously charged nucleus was obtained previously in~\cite{Pospelov:2000bq}. The values for $R$ come from experiments, while the values for $C$ are exact for the homogeneously charged nucleus and exponential distribution, and approximate for the Gaussian distribution. While the charge density distribution of a nucleus is much more complex than the three models considered here, the parameters enable us to capture the essential features, and the expressions for $C$ and $R$ are consistent within an order of magnitude. For the Gaussian charge distribution, the scale is determined by the variance $\ev*{r^2}^{1/2}=\sqrt{3/5} r_0$ of a sphere with radius $r_0$. In the exponential charge distribution, the scale $R$ corresponds to the proton charge radius.

\begin{table}[htb]
    \centering
    \caption{\label{tab:C_and_R}Values for the dimensionless constant $C$ and the atomic scale $R$ appearing in Eq.~\eqref{eq:sigma_chi_n}. $C$ for the homogeneously charged nucleus and exponential distributions are exact, Gaussian distribution is approximate, and $R$ values are extracted from experiments.}
    \begin{tabular}{l c c}
    \hline
         & $C$ & $R$ [fm] \\ \toprule 
        homogeneously charged nucleus & \, $\displaystyle \frac{144\pi}{25}$ \, & $1.2 \ A^{1/3}$ \\[0.3cm]
        Gaussian charge distribution & $\displaystyle \frac{32\pi}{25}$ & $\sqrt{\tfrac{3}{5}} 1.2 \ A^{1/3}$ \\[0.3cm]
        exponential charge distribution & $\displaystyle \frac{75\pi}{16}$ & $0.87 $ \\[0.3cm] \bottomrule
    \end{tabular}
\end{table}

Another non-zero electromagnetic contribution to the sexaquark-nucleus scattering comes from the sexaquark's charge radius. However, since this cross section scales as the fourth power of the charge radius~\cite{Pospelov:2000bq} and it has been argued that the sexaquark radius may be comparable to its Compton wavelength~\cite{McDermott:2018ofd} (0.1~fm), we expect the leading electromagnetic contribution to arise from the polarizability.

Our work focuses on non-relativistic dark matter scattering due to a non-renormalizable effective electromagnetic Hamiltonian. Other studies have classified the effective electromagnetic operators~\cite{Kavanagh:2018xeh}, explored the relativistic electromagnetic cross section of dark matter with nuclei, using various properties such as the dipole moment~\cite{Sigurdson:2004zp}, anapole moment~\cite{DelNobile:2014eta}, charge radius~\cite{Barger:2010gv}, and polarizability~\cite{Weiner:2012, Frandsen:2012db, Ovanesyan:2014fha}. \cite{Appelquist:2015} employed a cross section similar to ours in the context of a dark SU(4) theory using a parameter that varies over an order of magnitude, highlighting the effect of nuclear structure, which is analogous to our factor $C$.

In addition to the cross section arising from the electric field of a nucleus, we also investigated the elastic scattering with photons and the Casimir-Polder effect~\cite{Casimir:PhysRev.73.360}, which arises from the polarizability of both the dark matter particle and the nucleus, in Appendix~\ref{sec:Pospelov}. We find that both are significantly suppressed compared to the cross section presented in Eq.~\eqref{eq:sigma_chi_n}.

\subsection{\label{subsec:current_DD_limits}Current limits}

The most recent direct detection bounds on spin-independent dark matter interactions with nucleons have been obtained through a variety of targets. For the mass range of interest for the sexaquark (approximately 2~GeV), direct detection experiments located deep underground have set strong constraints on small cross sections. We make use of nuclear recoil bounds from CRESSTIII~\cite{CRESST:2019jnq, Cappiello:2023hza}, Migdal and nuclear recoil limits from CDEX~\cite{CDEX:2019hzn, CDEX:2021cll}, nuclear recoil, bremsstrahlung, and Migdal effect in CDMSlite~\cite{SuperCDMS:2018gro,SuperCDMS:2022kgp}, as well as the ionization signal from DarkSide50~\cite{DarkSide:2018bpj} and DAMIC~\cite{DAMIC:2020cut}. We also combine under the label XENON1T various bounds coming from the Migdal effect for ${m_\chi < 2}$~GeV~\cite{XENON:2019zpr}, ionization signals for ${3 < m_\chi < 6}$~GeV~\cite{XENON:2019gfn}, and nuclear recoil for larger dark matter mass~\cite{Aprile:2018dbl}. We also include limits on cosmic-ray boosted dark matter interacting through a vector mediator with ${m_V = 1}$~GeV (such as the $\phi$ meson) and cutoff mass ${\Lambda_V=0.5}$~GeV, for ${m_\chi < 4}$~GeV; these choices provide conservative limits on the sexaquark according to the original work~\cite{Alvey:2022pad}. The PandaX limit comes from cosmic-ray boosted dark matter at {PandaX-II} for ${m_\chi<0.1}$~GeV~\cite{PandaX-II:2021kai} and momentum-independent scattering through a heavy mediator at PandaX-4T for ${m_\chi < 2}$~GeV~\cite{PandaX:2023xgl}. We use Ref.~\cite{PandaX:2022xqx} to estimate the ceiling of this experiment. However, for larger cross sections, the overburden of underground laboratories is much greater than the dark matter interaction length, causing dark matter to lose energy through scattering with the atmosphere and rock overburden, resulting in kinetic energy below the detector threshold~\cite{Starkman1990}. This stopping power of the rock and atmosphere provides upper bounds for the constraints from direct detection experiments located underground~\cite{Witten1985, Starkman1990, Hooper:2018bfw, McKeen:2022poo}. Complementary surface-based experiments, such as the CRESST~\cite{CRESST:2017ues} and EDELWEISS~\cite{EDELWEISS:PhysRevD.99.082003} surface runs, provide sensitive bounds at an intermediate interaction strength, albeit with higher background rates. Finally, re-analyses of the rocket-based X-Ray Quantum Calorimeter (XQC) experiment~\cite{XQC:McCammon_2002, Mahdawi_2018} close the gap at large cross sections. Various techniques can be used to estimate the upper bounds, from the straight-line approximation~(e.g. \cite{Starkman1990}) to analytic approaches~\cite{Cappiello:2023hza} and Monte-Carlo codes~(e.g. \cite{Emken:2018run,Mahdawi_2018}). In this work, the upper limits we present employ the straight-line approximation, except for CRESSTIII and CRESST surface run, which are extracted using the analytic approach from \cite{Cappiello:2023hza}.

\begin{figure}[ttt]
    \centering
    \includegraphics[width=\columnwidth]{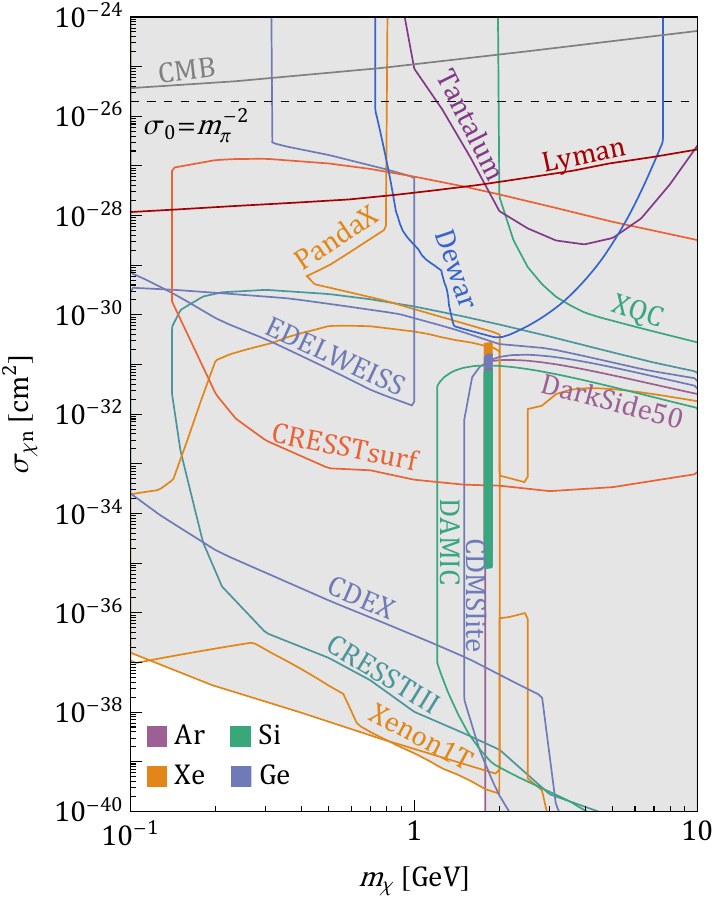}
    \caption{Experimental constraints on the spin-independent dark-matter nucleon cross section and expected cross section due to the estimated polarizability of the sexaquark. Experiments using the same active detector material are shown in the same color: orange~(xenon) for PandaX-II and PandaX-4T (together labeled as PandaX) and Xenon1T, green~(silicon) for XQC and DAMIC, turquoise~(calcium tungsten oxide) for CRESSTIII, blue~(germanium) for EDELWEISS, CDMSlite, and CDEX, red~(aluminium oxide) for CRESST surface, and purple~(argon) for DarkSide50. The grey region represents the combined ruled out parameter space. The small color-shaded regions indicate the nucleon cross section due to the electric polarizability of the sexaquark for commonly used target materials. The horizontal black dashed line represents the typical strong interaction strength.}
    \label{fig:all_bounds}
\end{figure}

In this context, we present up-to-date direct detection constraints on dark matter-nucleon interactions in Fig.~\ref{fig:all_bounds}, which are relevant for the sexaquark. The grey region is the combined excluded parameter space to which detectors are currently sensitive. We superimpose on it the polarizability scattering cross section obtained from Eq.~\eqref{eq:sigma_chi_n} for four main active detector materials, color-coded by material: argon, xenon, silicon, and germanium, assuming uniformly charged nuclei, which is a good approximation for the heavy nuclei employed in direct detection experiments. We also indicate the approximate scale appropriate for a strong cross section, $\sigma_0=m_\pi^{-2}$. 

Our analysis shows that the polarizability cross section for all four elements falls within the experimental reach of published dark matter exclusion limits; namely, that the polarizability cross section appears to be ruled out by DAMIC (silicon), DarkSide50 (argon), CDEX and {CDMS}lite (germanium), Xenon1T and PandaX-4T (xenon). The per-nucleon cross sections in these different materials are similar but differ by the material-dependent factor $Z^4 A^{-8/3}$ (Eq.~\ref{eq:sigma_chi_n}), which takes the approximate values 19 (Xe), 11 (Ge), 6 (Ar), and 5 (Si), such that xenon targets result in the largest cross section.  

This indicates that if the sexaquark possessed the polarizability we estimated and composed all of the dark matter, it would have been detected in these experiments. The cross section also falls within the experimental reach of CRESSTIII and CRESST surface run. The width and height of the polarizability cross section regions are due to the range of stable mass and uncertainty in the estimate of the polarizability, respectively; we take the polarizability to vary from $(0.01-1)\times$ the neutron polarizability. We note that for the sexaquark mass range we consider, for every choice of cross section within the range displayed on this plot, there are generally always at least two experiments/probes that claim to exclude such a cross section; this gives some protection against systematic mis-estimates of the sensitivity.

Using the estimate of the electric polarizability, we found that the sexaquark-nucleus cross section falls within the regime of strongly interacting dark matter, but remains below the saturation level of the nuclear area for targets typically used in dark matter experiments. It is also a few orders of magnitude below the typical strong interaction strength, $\sigma_0 = m_\pi^{-2}$. The total nucleon-sexaquark cross section is likely dominated by a strong force component, leading to an upward shift in the cross section. The polarizability cross section represents an approximate lower bound for the total cross section of sexaquarks with nuclei. As shown in Fig.~\ref{fig:all_bounds}, direct detection experiments such as XQC, dewar heating~\cite{xu2021constraints}, de-excitation of a tantalum isomer~\cite{Lehnert:PhysRevLett.124.181802}, the cosmic microwave background (CMB) and baryon acoustic oscillations~\cite{Buen_Abad_2022}, and the Lyman-alpha forest~\cite{Buen_Abad_2022} exclude all of the parameter space at large cross sections and relevant masses. Limits from the Milky Way satellite galaxies are comparable to the Lyman-alpha constraints~\cite{Buen_Abad_2022} and are not included. In the limits we are showing, the dewar heating assumes a repulsive Yukawa coupling with a heavy mediator, such as is the case in vector meson exchange of a strong interaction. In the low energy-transfer limit, the Yukawa coupling is momentum-transfer independent akin to our polarizability cross section, Eq.~\eqref{eq:sigma_chi_n}. 

Regardless of whether the nucleon cross section is dominated by the polarizability or strong force, our findings suggest that if sexaquarks made up all of dark matter (through strongly suppressed values of $A$ and $B$), they would have been observed in numerous underground and surface direct detection experiments.

In the scenario where a dark matter candidate constitutes a subcomponent of the total dark matter (${f_\chi = \Omega_\chi / \Omega_\text{DM} < 1}$), we kept the upper bound of each experiment unchanged, while the lower limit moves upward proportionally to the fraction of the total dark matter density that can be probed, represented by ${\log_{10} f_\chi}$~\cite{McKeen:2022poo}. Note that because the cross section scales as $\alpha_S^2$, changes in the polarizability modify the target cross section faster than variations in $f_\chi$ modify the detectability.

The straight-line approximation captures the average energy retained by dark matter particles as they cross the atmosphere and crust. From this, we can determine how much energy would be given to an active detector nucleus and compare this energy with the minimum experimental threshold~\cite{Starkman1990}. It behaves as a step function; either accepting or rejecting all wind dark matter particles instead of analyzing the cross section needed for the number of dark matter particles scattering within the experimental exposure to fall below one. Compared to other approaches, the straight-line approximation was found to overestimate the stopping power of the Earth~\cite{Mahdawi:2017utm, Emken:2018run}. For subcomponents of dark matter, fewer particles should be able to reach detectors during the running time, resulting in a slight lowering of the experimental ceiling~\cite{Cappiello:2023hza}. The effect appears minimal unless the experimental floor is very close to the ceiling and the straight-line approximation does not allow to implement it, so we did not account for this effect~\cite{Emken:2018run}. However, we emphasize that the Monte Carlo results of Ref.~\cite{Emken:2018run} suggest that the straight-line approximation should mildly underestimate the signal in any case, down to $f_\chi \sim 10^{-9}$.

\begin{figure*}[ttt]
\centering
\begin{minipage}[b]{0.5\textwidth}
        \centering
        \includegraphics[width=\textwidth]{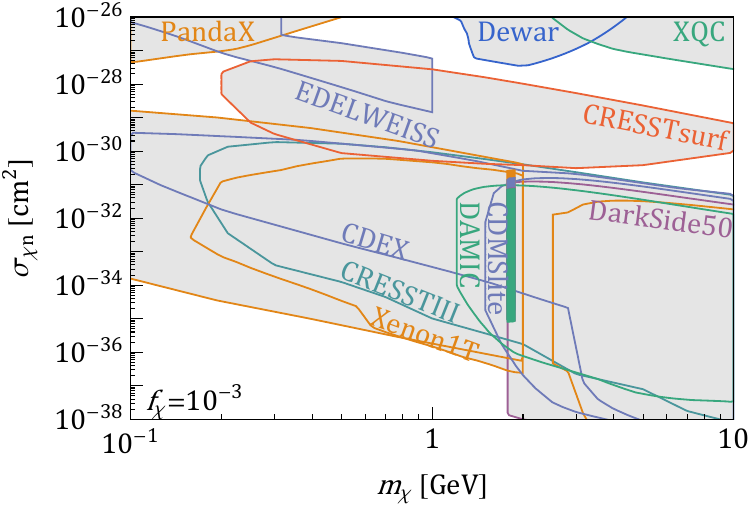} 
  \end{minipage}~~
  \begin{minipage}[b]{0.5\textwidth}
    \centering
    \includegraphics[width=\textwidth]{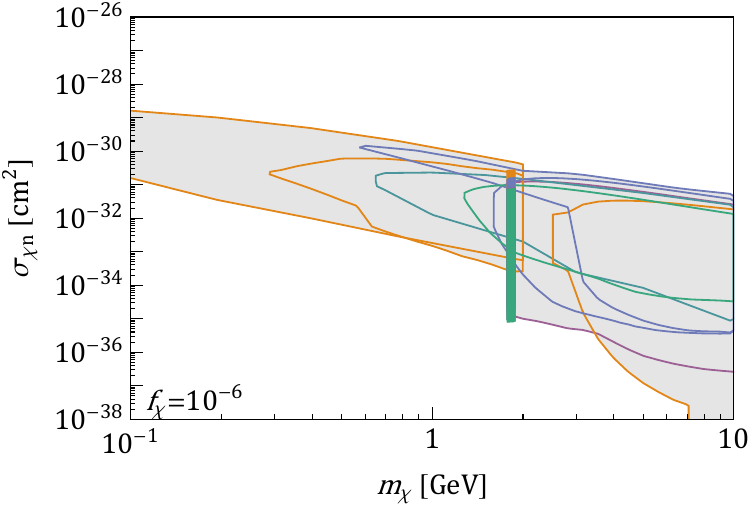}
  \end{minipage} 
\begin{minipage}[b]{0.5\textwidth}
        \centering
        \includegraphics[width=\textwidth]{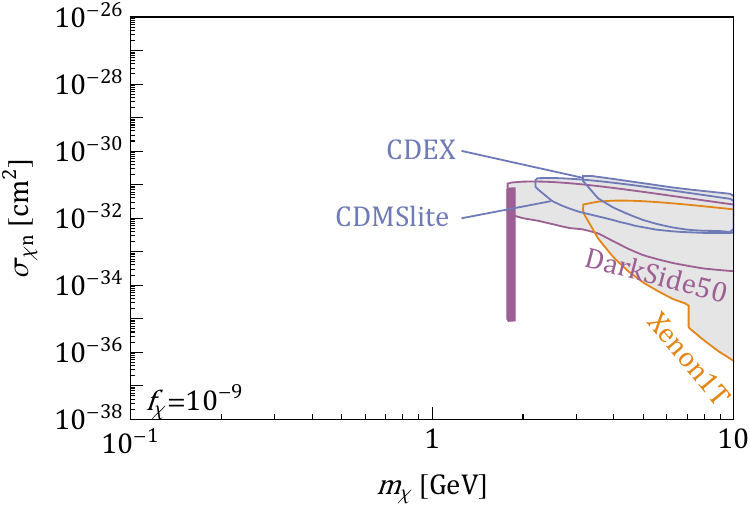} 
  \end{minipage}~~
  \begin{minipage}[b]{0.5\textwidth}
    \centering
    \includegraphics[width=\textwidth]{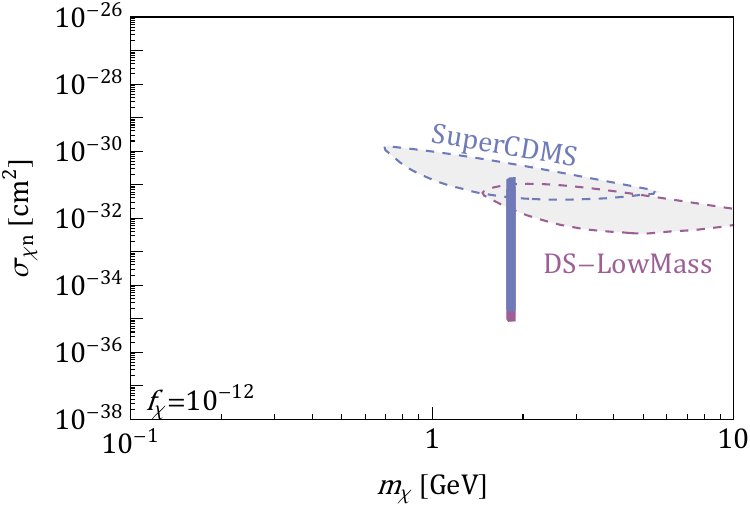}
  \end{minipage} 
  \caption{Direct detection constraints on the spin-independent dark matter-nucleon cross section for different dark matter fractions ${f_\chi = \Omega_\chi / \Omega_\text{DM}}$, namely $f_\chi = 10^{-3}$~(top left), $10^{-6}$~(top right), and $10^{-9}$~(bottom left), along with projections for the next generation of experiments aiming to reach $10^{-12}$~(bottom right). The dashed lines represent the projected exclusion limits from SuperCDMS Ge HV and DarkSide (DS)-LowMass, which are distinguished from the reported exclusion limits. The cross section due to the static electric polarizability of the sexaquark for commonly used target materials is indicated by the color-shaded regions.}
  \label{fig:small_x_section}
\end{figure*}

We present in Fig.~\ref{fig:small_x_section} the direct detection constraints on the electric polarizability cross section for a subcomponent of dark matter, ${f_\chi = \lbrace 10^{-3}, 10^{-6}, 10^{-9} \rbrace}$. We find that a gap at large cross sections rapidly opens when ${f_\chi < 10^{-3.5}}$ due to the upward shift of the CRESST surface experimental bound. If sexaquarks are a subcomponent of dark matter and their strong elastic cross section is large, current direct detection experiments would not be able to probe them due to the overburden. This is due to the lack of surface-based detectors. Furthermore, the detector's efficiency threshold could be lower than assumed in the original analysis, implying that the CRESST surface limits might not be highly reliable~\cite{Mahdawi_2018}. In this case, the gap at larger cross sections would already appear for ${f_\chi < 0.5}$, when the dewar bound shown in Fig.~\ref{fig:all_bounds} shifts upward. Our analysis shows that even with a fraction as low as ${f_\chi = 10^{-9}}$, direct detection experiments using argon as the target material can still exclude part of the parameter space of the polarizability cross section we obtained. We find that direct detection constraints are currently insensitive to a polarizability nucleon cross section for a smaller subcomponent, of the order of ${f_\chi \sim 10^{-10}}$. Still, we are optimistic in the projected reach of the next generation of experiments to probe subcomponents reaching down to ${f_\chi = 10^{-12}}$, such as DarkSide-LowMass~\cite{GlobalArgonDarkMatter:2022ppc} and SuperCDMS~Ge~HV~\cite{SuperCDMS:2016wui}, shown in the bottom right panel of Fig.~\ref{fig:small_x_section}. Additionally, proposed experiments such as CYGNUS-1000~\cite{Vahsen:2020pzb},  ALETHEIA~\cite{Liao:2021npo}, and SBC~\cite{Alfonso-Pita:2022akn} could also cover the phase space of interest. With such sensitivity, the upcoming direct detection effort will reach the neutrino floor, which is higher at $\mathcal{O}(1 \ \text{GeV})$ dark matter mass than at $\mathcal{O}(100 \ \text{GeV})$. We also find that DAMIC-M~\cite{deDominicis:2022nls} will be insensitive to a subcomponent ${f_\chi = 10^{-12}}$. Another proposal is to employ underground nuclear accelerators to up-scatter thermalized dark matter, greatly enhancing the reach of current dark matter experiments to strongly interacting sub-components of dark matter~\cite{McKeen:2022poo}.

\section{\label{sec:accumulation} Accumulation of (anti)sexaquarks in the Earth}

In the event that the sexaquark has a large elastic cross section (which may help hide it from direct detection in the case where it constitutes a fraction of the dark matter), sexaquarks may accumulate in the Earth and other astrophysical bodies through efficient capture and subsequent deceleration until they achieve thermal equilibrium. Such capture happens when the mean free path of the particle is significantly smaller than the width of the celestial body.

In this section, we explore the accumulation of (anti)sexaquarks in the Earth, with a focus on the near-surface crust, which will be an important ingredient in some constraints presented in Sec.~\ref{sec:SuperK}. Assuming that the rates of evaporation $\Gamma^\text{evap}$ and annihilation $\Gamma^\text{ann}$ per unit volume are sufficiently low, sexaquarks can attain an impressive overabundance. Our investigation addresses the accumulation of both symmetric and asymmetric dark matter within the Earth. Unlike previous studies which considered dark matter to be its own antiparticle (or with a negligible asymmetry), e.g.~\cite{Mack:2007xj, Kouvaris:2007ay, Bramante:2022pmn, Pospelov:2023mlz}, where the only annihilation rate is due to the self-annihilation, we extend our analysis to include both the self-annihilation channel ($S\bar{S} \to Y\bar{Y}$) and the annihilation of antisexaquarks with nucleons in the Earth ($\bar{S}\lbrace n,p \rbrace \to \bar{b}X$, which we will label in this section as $\bar{S}n$ with $n$ for nucleon). Even if sexaquarks constitute only a subcomponent of dark matter, their accumulation within the Earth can lead to a significantly larger density than the local dark matter density, owing to the large age of the Earth.

To estimate the populations of sexaquarks and antisexaquarks accumulating near the Earth's surface over the Earth's lifetime, we adopt the approach outlined in~\cite{Neufeld:2018slx, Pospelov:2023mlz}. Our calculations incorporate the PREM model for the Earth's density profile and the NRLMSISE-00 model for the atmosphere~\cite{PREM,NRLMSISE}. The effect of the traffic jam, deemed relevant only for dark matter masses above approximately 10~GeV~\cite{McKeen:2022poo, Leane:2022hkk}, is excluded. Denoting $\Gamma^\text{cap}$ as the capture rate, the evolution of the number of sexaquarks ($N_S$) and antisexaquarks ($N_{\bar{S}}$) within the Earth is described by
\begin{align}
    \dv{N_S}{t} &= f_S\Gamma^\text{cap} - N_S \Gamma^\text{evap} - N_S N_{\bar{S}} \Gamma^\text{ann}_{S\bar{S}} \ , \label{eq:N_S'}\\
    \dv{N_{\bar{S}}}{t} &= f_{\bar{S}} \Gamma^\text{cap} - N_{\bar{S}} \Gamma^\text{evap} - N_S N_{\bar{S}} \Gamma^\text{ann}_{S\bar{S}} - N_{\bar{S}} N_n \Gamma^\text{ann}_{\bar{S}n} \ . \label{eq:N_Sbar'}
\end{align}
$f_S$ and $f_{\bar{S}}$ are derived from the freeze-out abundance of sexaquarks and antisexaquarks and are general functions of $A, B, \ev{\sigma v}_{S \bar{S}}^\text{ann}$. The capture rate is expressed as $\Gamma^\text{cap} = f^\text{cap} P(x \geq N_0, R_\oplus / \ell_\text{MFP}) \Gamma^\text{cap}_\text{geom}$, where the capture fraction, $f^\text{cap}$, is estimated using the multi-scatter formalism~\cite{Neufeld:2018slx}, 
\begin{align}\label{eq:fcap}
    f^\text{cap} &= \frac{2}{(\pi N_0)^{1/2}} \ ,
\end{align}
where $N_0$ represents the mean number of scatterings needed to reduce the particle speed below the Earth's escape velocity $v_\text{es}$ and is given by
\begin{align}
    N_0 &= \frac{ \ln(v_\text{es}^2/v_\oplus^2)}{ \ln(1 - \bar{f}_\text{KE})} \ .
\end{align}
The mean kinetic energy transfer $\bar{f}_\text{KE}$ is defined as
\begin{align}
    \bar{f}_\text{KE} &= \frac{2m_S m_N}{(m_S + m_N)^2} \ .
\end{align}
We add to $\Gamma^\text{cap}$ a term $P(x \geq N_0, R_\oplus / \ell_\text{MFP})$ accounting for probability that the sexaquarks scatter at least $N_0$ times as they cross the Earth, given a Poissonian probability distribution with mean $R_\oplus/\ell_\text{MFP}$, the length of the sexaquarks' mean free path $\ell_\text{MFP}$ compared to the Earth radius $R_\oplus$. This term estimates the fraction of particles that can scatter at least $N_0$ times in the Earth before exiting so they can accumulate. The geometric capture rate, $ \Gamma^\text{cap}_\text{geom}$, is given by~\cite{Bramante:2022pmn}
\begin{align}
    \Gamma^\text{cap}_\text{geom} &= \pi R_{\oplus,\text{atm}}^2 \sqrt{\frac{8}{3\pi}} \frac{\rho_\chi v_d}{m_S} \left( 1 + \frac{3 v_\text{es}^2}{v_d^2} \right) \xi \ ,
\end{align}
where 
\begin{align}
    \xi &= \frac{1}{2v_d^2 + 3 v_\text{es}^2} \Bigg( v_d^2 e^{-3v_\oplus^2/2v_d^2} \nonumber\\
    &+ \sqrt{\frac{3\pi}{2}} \frac{v_d}{v_\oplus} \left[ v_\text{es}^2 + v_\oplus^2 + \frac{v_d^2}{3} \right] \text{Erf} \left( \sqrt{\frac{3}{2}} \frac{v_\oplus}{v_d} \Bigg) \right) \ .
\end{align}
The values for the Maxwell-Boltzmann velocity distribution, galactic dark matter velocity, and Earth's escape velocity are respectively $\lbrace v_d, v_\oplus, v_\text{es}\rbrace = \lbrace 270, 220, 11.2  \rbrace$~km/s. The Earth radius including the atmosphere, ${R_{\oplus,\text{atm}} = 6471}$~km, is 100~km larger than the Earth radius.

We expect our formulation for $\Gamma^\text{cap}$ to be generally a reasonable approximation, but it slightly undershoots the capture rate at large cross sections, and exhibits a more abrupt cutoff at small cross sections, in comparison to results obtained through Monte Carlo simulations~\cite{Bramante:2022pmn}. Nevertheless, it proves to be a reliable approximation for scattering cross sections above $10^{-35}~\text{cm}^2$ for the mass of the sexaquark.

Post-capture, particles undergo diffusion due to gravity and thermal gradients. Hydrostatic equilibrium states that if the particles are diffusing over a timescale much shorter than the Earth's lifetime, they can be in thermal equilibrium with their environment, which is a valid assumption for scattering cross sections larger than ${\sim 10^{-37}~\text{cm}^2}$~\cite{Acevedo:2023owd}. The distribution of these particles' number density as a function of the radius from the center of the Earth is approximated by the solution to the differential equation
\begin{align}
    \frac{\nabla n_S(r)}{n_S(r)} = - \frac{1}{T(r)} \left[ m_S g(r) + (\kappa + 1) \nabla T(r) \right] \ ,
\end{align}
where $\kappa \approx -1/[2(1+m_S/m_N)^{3/2}]$ is the thermal diffusion coefficient~\cite{Leane:2022hkk}, $g(r)$ is the gravitational acceleration at a height $r$, and we adopt the temperature profile of Ref.~\cite{Neufeld:2018slx}. From this point, we define ${G_S(r) = G_{\bar{S}}(r) = V_\oplus n_S(r)/N_S}$ as the radial profile function, where $V_\oplus$ is the Earth volume. The radial profile function is identical for sexaquarks and antisexaquarks given their equal mass and only depends on temperature and gravity gradients through the Earth. For sexaquarks, the number density remains approximately constant across the Earth, with a slight overdensity in the core and near the surface~\cite{Neufeld:2018slx, Moore_2021}. Their mass is not sufficiently heavy to all sink to the Earth' core, nor to all evaporate, preventing substantial depletion near the Earth surface.

To study thermal evaporation, we adopt the expression~\cite{Neufeld:2018slx, Pospelov:2023mlz}
\begin{align}\label{eq:evaporation}
    \Gamma^\text{evap} &= G_{S}(R_\text{LSS}) \frac{3 R_\text{LSS}^2 v_\text{LSS}}{R_\oplus^3} \left( 1 + \frac{v_\text{es}^2}{v_\text{LSS}^2} \right) e^{-v_\text{es}^2/v_\text{LSS}^2} \ ,
\end{align}
where $R_\text{LSS}$ and $v_\text{LSS}$ denote the Earth radius and sexaquark thermal velocity evaluated at the surface of last scattering (LSS). The LSS is the location $z_\text{LSS}$ where the last collision between dark matter and atoms occurs, providing the dark matter particle with a sufficient amount of energy in the upward direction to escape Earth's gravitational attraction. The optical depth defines the location of the LSS through~\cite{Neufeld:2018slx}
\begin{align}
    1 &= \int_{z_\text{LSS}}^\infty \dd{z} \sum_N n_N(z) \left( \frac{\mu_{\chi N}^2}{\mu_{\chi n}^2} A^2 \right)^2 \sigma_{\chi n}\ .
\end{align}
The annihilation rates are given by
\begin{align}
    \Gamma_{S\bar{S}}^\text{ann} &= \frac{4\pi \ev{\sigma v}_{S\bar{S}}^\text{ann}}{V_\oplus^2} \int_0^{R_\oplus} \dd{r} r^2 G_S^2(r) \ , \\
    \Gamma_{\bar{S}n}^\text{ann} &= \frac{4\pi \ev{\sigma v}_{\bar{S}n}^\text{ann}}{V_\oplus N_n} \int_0^{R_\oplus} \dd{r} r^2 G_{{S}}(r) n_n(r) \ .
\end{align}
Eqs.~\eqref{eq:N_S'} and \eqref{eq:N_Sbar'} are coupled differential equations and their full solution can only be evaluated numerically. Accounting for capture and evaporation, the number density of dark matter particles with a mass around a~GeV and large scattering cross section has been previously found to be approximately $10^{14}~\text{cm}^{-3}$~\cite{Neufeld:2018slx}. Works incorporating dark matter annihilation introduce the third factor shown on the right hand side of Eqs.~\eqref{eq:N_S'} and \eqref{eq:N_Sbar'}. In addition, we include the fourth term of Eq.~\eqref{eq:N_Sbar'} to capture possible annihilation of antisexaquarks against nucleons. We emphasize that for similar annihilation strengths, the last term of Eq.~\eqref{eq:N_Sbar'} will dominate over the third due to the factor of $N_n$ instead of $N_S$, i.e. the larger abundance of nucleons compared to sexaquarks in the Earth.

\begin{figure*}[ttt]
\centering
\begin{minipage}[b]{0.43\textwidth}
    \centering
    \includegraphics[width=\textwidth]{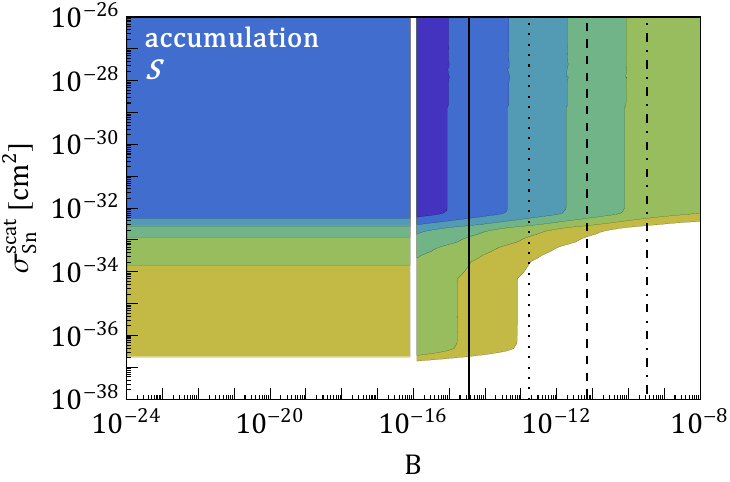} 
\end{minipage}~~
\begin{minipage}[b]{0.43\textwidth}
    \centering
    \includegraphics[width=\textwidth]{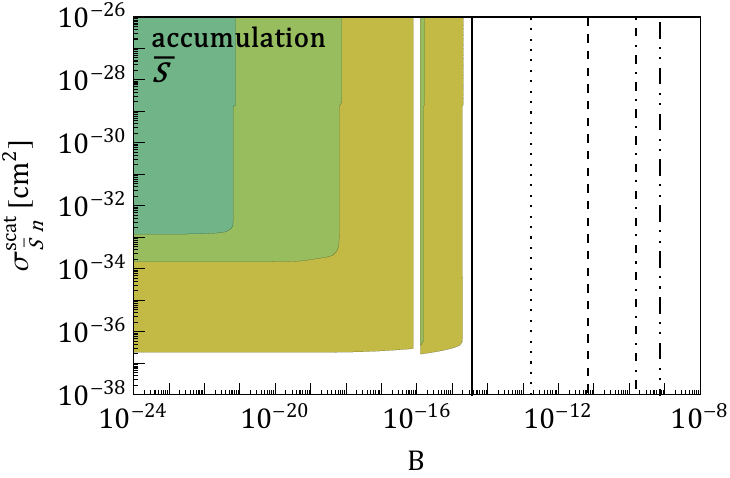}
\end{minipage}~~
\begin{minipage}[t]{0.14\textwidth}
    \centering
    \includegraphics[width=\textwidth]{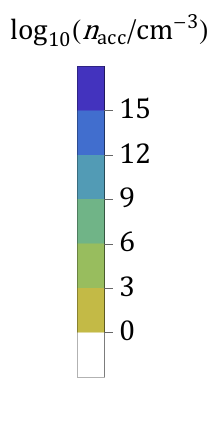}
\end{minipage}
  \caption{Number density of accumulated sexaquarks (left) and antisexaquarks (right) near the Earth crust (depth of 1~km) as a function of the annihilation cross section parameterized by $\ev*{\sigma v}_{\bar{S}n \to \bar{b}'X}^\text{ann} = Bm_\pi^{-2}$ and scattering cross section $\sigma_{Sn}^\text{scat}$ or $\sigma_{\bar{S}n}^\text{scat}$. These figures assume the annihilation cross section $\ev{\sigma v}_{S\bar{S}}^\text{ann}$ to be negligible, both for freeze-out and Earth accumulation, as well as $\ev{\sigma v}_{SX\to bb'}$ being negligible for freeze-out. For the sexaquarks, the plot shows only the capture and evaporation, while for the antisexaquarks, it further shows the depletion due to annihilation against nucleons. The black vertical lines are for 1860~MeV (anti)sexaquarks as all~(full), $10^{-3}$~(dotted), $10^{-6}$~(dashed), $10^{-9}$~(dot-dashed), and $10^{-12}$~(dot-dot-dashed) of dark matter. To the left-hand side of the white gaps, we take dark matter to be half in sexaquarks and half in antisexaquarks.}
  \label{fig:acc_SbarN}
\end{figure*}

In scenarios where the self-annihilation dominates over evaporation and annihilation against nucleons, the populations of sexaquarks and antisexaquarks evolve together as
\begin{align}
    N_S \Gamma^\text{ann}_{S\bar{S}} &\gg N_n \Gamma^\text{ann}_{\bar{S}n}, \Gamma^\text{evap} \\
    N_S(t) = N_{\bar{S}}(t) &= \sqrt{\frac{\Gamma^\text{cap}}{\Gamma^\text{ann}_{S\bar{S}}}} \tanh(\sqrt{\smash[b]{\Gamma^\text{cap} \Gamma^\text{ann}_{S\bar{S}}}}t) \ .
\end{align}
However, the self-annihilation rate is suppressed due to the factor ${n_S \leq 10^{14}~\text{cm}^{-3}}$, while the annihilation against nucleons has a term ${n_n \sim 10^{24}~\text{cm}^{-3}}$. For example, taking ${\ev{\sigma v}_{S\bar{S}}^\text{ann} = 10^{-27}~\text{cm}^3/\text{s}}$, self-annihilation only becomes the dominant annihilation channel for ${\ev{\sigma v}_{\bar{S}n}^\text{ann} < 10^{-25} m_\pi^{-2}}$. Furthermore, the importance of evaporation for dark matter with a mass less than 10~GeV is emphasized in various works~\cite{Bramante:2022pmn, Pospelov:2023mlz}. In a more realistic scenario, we take the self-annihilation $\Gamma^\text{ann}_{S \bar{S}}$ to be small compared to the evaporation and annihilation against nucleons. In this case, the populations are given by
\begin{align}
    N_S \Gamma^\text{ann}_{S\bar{S}} &\ll N_n \Gamma^\text{ann}_{\bar{S}n}, \Gamma^\text{evap} \\
    N_S(t) &= \frac{\Gamma^\text{cap}}{\Gamma^\text{evap}} [1 - e^{-\Gamma^\text{evap} t}] \ , \label{eq:NS_cap_evap}\\
    N_{\bar{S}}(t) &= \frac{\Gamma^\text{cap}}{N_n \Gamma^\text{ann}_{\bar{S}n}  + \Gamma^\text{evap}} [1 - e^{-(N_n \Gamma^\text{ann}_{\bar{S}n}  + \Gamma^\text{evap})t}] \ .\label{eq:NSbar_cap_evap_ann}
\end{align}
Figure~\ref{fig:acc_SbarN} illustrates the accumulation of sexaquarks and antisexaquarks. For $B > 10^{-16}$, it incorporates the freeze-out abundance factor $f_{S,\bar{S}}$ (shown in the second panel of Fig.~\ref{fig:freeze_out_abundance_fchi}), assuming that only $\ev{\sigma v}_{\bar{S}n}^\text{ann}$ influences the final abundance. For sexaquarks, a single factor of $B$ regulates the overabundance stemming from the freeze-out process. Conversely, antisexaquarks get two factors of $B$: one from freeze-out and another from the depletion of the abundance during accumulation. They cannot achieve a considerable overabundance compared to the dark matter wind through accumulation due to their rapid annihilation with nucleons in the Earth, shown as the white region on the right panel of Fig.~\ref{fig:acc_SbarN}.

At annihilation rates corresponding to $B = 10^{-16}$ or lower, sexaquarks are never coupled to the baryon-photon bath if no other process helps regulating the freeze-out abundance. Still, sexaquarks could in principle make up all or a substantial fraction of dark matter based on other processes occurring either during (e.g. the rearrangement of quarks during the QCD phase transition as described in Ref.~\cite{Farrar:2018hac}) or shortly after the QCD phase transition (e.g. by tuning the values of $A$ and $\ev{\sigma v}_{S \bar{S}}^\text{ann}$). These scenarios would imply a symmetric sexaquark and antisexaquark population. In this case, a substantial accumulation can occur as the depletion from the annihilation parameterized by $B$ is very small. In this context, we present in Figure~\ref{fig:acc_SbarN} the accumulation of sexaquarks and antisexaquarks for $B < 10^{-16}$, demonstrating that this scenario can still lead to a substantial (anti)sexaquark overabundance. The observed mismatch in contours at $B = 10^{-16}$ arises from the fact that freeze-out between the thin white gap and the full black line results in an overabundance of sexaquarks and antisexaquarks -- up to a thousand times more than the observed dark matter abundance, while to the left of $B = 10^{-16}$, we take ${f_S = f_{\bar{S}} = 0.5}$. If the process generating the relic (anti)sexaquark abundance yields a subcomponent of dark matter, the curves shown in Fig.~\ref{fig:acc_SbarN} can be scaled down accordingly.

For large values of $B$, evaporation becomes significant for nucleon scattering cross sections below $10^{-32}~\text{cm}^2$. As $B$ decreases below $10^{-16}$, the evaporation begins to dominate over the annihilation and for $B < 10^{-29}$, the annihilation rate becomes negligible compared to $t_\oplus^{-1}$ and can be omitted from Eq.~\eqref{eq:NSbar_cap_evap_ann}. The largest abundance occurs for scattering cross sections between $10^{-32}$ and $10^{-26}~\text{cm}^2$, where the evaporation rate is smallest due to the LSS being at or near the Earth surface, which is significantly colder than inner parts of the Earth, reducing the factors of $v_\text{LSS}$ in Eq.~\eqref{eq:evaporation}.

In the accumulation plots presented, even when the sexaquark population in the Solar neighborhood is symmetric (i.e. for $B \leq 10^{-11}$), the depletion of the antisexaquark population during the accumulation process results in an asymmetric Earth accumulation. Symmetry is restored for $B \leq 10^{-29}$, where virtually no antisexaquark annihilates against nucleons in the Earth. For sexaquarks with freeze-out, even at large values of $B$ (which lead to $f_S \sim 10^{-11}$ as discussed in Secs.~\ref{sec:abundance}-\ref{sec:freezeout}) a slightly larger abundance than the local dark matter abundance can be obtained. This becomes especially significant at large scattering cross sections, where the atmosphere and crust's stopping power otherwise prevents the penetration of the dark matter wind down to underground detectors.

The consideration of sexaquarks binding to nuclei is omitted in our analysis due to uncertainties regarding the dominant interaction channels. Specifically, it remains unclear whether the attractive two-pion channel prevails over the repulsive single vector meson channel~\cite{Farrar:2003gh}.

\section{\label{sec:SuperK}Antisexaquark annihilation signatures}

As discussed in Sec.~\ref{sec:abundance}, in cosmological scenarios where a large (anti)sexaquark abundance is achieved by an early decoupling from chemical equilibrium with the baryons (i.e. at temperatures $\gtrsim 40$~MeV), and where $S\bar{S}$ annihilation does not subsequently deplete the relic abundance, we would expect the sexaquark component of dark matter to be approximately symmetrically distributed between sexaquarks and antisexaquarks. In particular, as we showed previously, obtaining the full dark matter density in (anti)sexaquarks (without invoking a very large sexaquark chemical potential at the end of the quark-hadron crossover) typically requires suppression of all the depletion processes, leading to a significant relic antisexaquark abundance. Because antisexaquarks carry antibaryon number and can annihilate against nuclei, this opens up the prospect of new searches that exploit the very large density of nuclei in planets and stars, most obviously the Earth.

In this section, we explore how to detect the presence of antisexaquarks that have accumulated in the Earth or pass through as part of the Galactic dark matter wind. We consider both the wind antisexaquark population and the thermalized and accumulated antisexaquarks discussed in Sec.~\ref{sec:accumulation} and Fig.~\ref{fig:acc_SbarN}, and exploit data from the Super-Kamiokande neutrino detector. Due to its long running time and good energy resolution, Super-Kamiokande has high sensitivity to such signals. The wind and accumulated antisexaquarks are complementary in their reach for the scattering cross section; the wind population has a high-cross section cutoff due to the stopping power of the atmosphere and crust, while the accumulated population has a low-cross section cutoff due to the finite size of the Earth compared to the antisexaquark's mean free path. 

The annihilation of dark matter carrying antibaryon number in Super-Kamiokande, leading to induced nucleon decay, has been previously discussed in the literature, e.g.~in the context of hylogenesis models~\cite{Davoudiasl:2010am}, as well as a first study for the sexaquark~\cite{Farrar:2004qy}. There are also other works which have considered the related signal from dark matter self-annihilation (e.g.~
\cite{McKeen:2023ztq, Pospelov:2023mlz}). In this scenario, looking at the Galactic wind dark matter annihilating in or near the detector does not result in a large signal and thus cannot probe a wide range of annihilation cross sections or dark matter subcomponents. In our case, because one of the initial-state particles is an abundant nucleon in or near the detector, we show that a Galactic wind population of antisexaquarks could give rise to a striking signal.

\subsection{Annihilations within Super-Kamiokande}

For both the transient population of antisexaquarks composing the dark matter wind and the accumulated population inside the Earth, the strong annihilation cross section $\bar{S}\lbrace n,p \rbrace \to \bar{b} X$ would result in the annihilation of nucleons inside a detector and potentially leave a detectable signal~\cite{Berger:2023ccd}. This section relies on the assumption that the reaction rate of $\bar{S} \lbrace n, p\rbrace  \to \bar{b} X$ is non-zero. Most annihilation channels of an antisexaquark with a proton or a neutron would produce at least one energetic neutral pion above the Cherenkov threshold, for example through $p \bar{S} \to \overline{\Xi^-} \pi^0$ or $n \bar{S} \to \overline{\Xi^0} \pi^0$. The first pion would have up to a GeV of kinetic energy, and all subsequent produced pions would travel faster than light in the ultrapure water of the detector. The full range of expected final states is presented in Table~\ref{tab:finalstates_SK}. 

Super-Kamiokande is a large neutrino observatory with a total exposure of $450$~kton$\cdot$years~\cite{Super-Kamiokande:2020wjk}. If such an annihilation event occurs inside Super-Kamiokande, it would first produce two photon rings for each neutral pion (up to three in channels with large branching ratios, as described in Table~\ref{tab:finalstates_SK}) and would have been tagged with efficiency of $\sim 50\%$, although not necessarily reconstructed~\cite{Kearns:private_comm}. The heavy antihyperon would then decay to an antiproton or antineutron, injecting another GeV of energy in the detector, and this antinucleon would annihilate with another nucleon within the detector. Since the decay timescale of the antihyperon is $\sim 10^{-10}$~s, it would travel at most a few centimeters before decaying, such that the full decay chain is likely to be contained within the detector volume.

As the antisexaquarks are non-relativistic, the events would appear to have invariant mass equal to the sexaquark plus the nucleon mass ($\sim 3$~GeV), and small momentum ($\sim 2$~MeV), resembling a heavy particle decaying at rest. Compared to previous analyses undertaken by Super-Kamiokande, the antisexaquark-nucleon annihilation has a larger center-of-mass energy than $n-\bar{n}$ oscillations. The final state particles would resemble those from a neutrino neutral-current interaction, although they would be emitted approximately isotropically rather than forward. Due to these differences, the expected signature of an antisexaquark annihilation would be unique, with virtually no background.

In the absence of a dedicated analysis, we could seek to repurpose searches for nucleon and di-nucleon decay~\cite{Super-Kamiokande:2015jbb, Super-Kamiokande:2018apg, Super-Kamiokande:2020wjk}, and neutron-antineutron oscillations~\cite{Super-Kamiokande:2011idx, Super-Kamiokande:2020bov}, in part since all decay chains stemming from $\bar{S} \lbrace n, p \rbrace$ would produce an antinucleon that would annihilate with a nucleon, which have already been searched in these dedicated studies. However, a concern is whether the additional particles in the final state would cause antisexaquark events to fail the cuts for these searches, e.g. by having too many Cherenkov rings from $\pi^0$ or Michel electrons from $\pi^\pm$. A simpler approach is just to require that the rate of antisexaquark events is smaller than the relevant backgrounds, before cuts.

\begin{table}[htt]
    \centering
    \caption{\label{tab:finalstates_SK}Possible final states of antisexaquark annihilation with nucleons. We neglect sub-percent branching ratios and assume $K^0$ are $K^0_s$. BR stands for branching ratio and refers to the decay of the prompt products of annihilation; we do not attempt to estimate the branching ratios of $p\bar{S}$ and $n\bar{S}$ into these various products.}
    \begin{tabular}{l l l l l}
        \hline
        initial \;\; & products \;\; & $\bar{\Xi},\bar{\Sigma}$ decay \;\; & $\bar{\Lambda}, K^0$ decay \;\; & BR \\ \hline
        $p \bar{S}$ & $\overline{\Xi^-} \pi^0$ & $\overline{\Lambda} \pi^+ \pi^0$ & $\bar{p} \pi^+ \pi^+ \pi^0$ & $64\%$\\
        & & & $\bar{n} \pi^0 \pi^+ \pi^0$ & $36\%$\\ \hline
         & $\overline{\Xi^0} \pi^+$ & $\overline{\Lambda} \pi^0 \pi^+$ & $\bar{p} \pi^+ \pi^0 \pi^+$ & $64\%$\\
        & & & $\bar{n} \pi^0 \pi^0 \pi^+$ & $36\%$\\ \hline 
         & $\overline{\Sigma^-} K^0$ & $\bar{n} \pi^+ K^0$ & $\bar{n} \pi^+ \pi^0 \pi^0$ & $31\%$\\
         & & & $\bar{n} \pi^+ \pi^+ \pi^-$ & $69\%$\\ \hline
         & $\overline{\Sigma^0} K^+$ & $\bar{\Lambda} \gamma K^+$ & $\bar{p} \pi^+ \gamma K^+$ & $64\%$\\
         & & & $\bar{n} \pi^0 \gamma K^+$ & $36\%$\\ \hline
         & $\overline{\Lambda} K^+$ & & $\bar{p} \pi^+ K^+$  & $64\%$\\
         & & & $\bar{n} \pi^0 K^+$ & $36\%$ \\ \hline
         $n \bar{S}$ & $\overline{\Xi^0} \pi^0$ & $\bar{\Lambda} \pi^0 \pi^0$ & $\bar{p} \pi^+ \pi^0 \pi^0$ & $64\%$\\
         & & & $\bar{n} \pi^0 \pi^0 \pi^0$ & $36\%$\\ \hline
         & $\overline{\Xi^-} \pi^-$ & $\bar{\Lambda} \pi^+ \pi^-$ & $\bar{p} \pi^+ \pi^+ \pi^-$ & $64\%$\\
         & & & $\bar{n} \pi^0 \pi^+ \pi^-$ & $36\%$\\ \hline
         & $\overline{\Sigma^0} K^0$ & $\bar{\Lambda} \gamma K^0$ & $\bar{p} \pi^+ \gamma \pi^0 \pi^0$ & $20\%$ \\
         & & & $\bar{p} \pi^+ \gamma \pi^+ \pi^-$ & $44\%$ \\
         & & & $\bar{n} \pi^0 \gamma \pi^0 \pi^0$ & $11\%$\\
         & & & $\bar{n} \pi^0 \gamma \pi^+ \pi^-$ & $25\%$\\ \hline
         & $\overline{\Sigma^+} K^+$ & & $\bar{p} \pi^0 K^+$  & $52\%$\\
         & & & $\bar{n} \pi^- K^+$ &  $48\%$\\ \hline 
         & $\overline{\Lambda} K^0$ & $\bar{p} \pi^+ K^0$ & $\bar{p} \pi^+ \pi^0 \pi^0$ & $20\%$ \\
         & & & $\bar{p} \pi^+ \pi^+ \pi^-$ & $44\%$\\
         & & $\bar{n} \pi^0 K^0$ & $\bar{n} \pi^0 \pi^0 \pi^0$ & $11\%$ \\
         & & & $\bar{n} \pi^0 \pi^+ \pi^-$ & $25\%$ \\ \hline
    \end{tabular}
\end{table}

Super-Kamiokande has a rate of multi-GeV multi-ring atmospheric neutrino events of $\sim 1$ per day for energies above 1.33~GeV~\cite{Super-Kamiokande:2020sgt}. This corresponds to roughly 5000 events over the detector's total exposure; to estimate a bound we require less than 1000 antisexaquark events over the same exposure, corresponding to a rate of 0.2 events/day. A more sophisticated analysis could almost certainly significantly improve this bound; for example, even without modeling the signal in depth, Super-Kamiokande has searched for seasonal correlations in its dataset and found none~\cite{Super-Kamiokande:2015qek}, which could potentially place limits on the fraction of the background originating from wind antisexaquarks. Searches for $nn\rightarrow \pi^0 \pi^0$ over 282.1~kton$\cdot$years found zero events with a total invariant mass above 2 GeV~\cite{Super-Kamiokande:2015jbb}, so the limit could potentially be up to around three orders of magnitude better than our estimate.

In order to assess the sensitivity of Super-Kamiokande to a relic antisexaquark population, we estimate in this section the rate of annihilation events occurring in the detector. The signal is controlled directly by the $B$ parameter described earlier, but the relic antisexaquark abundance can also be depleted by the processes controlled by $A$ and $\ev{\sigma v}^\text{ann}_{S \bar{S}}$ in the early universe. We will focus initially on results where this extra depletion is negligible, which provides an upper bound on the potential signal rate from antisexaquark-baryon annihilation; the rate can be scaled downward appropriately in the case of a lower antisexaquark abundance.  

The number of annihilated nucleons due to the transient antisexaquark population is given by
\begin{align}\label{eq:NSuperK}
        N_{n,\text{ann}}^\text{wind} (t_E) &\approx N_{n,\text{SK}}^0 \ev*{\sigma v}_{\bar{S}n}^\text{ann} t_E n_{\bar{S},0}  \nonumber \\
        &\times \exp(-\frac{\ev*{\sigma v}_{\bar{S}n}^\text{ann}}{v_{\bar{S}}}  \int^{\infty}_{z_\text{det}} \dd{z} \sum_A n_A(z) )  \ ,
\end{align}
where $N_{n,\text{SK}}^0$ is the number of nucleons in the detector at the beginning of the experiment lifetime, ${\ev*{\sigma v}_{\bar{S}n}^\text{ann} = B m_\pi^{-2}}$ is the annihilation rate, $t_E$ is the running time, $n_{\bar{S},0}$ is the local number density of antisexaquarks, and $v_{\bar{S}}$ is the velocity of antisexaquarks at the detector's depth. The factor on the second line of Eq.~\eqref{eq:NSuperK} accounts for the depletion of the antisexaquark flux as they cross the Earth atmosphere and crust due to annihilation with ambient nucleons. This expression gives a range of annihilation rates $\ev*{\sigma v}_{\bar{S}n}^\text{ann}$ over which antisexaquarks can annihilate a noticeable number of nucleons in a detector. For smaller cross section, antisexaquarks go through the detector without interacting, while for larger cross sections, antisexaquarks annihilate in the atmosphere and the crust before reaching the detector. 

\begin{figure*}[ttt]
\centering
\begin{minipage}[b]{0.45\textwidth}
    \centering
    \includegraphics[width=\textwidth]{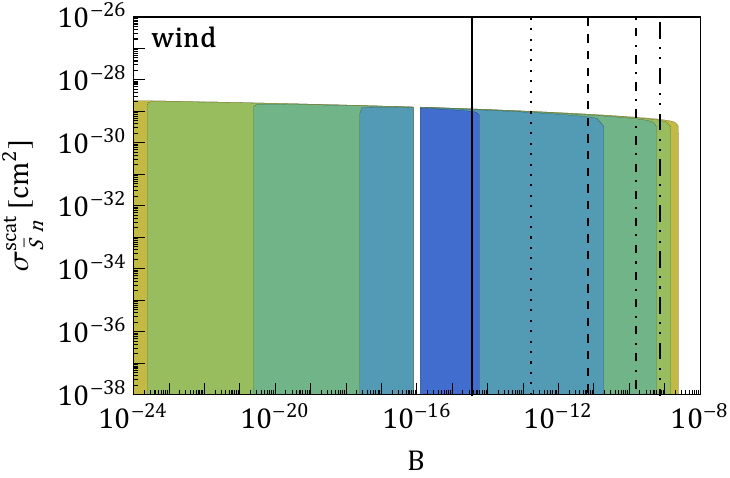} 
\end{minipage}~~
\begin{minipage}[b]{0.45\textwidth}
    \centering
    \includegraphics[width=\textwidth]{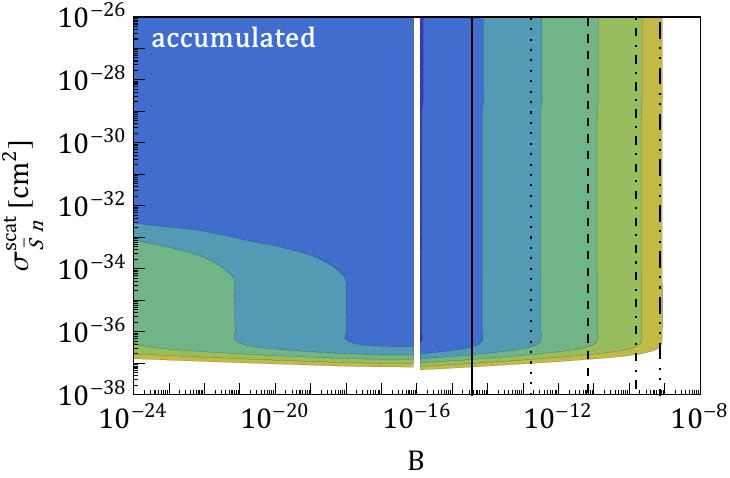}
\end{minipage}~~
\begin{minipage}[t]{0.1\textwidth}
    \centering
    \includegraphics[width=\textwidth]{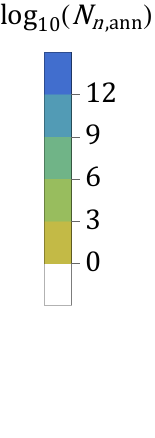}
\end{minipage}
    \caption{Estimated number of nucleons annihilated in the Super-Kamiokande detector due to (left) antisexaquarks from the dark matter wind and (right) accumulated antisexaquarks. This number depends on the annihilation rate parameterized by $\ev*{\sigma v}_{\bar{S}n}^\text{ann} = Bm_\pi^{-2}$ and on the scattering cross section $\sigma_{\bar{S}n}^\text{scat}$. The black vertical lines are for 1860~MeV antisexaquarks as all~(full), $10^{-3}$~(dotted), $10^{-6}$~(dashed), $10^{-9}$~(dot-dashed), and $10^{-12}$~(dot-dot-dashed) of dark matter. To the left-hand side of the white gaps, we take dark matter to be half in antisexaquarks.}
    \label{fig:superK}
\end{figure*}

In addition to the annihilation cross section, antisexaquarks may scatter as they cross the atmosphere and crust, such that the velocity and hence the flux at the depth of Super-Kamiokande may be reduced. We follow common literature (e.g.~\cite{Starkman1990,Emken:2018run}) to include this factor through
\begin{align}
    v_{\bar{S}} &= v_\chi \exp(- \sigma_{\bar{S} n}^\text{scat} \int^{\infty}_{z_\text{det}} \sum_A n_A(z) \frac{\mu_{\bar{S} A}^4}{m_n m_{\bar{S}} \mu_{\bar{S} n}^2} z) \ ,
\end{align}
where $v_\chi \sim 220$~km/s is the galactic dark matter velocity. $n_{\bar{S},0}$ is a function of $\sigma_{\bar{S}n}^\text{scat}$, but also $\ev{\sigma v}_{S\bar{S}}^\text{ann}$ and $\ev{\sigma v}_{SX}^\text{ann}$ since the fraction of dark matter composed of antisexaquarks is determined through freeze-out. For a typical dark matter detector where we consider scattering with an energy above the experimental threshold, we would cut all dark matter particles for which the velocity at the detector's depth does not result in a sufficiently large kinetic energy. However, in the context of annihilation, we have no such concerns. The left panel of Fig.~\ref{fig:superK} presents the number of expected events in the Super-Kamiokande detector as a function of the nucleon scattering cross section and the annihilation cross section $\bar{S}\lbrace n,p \rbrace \to \bar{b}X$, parameterized by $\ev*{\sigma v}_{\bar{S}n}^\text{ann} = B m_\pi^{-2}$. The contours for different sexaquark masses are essentially identical. For larger values of $B \gtrsim 10^{-16}$, we assume that $B$ controls the relic abundance. For very small values of $B$, this process never equilibrates in the early universe. As argued previously, in this case the relic abundance can be determined by $A$ and/or $\ev{\sigma v}_{S \bar{S}}^\text{ann}$, or by processes occurring during the QCD phase transition. Thus for $B \lesssim 10^{-16}$ we assume the sexaquarks are symmetric and constitute 100\% of the dark matter density; the presented rate can be scaled down accordingly if either of these assumptions is violated. Even though the scattering cross section range shown is constrained by dark matter direct detection experiments (Fig.~\ref{fig:all_bounds}) for $f_\chi = 1$, the bounds from antisexaquarks annihilating in Super-Kamiokande provide an additional independent check.

In addition to annihilated nucleons due to the transient antisexaquark population, there may also be a population of thermalized antisexaquarks which are accumulated near the surface of the Earth, as described in Sec.~\ref{sec:accumulation}. The number of annihilated nucleons due to the accumulated antisexaquark population is given by
\begin{align}\label{eq:NSuperK_acc}
    N_{n,\text{ann}}^\text{acc} (t_E) \approx N_{n,\text{SK}}^0 \ev*{\sigma v}_{\bar{S}n}^\text{ann} t_E n_{\bar{S},\text{acc}} \ ,
\end{align}
where $n_{\bar{S},\text{acc}}$ is the number density of accumulated antisexaquarks at the depth of Super-Kamiokande. This equation depends on the cross section $\ev{\sigma v}_{\bar{S}n}^\text{ann}$ in three ways: once in the freeze-out calculation to determine the fractional relic abundance of antisexaquarks, $f_{\bar{S}}$ (although for $B \lesssim 10^{-16}$ we again assume $f_{\bar{S}}=0.5$; the rate can be rescaled accordingly for smaller $f_{\bar{S}}$), once in the accumulation through the annihilation rate $\Gamma_{\bar{S}n}^\text{ann}$, and once in the annihilation in the Super-Kamiokande detector explicitly in Eq.~\eqref{eq:NSuperK_acc}. The right panel of Fig.~\ref{fig:superK} presents the resulting number of annihilated nucleons due to accumulated antisexaquarks.

Comparing annihilation rates due to the antisexaquark wind and accumulated antisexaquarks, the accumulated population dominates at large scattering cross sections ($>10^{-34}~\text{cm}^2$), where the stopping power of the Earth's crust to the wind is significant. On the other hand, the wind population dominates at small scattering cross section ($<10^{-37}~\text{cm}^2$), where the long mean free path implies that very few antisexaquarks are captured by the Earth.

Combining the effects of the two populations (wind and accumulated), let us examine the limiting case where the cross sections parameterized by $A$ and $\ev{\sigma v}_{S\bar{S}}^\text{ann}$ are negligible, and where as previously we assume the full dark matter abundance is obtained (and is symmetric, $f_S=f_{\bar{S}}$) when $B < 10^{-16}$ and so chemical equilibrium is never attained after the QCD phase transition. In this case, we would expect more than 1 event in Super-Kamiokande over the range $10^{-27} < B < 10^{-9}$ (with the lower limit scaling linearly with the number of events, e.g. there are $\sim 10^3$ events for $B\sim 10^{-24}$, and so on). This calculation assumes ideal conditions, i.e. perfect efficiency for tagging and reconstructing, and is intended as an order-of-magnitude estimate of the range of $B$ that can be excluded.

This limit is independent of the elastic scattering cross section. In the case of elastic scattering cross sections above ${10^{-32}~\text{cm}^2}$, the range of $B$ that would produce ${\sim 1000}$ events in Super-Kamiokande extends to ${10^{-39} < B < 10^{-9}}$, while for cross sections exceeding ${10^{-37}~\text{cm}^2}$ (note that our polarizability cross section estimate satisfies this criterion), the range is ${10^{-28} < B < 10^{-9}}$. These results are summarized in Table~\ref{tab:SK_summary}. The lower ends of all these ranges correspond to a far stronger suppression to the effective sexaquark-baryon coupling than has been envisaged in any source, to our knowledge (see e.g.~the discussion in Ref.~\cite{Farrar:2023wta}). The unconstrained region with $B > 10^{-9}$ corresponds to the plateau where the sexaquark abundance is less than $10^{-10}$ of the dark matter density and the antisexaquark abundance is suppressed even further; this region seems experimentally viable, but $S,\bar{S}$ would contribute only a tiny fraction of the dark matter. Comparing the two panels of Fig.~\ref{fig:superK}, Super-Kamiokande is more sensitive to large values of $B$ for the Galactic dark matter wind than the accumulated population, since as antisexaquarks are captured their abundance is depleted by annihilation with nearby nucleons.

If we relax the assumption that $A$ is small, the effect of a larger $A$ is to deplete relic abundance $f_\chi$ following the same equilibrium trajectory as the case of large $B$ (see Sec.~\ref{sec:abundance}). As shown in Fig.~\ref{fig:freeze_out_AB}, this has little effect for $B \gtrsim 10^{-9}$ but will suppress the signal for smaller values of $B$. However, because the constraints for smaller values of $B$ are already so strong, eliminating them requires the severe suppression of antisexaquark relic abundance found for $A \gtrsim 10^{-14}$, where $f_\chi \lesssim 10^{-10}$.

A more promising way to evade this constraint while achieving a non-negligible contribution to $f_\chi$ is to annihilate away the relic sexaquark abundance via $\ev{\sigma v}_{S\bar{S}}^\text{ann}$. As discussed in Secs.~\ref{sec:abundance}-\ref{sec:freezeout}, this can lead to $f_\chi$ as large as $10^{-3}$ (for a decoupling temperature $T_*$ up to 150~MeV, corresponding to small $A$ and $B$) with a very suppressed antisexaquark abundance. In this specific case, suppressing the antisexaquark abundance by a $\mathcal{O}(1)$ factor (relative to sexaquarks) requires $\ev{\sigma v}_{S\bar{S}}^\text{ann} \gtrsim 10^{-22}~\text{cm}^3/\text{s}$ (as shown in Fig.~\ref{fig:freeze_out_abundance_fchi}); there will be limits on the residual antisexaquark annihilation, but above this cross-section threshold, the antisexaquark abundance and hence the annihilation signal is exponentially sensitive to the cross section, as is usual for asymmetric dark matter.

\begin{table}[ht]
    \centering
    \caption{\label{tab:SK_summary}Approximate lower reach of Super-Kamiokande to the annihilation rate $\ev{\sigma v}_{\bar{S}n}^\text{ann}$ parameterized by $B$ to see at least 0.2 annihilation per day of an antisexaquark in the Galactic dark matter wind and in an accumulated population, assuming sexaquarks are all of dark matter. To either go from the given rate (0.2~events/day) to a different signal rate, or decrease the fraction of dark matter in (anti)sexaquarks, scale the number in the table accordingly.
    }
    \begin{tabular}{l c c}
         \hline
         $\sigma_{\bar{S}n}^\text{scat}~[\text{cm}^2]$ & \;\; wind \;\; & accumulated \\[0.1cm] \hline
         $> 10^{-29}$ & $-$ & $10^{-39}$ \\
         $10^{-32} - 10^{-29}$ \; & $10^{-24}$ & $10^{-39}$ \\
         $10^{-37} - 10^{-32}$ & $10^{-24}$ & $10^{-28}$ \\
         $<10^{-37}$ & $10^{-24}$ & $-$ \\
         \hline
    \end{tabular}
\end{table}

\subsection{Other annihilation constraints}

We can also expand our analysis beyond the antisexaquarks annihilating within Super-Kamiokande, to include those annihilating anywhere within the Earth that produce neutrinos detectable by the detector. When antisexaquarks annihilate with either a proton or neutron, they typically form a strange antibaryon. The decay chain of this baryon usually includes a charged pion, which subsequently decays into a muon neutrino. Our focus is primarily on antisexaquarks annihilating beneath the detector's location ($r < R_\text{SK}$), ensuring that all resulting events are upward-moving. This distinction would be crucial for differentiating them from cosmic rays. The energy of these neutrinos could be as high as $\mathcal{O}(\text{GeV})$ if they were produced directly in the annihilation, or via a rapid decay or annihilation of a charged pion shortly after its production; however, they could also appear at significantly lower energy if they originated from pions that stopped in the Earth prior to decaying. In Appendix~\ref{app:Earthannihilations} we explore the possible Super-Kamiokande signature of these neutrinos in the optimistic case where they have $\mathcal{O}(\text{GeV})$ energy, and show that the number of events is expected to be around 7 orders of magnitude below that discussed in the previous subsection (at lower energies we would expect a suppressed cross section for neutrino detection).

We further investigated the accumulation of (anti)sexaquarks and their subsequent $S\bar{S}$ annihilation in Super-Kamiokande, assuming the other thermally-averaged annihilation cross sections parameterized by $A$ and $B$ are strongly suppressed (and thus the sexaquark asymmetry is determined at ${T_\ast = 150}$~MeV). If $\ev{\sigma v}_{S\bar{S}}^\text{ann}$ sets the freeze-out abundance as well as the accumulation of (anti)sexaquarks in the Earth, Super-Kamiokande could constrain this annihilation rate up to ${3 \times 10^{-22}~\text{cm}^3/\text{s}}$ for elastic scattering cross sections above $10^{-33}~\text{cm}^2$, corresponding to ${f_{\bar{S}} \sim 10^{-9}}$ (and ${f_S \sim 10^{-3}}$). In the scenario where $f_S$ were somehow fixed to 1 at the end of the quark-hadron transition, we would expect to have sensitivity to $\ev{\sigma v}_{S\bar{S}}^\text{ann}$ up to ${3 \times 10^{-25}~\text{cm}^3/\text{s}}$, again corresponding to ${f_{\bar{S}} \sim 10^{-9}}$. If $\ev{\sigma v}_{S\bar{S}}^\text{ann}$ is too weak to fully deplete the antisexaquarks in the early universe, leading to a symmetric $f_S\approx f_{\bar{S}}$ population, and assuming the annihilation products produce a distinct and distinguishable signature, we can constrain a wide range of values for $\ev{\sigma v}_{S\bar{S}}^\text{ann}$ for elastic scattering cross sections above $10^{-33}~\text{cm}^2$. Specifically, where $f_S=f_{\bar{S}}=0.5$, the cross section sensitivity can be as small as  $\ev{\sigma v}_{S\bar{S}}^\text{ann} = 10^{-46}~\text{cm}^3/\text{s}$. These constraints coming from accumulated (anti)sexaquarks are much stronger than what can be obtained for the Galactic wind population because of the sparse density of (anti)sexaquarks, as presented in Ref.~\cite{Arguelles:2019ouk} for a symmetric $f_\chi=1$ population. We could also look for neutrinos from $S\bar{S}$ annihilating in the Earth, and in this case the neutrinos could be somewhat higher in energy than in the case of $\bar{S}b$ annihilation, although as previously we would expect the event rate to be lower compared with annihilation within the detector volume.

Annihilation between sexaquarks and antisexaquarks could provide a classic indirect detection signal. The thermal relic cross section ${\ev{\sigma v}_{S\bar{S}}^\text{ann} \approx 2 \times 10^{-26}~\text{cm}^3 /\text{s}}$ is ruled out for a $\sim 2$~GeV relic with the full dark matter density annihilating through hadronic channels (e.g.~\cite{Leane:2018kjk}), and corresponds to freeze-out in a regime where the sexaquark is still quite symmetric, as discussed in Secs.~\ref{sec:abundance}-\ref{sec:freezeout}. Of course, larger cross sections are viable for $f_{\bar{S}} \ll 1$. As has been assumed in previous studies and seems natural, this suggests that either the dominant depletion processes at freeze-out should involve (anti)sexaquarks interacting with more abundant Standard Model states, rather than annihilation between sexaquarks and antisexaquarks, or that ${\ev{\sigma v}_{S\bar{S}}^\text{ann}}$ must be large enough to efficiently deplete $f_{\bar{S}}$.

Additional bounds from the Earth's heat flow could also be derived~\cite{Mack:2007xj, Bramante:2019fhi}. The limits obtained by these works require the annihilation products to deposit their energy in the Earth, i.e. this excludes neutrinos as the final state particles, and so is complementary to Super-Kamiokande bounds.

There is a possible signal from sexaquark self-interactions via meson exchange, which could have a QCD-scale cross section. The current upper boundary for the self-interaction cross section $\sigma^\text{s.i}$ per dark matter mass $m_\chi$ is ${\sigma^\text{s.i}/m_\chi <0.1-1~\text{cm}^2}$/g~\cite{Bechtol:2022koa}, which (using the sexaquark mass for $m_\chi$) converts into ${\sigma^\text{s.i} = C m_\pi^{-2}}$, with ${C \lesssim 10-100}$. Given the uncertain self-interaction cross section, it is difficult to see how this channel could provide a robust constraint even for $f_\chi=1$, and self-interaction constraints for a subdominant component are much weaker or non-existent.

\section{\label{sec:summary}Summary}

We have explored the implications of the hypothesis stable sexaquarks could exist, focusing on exploring whether sexaquarks could compose a large fraction of the dark matter. We investigated the abundance of (anti)sexaquarks through freeze-out and their potential detection signals, focusing on direct detection experiments and the annihilation of antisexaquarks, including an accumulated population in the Earth. We explored a range of possibilities for the elastic scattering cross section, $\sigma_{Sn}^\text{scat}$, as well as three thermally averaged cross section times velocity, $\ev{\sigma v}_{b b' \to SX}^\text{ann}$, $\ev{\sigma v}_{\bar{S} b \to \bar{b}'X}^\text{ann}$, and $\ev{\sigma v}_{S \bar{S}}^\text{ann}$ with the first two parameterized by $A$ and $B$ times $m_\pi^{-2}$, respectively. We varied these interactions separately to sweep the allowed parameter space.

To obtain a sexaquark relic density that matches the observed dark matter abundance, processes that maintain the sexaquark chemical potential $\mu_S=2\mu_B$ must freeze out at a high temperature, where chemical potentials are negligible and a nearly-symmetric population of (anti)sexaquarks is obtained. This requires a severe suppression of all strong number-changing interactions that allow (anti)sexaquarks to exchange baryon number with other particles. Specifically, if either $A \gtrsim 10^{-15}$ or $B\gtrsim 10^{-8}$, the sexaquark number density is driven to around $10^{-11}$ of the total dark matter density, with antisexaquarks being further suppressed. This value remains almost unchanged for higher values of $A$, $B$; we have presented a simple analytic estimate for the value of the abundance at this plateau in Appendix~\ref{sec:scaling}. However, for smaller values of $A$ and $B$, the freeze-out abundance increases steeply with decreasing cross sections; in the case where the process parameterized by $A$ freezes out last, we have shown analytically that the final relic abundance is approximately proportional to $A^{-m_S/B_S}$ where $B_S$ is the binding energy of the sexaquark and $m_S$ is its mass. The transition between these two regimes is tied to the baryon chemical potential becoming significant (i.e. the baryon and antibaryon number densities diverging). Considering the high-temperature regime where chemical potential effects are negligible, we found that preventing the onset of chemical equilibrium after the QCD crossover requires $A \lesssim 10^{-19}$ and $B\lesssim 10^{-17}$ (in contrast to Ref.~\cite{Farrar:2023wta} which found a more modest suppression of only $A\lesssim 10^{-13}$ was required; we have identified the source of this difference as an error in a cross section adopted by Ref.~\cite{Farrar:2023wta}). This suppression for $A$ is quite severe and comes close to saturating previous estimates of the possible suppression from wavefunction factors and tunneling ~\cite{Farrar:2023wta}; it may be unrealistically small (especially given the likely presence of representation mixing due to flavor symmetry breaking).

Nonetheless, we considered scenarios where $A$ and $B$ are very small and a large symmetric (anti)sexaquark relic abundance is generated, both with and without subsequent depletion of $\bar{S}$ by $S\bar{S}$ annihilation. In the case where $S\bar{S}$ annihilation is efficient enough to exponentially deplete $\bar{S}$ abundance (requiring $\langle \sigma v\rangle^\text{ann}_{S\bar{S}} \gtrsim 10^{-22}$~cm$^3$/s), we find that the final sexaquark abundance is always less than $10^{-3}$ of the dark matter, if we start from an equilibrium configuration at $T_*\lesssim 150$~MeV. Achieving the full dark matter abundance in such a scenario would thus require a very enhanced yield of sexaquarks from the QCD phase transition relative to equilibrium expectations. If the $S\bar{S}$ annihilation is less efficient, we predict a symmetric relic antisexaquark population in the late universe.

We presented an estimate for the elastic scattering of sexaquarks with nucleons due to the sexaquark's electric polarizability (there may also be larger contributions from meson exchange) and found that the resulting cross section is within the experimental excluded region. Combining current direct-detection limits, we found that increasing the cross section did not allow us to escape these bounds. However, a $10^{-3}$ or smaller subcomponent of dark matter with a large scattering cross section could be viable.

If antisexaquarks are present with an abundance ${f_{\bar{S}} \gtrsim \mathcal{O}(10^{-12})}$, it becomes favorable to look at their annihilation signature rather than their scattering in direct detection detectors. Taking the Super-Kamiokande detector as a benchmark, and accounting for both the dark matter wind and the possible accumulation of antisexaquarks in the Earth, we found that in the case where antisexaquarks have an abundance similar to that of sexaquarks (requiring, among other factors, that $B\lesssim 10^{-9}$), an annihilation signal comparable to the relevant background rate at Super-Kamiokande can be obtained, for values of $B$ down to $10^{-24}$. For elastic per nucleon scattering cross sections exceeding ${10^{-37}~\text{cm}^2}$ (which holds for our polarizability estimate), the minimum testable value of $B$ is lowered further, by $4-15$ orders of magnitude depending on the scattering cross section.

Taken together, our results appear to strongly disfavor sexaquarks as all of dark matter or even as a substantial subcomponent (more than $\sim 10^{-3}$). However, a few possibilities remain open. First, either ${A \gtrsim 10^{-14}}$ or ${B \gtrsim 10^{-9}}$ lead to ${f_S \sim 10^{-11}}$ and an even smaller component of antisexaquarks. Our constraints using scattering in direct detection experiments and annihilation in Super-Kamiokande are not sensitive to such scenarios. Alternatively, there is a continuum of scenarios with ${A \lesssim 10^{-15}}$, ${B \lesssim 10^{-10}}$ where $S\bar{S}$ self-annihilations efficiently deplete the antisexaquark abundance in the early universe, leading to ${f_S \sim 10^{-3}-10^{-11}}$; this scenario requires a minimal self-annihilation cross section of ${10^{-22}~\text{cm}^3/\text{s}}$ (for ${f_S \sim 10^{-3}}$; this lower bound will increase roughly as the inverse of $f_S$). Direct detection also sets bounds on the elastic scattering cross section in this case, e.g.~requiring ${\sigma_{S n}^\text{scat} \approx 10^{-28}~\text{cm}^2}$ for ${f_S \approx 10^{-3}}$ (with a wider range of possibilities for lower $f_S$). Scenarios where the antisexaquark abundance is not strongly depleted in the early universe (relative to the sexaquarks) require an extremely tiny value of $B$ to survive Super-Kamiokande limits, which does not naively appear plausible. 

In order to obtain $f_S=1$, all of the following would need to be true:
\begin{itemize}
    \item There are either unaccounted-for errors in the constraints from multiple direct-detection experiments, OR the elastic scattering cross section between nucleons and the sexaquarks is strongly suppressed (${\sigma_{Sn}^\text{scat} < 10^{-41}~\text{cm}^2}$). This cross section is several orders of magnitude below what we would expect from purely electromagnetic interactions, for a sexaquark radius as small as $0.15$~fm.
    
    \item Either antisexaquark-proton annihilation is suppressed by at least 24 orders of magnitude relative to a QCD-scale cross section (to avoid signals comparable to the background rate in Super-Kamiokande), OR the following must all be true:
    \begin{itemize}
        \item Sexaquark-baryon interactions must be suppressed by large factors ${A < 10^{-19}}$, ${B < 10^{-17}}$;
        \item The net yield of sexaquarks (i.e.~sexaquarks minus antisexaquarks) from the quark-hadron transition must exceed expectations from the equilibrium calculation in the hadronic phase by three (or more) orders of magnitude;
        \item Self-annihilations or a similar process must be sufficiently strong to exponentially deplete antisexaquarks in the early universe, requiring ${\ev{\sigma v}^\text{ann}_{S\bar{S}} \gtrsim 10^{-25}~\text{cm}^3/\text{s}}$. (This criterion is not hard to envisage; the others are more difficult.)
    \end{itemize}
\end{itemize}

To our knowledge, the sexaquark is the first six-quark bound state to be proposed as a dark matter candidate. Other dark matter candidates carrying baryon number may undergo a freeze-out process similar to what we laid out, even though their exact quantum numbers may differ, to the degree that interactions with the baryons enforce a constraint on their chemical potential up to the point where they decouple. In the scenario where these candidates' baryon-number-transferring interactions decouple before the chemical potential becomes important, and they do not then efficiently self-annihilate (or do so in a way that yields a residual population carrying antibaryon number), the non-negligible dark matter population would leave a distinct annihilation signature in Super-Kamiokande.

We hope that our work encourages future lattice QCD calculations to shed light on the sexaquark's electromagnetic structure and confirm the polarizability estimate we have used. Additionally, lattice QCD can be used to investigate the interaction and decay channels of the sexaquark, providing further insight into its properties.

Expanding beyond the scope of sexaquarks, the equations we have derived for the interaction of a dark matter candidate with Standard Model particles in Eq.~\eqref{eq:sigma_chi_n} and Appendix~\ref{sec:Pospelov} hold true for all candidates with non-zero polarizability, independent of the specific underlying model. 

\section*{Acknowledgements}

We thank Glennys Farrar and Robert L. Jaffe for helpful conversations regarding the sexaquark, Edward Kearns for thoughtful discussions on Super-Kamiokande, as well as Maxim Pospelov and Pedro L.S. Lopes regarding dark matter electromagnetic interactions. We acknowledge beneficial discussions with William Detmold and Phiala E. Shanahan on lattice QCD. In addition, we express our thanks to Christopher V. Cappiello and David E. Morrissey for supplying us with experimental bounds for some of the direct detection experiments. We acknowledge helpful discussions with Michael Fedderke and Junwu Huang regarding signatures in neutron stars, with Joe Bramante regarding dark matter capture in the Earth, and regarding the sexaquark generally with the particle theory group of the Institute for Fundamental Science at the University of Oregon.

This work was supported by the U.S. Department of Energy, Office of Science, Office of High Energy Physics of U.S. Department of Energy under grant Contract Number DE-SC0012567, and by the Simons Foundation (Grant Number~929255, TRS). MM is supported by the Fonds de Recherche du Qu\'ebec -- Nature et Technologies~(FRQNT) doctoral research scholarship (Grant No.~305494) and the Natural Sciences and Engineering Research Council~(NSERC) Canada Postgraduate Scholarship - Doctoral (Grant No.~577851). TRS thanks the Aspen Center for Physics, which is supported by National Science Foundation grant PHY-2210452, for hospitality during the completion of this work.

\appendix

\section{\label{sec:group}Representation theory}

Group theory provides a powerful framework for interpreting the symmetries and transformations of physical systems. By examining the group properties of quark bound states, we can gain insights into their structure and behavior, and make predictions about their interactions with other particles. In this work, we review Ref.~\cite{Jaffe1977}'s treatment of group theory to investigate the properties of the sexaquark in a qualitative manner. To construct the six-quark state, we must build it as an $s$-wave SU(3)${}_c$ color singlet with an overall symmetric wavefunction and spin zero. The interesting structure arises when we examine the physical spinless combinations, which give rise to three possible irreducible representations of SU(3)${}_f$ flavor: the \textbf{1}, \textbf{27}, and \textbf{28}. In the SU(3)${}_f$ flavor symmetry limit, the sexaquark is the singlet, and in the literature it has frequently been treated as pure flavor-singlet as well as being color-singlet and spin zero.

However, the flavor symmetry is broken due to the significant mass gap between the light up and down quarks and the heavier strange quark. As a result, we expect the sexaquark-like state to have substantial admixtures of higher irreducible representations with the same quantum numbers $J^{PC}$ (representations that contain an isosinglet, strong-hypercharge zero, and spin-zero state). Admixtures with the {\bf 27} and {\bf 28} representations would be expected to modify the couplings of the sexaquark to other hadronic states non-trivially, compared to the case where the sexaquark is a pure flavor singlet~\cite{Jaffe:private_comm}. In particular, as discussed in Sec.~\ref{sec:properties} the $S\Lambda \Lambda$ coupling must be quite suppressed to avoid constraints from hypernuclei decays, but current arguments favoring a highly suppressed coupling assume the $S$ is a pure flavor singlet~\cite{Farrar:2023wta}. This argument is one motivation for our choice to treat this coupling as a free parameter.

A comparable scenario is observed with the $\omega$ and $\phi$ vector mesons. These mesons possess an isospin of 0 and are mixed states of the \textbf{1} and \textbf{8} representations to diagonalize the strange quark content. The $\phi$ meson is nearly a pure state of ${s \bar{s}}$, whereas the $\omega$ meson contains hardly any strange quark. In the limit of flavor symmetry, the masses of the two mesons should be approximately equal, but the symmetry breaking leads to the $\phi$ meson having a larger mass, and the two mesons, one belonging to the octet and the other being a singlet, become mixed~\cite{Zweig:1964jf}.

\section{\label{sec:scaling}Scaling of the freeze-out abundance}

In this appendix, in order to better understand the features of Fig.~\ref{fig:freeze_out_abundance_fchi}, we develop an analytic understanding of the temperature dependence of the sexaquark abundance at freeze-out, $n_S (T_f)$. For purposes of this appendix we assume that the freeze-out is controlled by the breakup process $SX\leftrightarrow bb'$ where collisions with mesons convert the sexaquark into two baryons (with a rate parameterized by $A$ in the main text). Then the chemical decoupling happens at the temperature $T_f$ where the breakup rate $\Gamma_{SX\to bb'}(T_f)$ from Eq.~\eqref{eq:<sigma_v>} falls below the Hubble rate $H(T_f)$. If we consider a single species $b=b'$ which we take to be the $\Lambda$ hyperon, the freeze-out temperature is approximately
\begin{align}\label{eq:T_freeze_out}
    T_f &\approx \frac{-B_S}{\ln(\frac{1.66 g_\ast^{1/2} 10~\text{MeV}^{1/2} m_S^{3/2}}{m_\text{Pl} m_\Lambda^3 A m_\pi^{-2}})} \ ,
\end{align}
where as previously, we define the binding energy as ${B_S = 2 m_\Lambda - m_S}$. Since both the denominator and numerator are negative, we have ${T_f>0}$. On the right hand side, we take $100$~MeV as a benchmark temperature to solve for the freeze-out and find that the approximation is somewhat insensitive to the choice of temperature, as long as it is in the regime of $1$ to $100$~MeV. We define $X \equiv \frac{H(100~\text{MeV}) m_S^{3/2}}{m_\pi^{-2} m_\Lambda^3 1000~\text{MeV}^{3/2}} = \frac{1.66 g_\ast^{1/2} 10~\text{MeV}^{1/2} m_S^{3/2}}{m_\pi^{-2} m_\Lambda^3m_\text{Pl} }$, then $T_f = -B_S / \ln( X/A)$ to simplify the notation below. Using this expression, we obtain the sexaquark abundance at freeze-out using ${n_S^\text{fo} (T_f) \approx A m_\pi^{-2} (n_\Lambda^\text{eq})^2 / H(T_f)}$. We take Eq.~\eqref{eq:number_density_equilibrium} for the $\Lambda$ number density $n_\Lambda^\text{eq}$ and insert Eq.~\eqref{eq:mu_b} as its chemical potential.

At high temperatures where the chemical potential is negligible, the lambda hyperons' equilibrium abundance resembles the sexaquarks' from Eq.~\eqref{eq:number_density_highT} with the replacement ${m_S \to m_\Lambda}$ and an overall factor of 2 due to the different number of degrees of freedom. Then, with the help of Eq.~\eqref{eq:T_freeze_out} the sexaquark freeze-out abundance is
\begin{align}
    n_S^\text{fo high T}(T_f) &\approx \frac{m_S^3}{2\pi^3}\frac{ \left( \frac{A}{X} \right)^{-m_S/B_S} }{\left( \ln(\left[\frac{A}{X}\right]^{m_S/B_S}) \right)^{3/2}} \ .
\end{align}
The $A$ dependence of the sexaquark abundance originates almost exclusively from ${A^{-m_S/B_S}}$. The large negative index for $A$, since $m_S \gg B_S$, gives rise to a rapid evolution in $f_S, f_{\bar{S}}$ as a function of $A$ at sufficiently small values of $A$, as shown in Fig.~\ref{fig:freeze_out_abundance_fchi}. 

For low temperatures, the chemical potential's dependence on the temperature is non-trivial and the scaling on $A$ is due to a fine tuning of all terms contributing to the freeze-out abundance, 
\begin{align}
    n_S^\text{fo low T} (T_f) &\sim  \frac{Y_{B}^2 m_S^6 \left( \ln( \left[\frac{A}{X} \right]^{m_S/B_S}) \right)^{-9/2}}{\left( \frac{A}{X} \right)^{m_S/B_S} \left( \sum_b m_b^{3/2} \left( \frac{A}{X}  \right)^{-m_b/B_S} \right)^2}  \ .
\end{align}
The dependence on $A$ is very limited and the overall number density is suppressed by the two factors of $Y_{B}$.

\section{\label{app:asymmetric}Asymmetric dark matter with baryon number}

Asymmetric dark matter models sometimes justify an asymmetry in the number density of a dark matter particle $\chi$ compared to its antiparticle $\bar{\chi}$ by the baryon asymmetry, e.g. \cite{Kaplan:2009ag, Zurek:2013wia}. Let us give this dark matter particle the chemical potential $\mu_\chi = \alpha \mu_B$ and its yield giving the full dark matter abundance today $Y_\text{DM}$. If dark matter does not have number-changing interactions with baryons, then the relation
\begin{align}
    \sum_b (Y_b^\text{eq} - Y_{\bar{b}}^\text{eq}) + \alpha (Y_\chi^\text{eq} - Y_{\bar{\chi}}^\text{eq}) &= Y_B + Y_\text{DM}
\end{align}
must hold as hadrons and dark matter are formed from the quark-gluon plasma at $T \sim T_\ast$. Then, for lower temperatures $T \lesssim T_\ast$ they evolve separately as
\begin{align}
    \sum_i (Y_b^\text{eq} - Y_{\bar{b}}^\text{eq}) &= Y_B \ , \label{eq:YB}\\
    \alpha (Y_\chi^\text{eq} - Y_{\bar{\chi}}^\text{eq}) &= Y_\text{DM} \ .\label{eq:YDM}
\end{align}
The individual baryon asymmetries of Eq.~\eqref{eq:YB} evolve in a similar fashion as what was shown in Fig.~\ref{fig:asymmetry}, with protons and neutrons dominating the asymmetry at late times. For dark matter, if it remains coupled to the photon bath until later times the symmetric component will annihilate, leaving only the asymmetric $Y_\text{DM}$ fraction.

\begin{figure}[htt]
    \centering
    \includegraphics[width=\columnwidth]{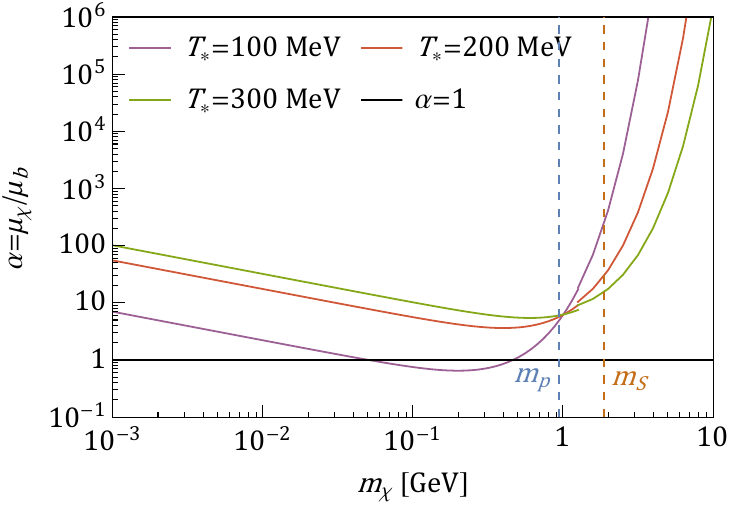}
    \caption{Dark matter to baryon chemical potential ratio required to obtain the full dark matter abundance for asymmetric baryonic dark matter. The curves show different choices for the temperature $T_*$ at which the abundances are initialized. The vertical lines are key masses; the proton's (blue) and the sexaquark's (brown).}
    \label{fig:alpha}
\end{figure}

Fig.~\ref{fig:alpha} shows the required dark matter-to-baryon chemical potential ratio $\alpha$ required to obtain the full dark matter abundance, for three choices of the hadronization temperature. For simplicity, we chose the number of dark matter degrees of freedom $g_\chi = 1$ as appropriate for a scalar (with an additional degree of freedom for the anti-dark-matter); modestly larger values of $g_\chi$ would not change our qualitative results. We present the ratio $\alpha$ for a wide range of dark matter masses, some of which would be relativistic at the temperature $T_\ast$ they would be formed.

A particle with a similar mass to the sexaquark would need to have $\mu_\chi \approx 10 \mu_B$ to account for all of dark matter. A smaller value results in the under-production of dark matter, while a larger value, in the over-production.

We can solve Eqs.~\eqref{eq:YB}-\eqref{eq:YDM} to obtain
\begin{align}
    \frac{\alpha^2 n_\chi^{\mu_B=0}(T_*)}{\sum_b n_b^{\mu_B=0}(T_*)} \frac{m_\chi}{m_p} &= \frac{\Omega_\chi}{\Omega_B} \\
    \frac{\alpha^2 m_\chi^{3/2} e^{-m_\chi/T_*}}{2 \sum_b m_b^{3/2} e^{-m_b/T_*}} \frac{m_\chi}{m_p} &= \frac{\Omega_\chi}{\Omega_B}\ , \label{eq:OmegaDM}
\end{align}
which is valid when $T_*$ is large enough that the chemical potentials are still small relative to the masses.

\section{\label{sec:Pospelov}Cross sections for elastic scattering of sexaquarks}

In this appendix we calculate several cross sections for the nucleus to scatter via its electric or magnetic polarizabilities.

\subsection{Scattering on a uniformly charged nucleus}

Let us compute the cross section for a sexaquark to scatter on a nucleus through its electric polarizability, modeling the nucleus as a uniformly charged sphere. The electric field generated by the nucleus with a spherical homogeneous charge distribution of radius $R$ and charge $Ze$ is given by
\begin{align}\label{eq:Efield}
    \vec{E}(\vec{r}) &= \frac{Z e}{4\pi } \left[ \Theta(R - \abs{\vec{r}}) \frac{\abs{\vec{r}}}{R^3} + \Theta(\abs{\vec{r}}-R) \frac{1}{\vec{r}^{\, 2}} \right] \hat{r} \ .
\end{align}
The matrix element is then
\begin{align}
    A_{fi} &= \mel{\vec{p}_f}{-\frac{1}{2} \alpha_S \vec{E}^2}{\vec{p}_i} \label{eq:alphaE2}\\
    &= -\frac{\alpha_S Z^2 \alpha}{2} \int \dd[3]{\vec{r}} e^{i (\vec{p}_i - \vec{p}_f) \cdot \vec{r}} \nonumber \\ 
    &\qquad \times \left[ \Theta(R - \abs{\vec{r}}) \frac{\vec{r}^{\,2}}{R^6} + \Theta(\abs{\vec{r}}-R) \frac{1}{\vec{r}^{\,4}} \right] \\
    &= -\frac{\alpha_S Z^2 \alpha}{2} \frac{\pi}{ q^5 R^6} \left[ -q^6 R^6 \left(\pi -2 \text{SI}\left(q R\right)\right) \right. \nonumber \\
    &\qquad \left. +2 q R \left(q^4 R^4-2 q^2 R^2 +12\right) \cos(q R) \right. \nonumber \\
    &\qquad \left. +2 \left(q^4 R^4+6 q^2 R^2-12 \right) \sin (q R)\right] \ ,
\end{align}
where SI is the sine integral and we defined $\vec{q} \equiv \vec{p}_i - \vec{p}_f$. In the limit where the deflection of the sexaquark particle is small, we get $q \approx \mu_{S N} v_S \cos(\theta)$. Expanding the matrix element for small $q R$, we get
\begin{align}
    A_{fi} &= -\frac{\alpha_S Z^2 \alpha}{2} \frac{2\pi}{ R } \left[ \frac{12}{5}- \frac{\pi q R}{2} +\mathcal{O} \left(\left[q R\right]^2\right) \right] \ .
\end{align}
We can now find the per-nucleus cross section to be
\begin{align}
    \sigma_{S N}^\text{scat} &= \frac{1}{v_S} \int \frac{\dd[3]{\vec{p}_f}}{(2\pi)^3} \ 2\pi \delta(E_f - E_i) \ \abs{A_{fi}}^2 \\
    &= \frac{144 \pi}{25} \mu_{S N}^2 Z^4 \alpha^2 \frac{\alpha_S^2 }{R^2} \ .
\end{align}

In order to determine the nucleus cross section doe to a different charge distribution, one needs to extract the electric field arising from the charge distribution, square it, and insert it in Eq.~\eqref{eq:alphaE2}.

\subsection{\label{sec:CasimirPolder}Other contributions to the cross section with nuclei from its intrinsic electric polarizability}

In addition to the interaction between a sexaquark and the electric field of a nucleus, there are further contributions from the coupling of the sexaquark to the electric or magnetic static polarizability of the nucleus, denoted $\alpha_N$ and $\beta_N$, respectively. These additional terms in the potential were first derived by Casimir and Polder~\cite{Casimir:PhysRev.73.360}, as well as Feinberg and Sucher~\cite{Feinberg:doi:10.1063/1.1669611,Feinberg:PhysRevA.2.2395}, and result from the exchange of two photons between two neutral spinless particles that are separated by a distance $r$ much larger than their respective sizes:
\begin{align}\label{eq:CasimirPolder}
    V(r) &= -\frac{1}{4\pi r^7} \left( 23 \alpha_S \alpha_N + 23 \beta_S \beta_N - 7 \alpha_S \beta_N \right. \nonumber \\
    &\qquad \left. - 7 \beta_S \alpha_N \right) \ .
\end{align}
The condition $r \gg \lambda$ signifies that the separation between the nucleus and the dark matter particle must be much greater than the wavelength $\lambda$ required to excite the nucleus from its ground state to the first excited state~\cite{Feinberg:doi:10.1063/1.1669611, Buhmann:Dispersion}. 

At distances large compared to the size of the particles, the induced electromagnetic force is generally attractive.

The electric-electric and magnetic-magnetic polarizability interactions are attractive, while the mixed polarizability interactions are repulsive. Here we only considered the terms with a factor of $\alpha_S$, the dark matter electric polarizability, i.e. the first and third terms. The leading order term in the cross section for these effects are
\begin{align}
    \sigma_{S n}^{\alpha\alpha} &\approx \frac{529}{16\pi} \mu_{S n}^2 \frac{\alpha_1^2 \alpha_2^2}{A^2 r_0^8} 
\end{align}
and
\begin{align}
    \sigma_{S n}^{\alpha\beta} &\approx \frac{49}{16\pi} \mu_{S n}^2 \frac{\beta_1^2 \alpha_2^2}{A^2 r_0^8} \ .
\end{align}
Here too we expanded for $qr \ll 1$. $r_0$ is the lengthscale corresponding to the smallest distance of approach of the two polarizable particles. 

We next sought to determine the finite-temperature corrections to the Casimir-Polder interaction~\cite{McLachlan:1963}, giving rise to the potential
\begin{align}
    V(r) = -3 \frac{k_B T}{r^6} \alpha_S \alpha_N \ ,
\end{align}
which decays with a gentler slope than the zero-temperature potential (Eq.~\eqref{eq:CasimirPolder}) and is valid at shorter distances. This potential is valid for the distance between the nucleus and dark matter satisfying ${r \gg (k_B T)^{-1}}$~\cite{Boyer:PhysRevA.11.1650}. Here, the cross section between a polarizable atom and dark matter becomes
\begin{align}
    \sigma_{S n}^{\alpha\alpha, T\not = 0} &= 16 \pi \mu_{S n}^2 \frac{(k_B T)^2}{A^2 r_0^6} \alpha_S^2 \alpha_N^2 \ .
\end{align}
These cross sections are negligible because they are only valid for $r_0$ large and scale as $R^{-n}$ with $n \geq 6$, whereas the cross section due to the electric field of a nucleus scales as $R^{-2}$. Thus, the sexaquark dark matter self-interactions due to the polarizability are highly suppressed.

\subsection{\label{sec:photon}Cross section for scattering with photons from the electric polarizability}

To study the scattering of a photon due to the electric polarizability of the sexaquark, we use the same effective Hamiltonian as before, but we modify to describe the electric field of a single photon in terms of raising and lowering operators instead of the electric field of a whole nucleus. The formalism yields two photon-number violating terms, corresponding to the absorption or emission of two photons, and a sole photon-number conserving operator, representing the elastic scattering of a single photon. Specifically, this operator can be seen as the absorption of a first photon and the simultaneous emission of a second photon~\cite{Berestetskii:1982qgu}.

The cross section for the interaction between a sexaquark at rest and a photon with energy $E_\gamma$ due to the electric polarizability can be expressed as
\begin{align}\label{eq:sigma_chi_photon}
    \sigma_{S\gamma}^\text{scat} &=\frac{ 4 }{3\pi} \alpha_S^2 m_S E_\gamma^4 \frac{ (m_S+E_\gamma)^2}{ (m_S + 2 E_\gamma)^3} \\
    &\propto \alpha_S^2 \begin{cases}
         m_S E_\gamma^3 & m_S \ll E_\gamma \\
         E_\gamma^4 & m_S \gg E_\gamma
    \end{cases}
    \ .
\end{align}
Notably, for a large-mass dark matter particle, the cross section scales in the same way as Rayleigh scattering. It is worth noting that the photon-sexaquark cross section due to polarizability scales with the photon energy, while the nucleon-sexaquark cross section is momentum-independent, at least at low energies. It is also significantly smaller than the cross section between a sexaquark and a nucleus obtained in Sec.~\ref{sec:Epol}.

\section{\label{app:kinetic}Sexaquark kinetic equilibrium}

In Secs.~\ref{sec:abundance}-\ref{sec:freezeout}, we made the assumption that (anti)sexaquarks shared a common temperature with the thermal bath. To justify this assumption, it is sufficient to show that sexaquarks remain in kinetic equilibrium with the baryon-photon fluid throughout the freeze-out process, even after departing from chemical equilibrium with Standard Model particles. The kinetic decoupling happens at the temperature $T_\text{kd}$ where the momentum transfer rate $\Gamma(T_\text{kd})$ falls below the Hubble rate $H(T_\text{kd})$.

\begin{figure}[ttt]
    \centering
    \includegraphics[width=\columnwidth]{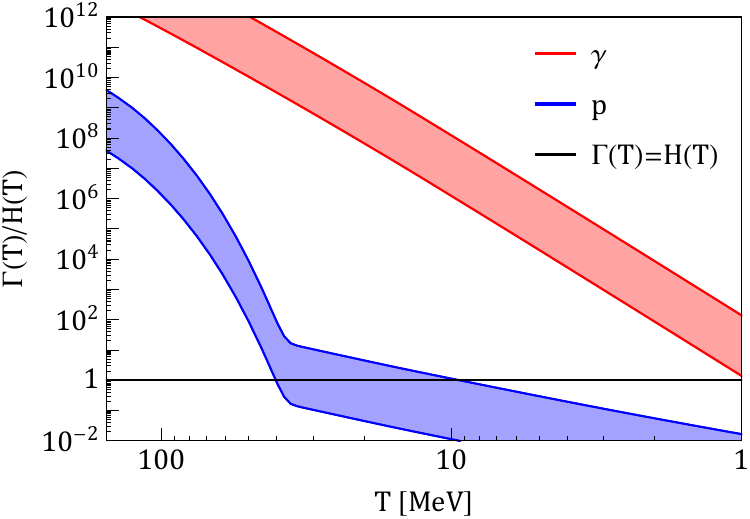}
    \caption{Interaction rate of sexaquarks with photons (red) and non-relativistic protons (blue) compared to the Hubble rate. Sexaquarks are kinetically decoupled when both rates divided by the Hubble rate fall below unity.}
    \label{fig:kin_decoupling}
\end{figure}

The sexaquark polarizability leads to interactions with photons and charged particles. We estimated the kinetic decoupling of sexaquarks by focusing on their interaction with photons and protons, which remained in thermal equilibrium with the bath in the temperature range of interest, by using the cross sections in Eqs.~\eqref{eq:sigma_chi_photon} and \eqref{eq:sigma_chi_n} (exponential charge distribution). The main contributions come from photons, whose large density outweighs the small suppressed cross section, and non-relativistic protons. We approximate the two contributions as~\cite{Feng:2010zp} 
\begin{align}
    \Gamma_\gamma^\text{kin} (T_\text{kd}) &\sim n_\gamma \ev{\sigma v}_{S\gamma}^\text{scat} \frac{T_\text{kd}}{m_S} \ , \\
    \Gamma_p^\text{kin} (T_\text{kd}) &\sim n_p^\text{eq} \ev{\sigma v}_{S p}^\text{scat} \frac{\mu_{S p}^2}{m_S m_p} \ ,
\end{align}
where the thermally-averaged elastic scattering cross sections times the velocity are given by
\begin{align}
    \ev{\sigma v}_{S\gamma}^\text{scat} &= \frac{1}{\pi^2 n_\gamma} \int \dd{E_\gamma} E_\gamma^2 f_\gamma \sigma_{S\gamma}^\text{scat} \ ,\\
    \ev{\sigma v}_{Sp}^\text{scat} &= \sigma_{Sp}^\text{scat} 2^{3/2} \sqrt{\frac{T}{\pi m_S}} \left( 1 + \frac{T}{m_S} \right) \ ,
\end{align}
with $f_\gamma$ the phase space occupancy of photons. Fig.~\ref{fig:kin_decoupling} shows the interaction rate of sexaquarks with photons and protons in the bath. The width of the red and blue lines corresponds to the range of sexaquark masses (1860 to 1890~MeV) and polarizabilities from Sec.~\ref{sec:Epol}. We find that sexaquarks remain kinetically coupled to photons until sub-MeV temperatures and to protons until 40~MeV to 1~MeV, thus the assumption to use a single temperature for all the thermal equilibrium number densities in Secs.~\ref{sec:abundance}-\ref{sec:freezeout} is justified. Scattering on electrons and positrons may also be important but the argument here is already sufficient to show that kinetic equilibrium is maintained.

\section{\label{app:Earthannihilations}Neutrinos from annihilations in the Earth}

In this appendix we explore the potential signature of antisexaquarks annihilating in the Earth below Super-Kamiokande's depth through the production and subsequent possible detection of neutrinos. We make use of the inclusive charge-current neutrino-nucleus cross section, $\sigma_{\nu N}/E_\nu \sim 0.7 \times 10^{-38}~\text{cm}^2/ \text{GeV}$~\cite{ParticleDataGroup:2020ssz} and take the neutrinos to have $\mathcal{O}(\text{GeV})$ energy.

The number of elastic neutrino-nucleus scatterings in the Super-Kamiokande detector due to antisexaquarks in the Galactic dark matter wind is
\begin{align}\label{eq:nu_SK}
    N_{\nu N,\text{scat}}^\text{wind} &\approx N_{N,\text{SK}}^0 \ev*{\sigma v}_{\bar{S}n}^\text{ann} t_E n_{\bar{S},0} N_{n,\oplus}^0 \frac{\sigma_{\nu N}}{A_\oplus} \nonumber \\
    &\times e^{-\ell \sigma_{\nu N} n_N } \nonumber \\
    &\times \Bigg[\exp(-\frac{\ev*{\sigma v}_{\bar{S}n}^\text{ann}}{v_{\bar{S}}} \int^{\infty}_{z_\text{det}} \dd{z} \sum_A n_A(z) ) \nonumber \\
    & - \exp(-\frac{\ev*{\sigma v}_{\bar{S}n}^\text{ann}}{v_{\bar{S}}} \int^{\infty}_{2R_\oplus} \dd{z} \sum_A n_A(z) )\Bigg] \ ,
\end{align}
where $A_\oplus$ is the surface area of the Earth shell at the depth of Super-Kamiokande, $N_{N,\text{SK}}^0$ is the number of oxygen nuclei in Super-Kamiokande, and $N_{n,\oplus}^0$ is the number of nucleons in the Earth, below the detector's depth. The exponential factor on the second line accounts for the neutrinos that interact with nuclei in the Earth before reaching Super-Kamiokande, while the one on the third and fourth lines removes the region of parameter space where antisexaquarks would have annihilated in the atmosphere or crust above the detector. The distance $\ell$ traveled by the neutrinos approaches $2 R_\oplus/\pi$, which is the average distance between any uniformly positioned location in a sphere of radius $R_\oplus$ and a point on the surface of this sphere. We assume that the neutrinos are produced isotropically from any point below the detector. The reduction in the antisexaquark flux from annihilation above the Super-Kamiokande detector is negligible (fraction of antisexaquarks lost is $< 10^{-12}$). Similarly, the reduction in the neutrino flux from interaction in the Earth before reaching Super-Kamiokande is very small, $< 10^{-6}$. Here we are assuming that the neutrinos are produced as muon neutrinos and do not oscillate during their journey to the detector.

In addition, the expected accumulated population of antisexaquarks at depths greater than Super-Kamiokande would lead to
\begin{align}
    N_{\nu N,\text{scat}}^\text{acc} (t_E) &\approx N_{N,\text{SK}}^0 \ev{\sigma v}_{\bar{S}n}^\text{ann} t_E n_{\bar{S},\text{acc}} N_{n,\oplus}^0 \frac{\sigma_{\nu N}}{A_\oplus} \nonumber \\
    &\times e^{-\ell \sigma_{\nu N} n_N} \ ,
\end{align}
similarly to the Galactic wind population in Eq.~\eqref{eq:nu_SK}, although without the factors accounting for the loss of antisexaquarks in their path to reach deeper depths than the detector.

\begin{figure*}[ttt]
\centering
\begin{minipage}[b]{0.45\textwidth}
    \centering
    \includegraphics[width=\textwidth]{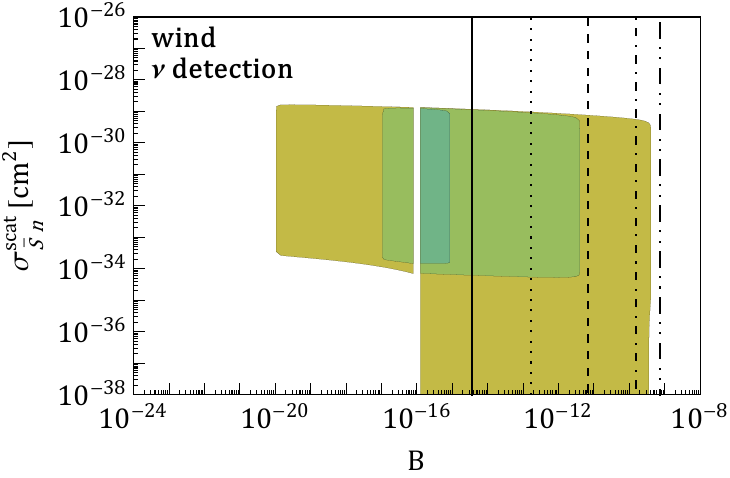} 
\end{minipage}~~
\begin{minipage}[b]{0.45\textwidth}
    \centering
    \includegraphics[width=\textwidth]{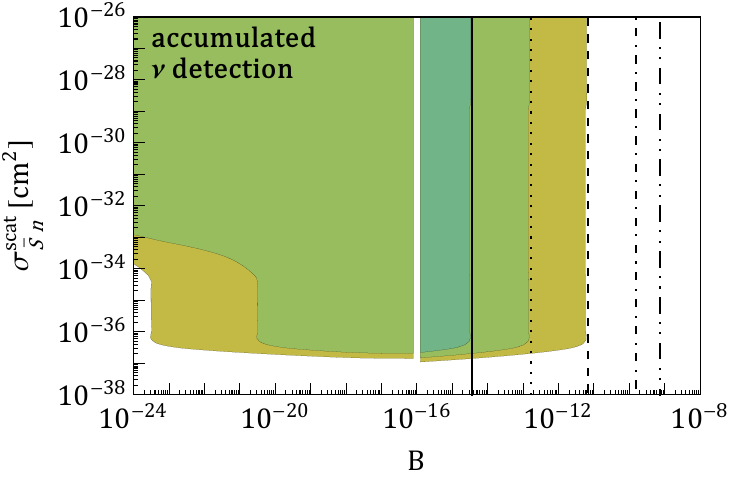}
\end{minipage}~~
\begin{minipage}[t]{0.1\textwidth}
    \centering
    \includegraphics[width=\textwidth]{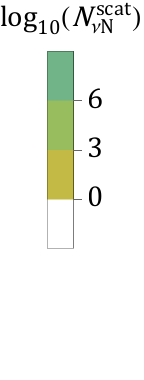}
\end{minipage}
    \caption{Estimated number of neutrinos interacting with nuclei in the Super-Kamiokande detector due to (left) antisexaquarks from the dark matter wind and (right) accumulated antisexaquarks annihilating with nucleons in the Earth. This number depends on the annihilation rate parameterized by $\ev*{\sigma v}_{\bar{S} n}^\text{ann} = B m_\pi^{-2}$ through the freeze-out abundance, but also that a too large cross section depletes the antisexaquark flux before reaching the detector, while a too small cross section results in a too long mean free path. It also depends on the scattering cross section $\sigma_{\bar{S} n}^\text{scat}$; if too large, the stopping power of the Earth crust reduces the flux. The black vertical lines are for 1860~MeV antisexaquarks as all~(full), $10^{-3}$~(dotted), $10^{-6}$~(dashed), $10^{-9}$~(dot-dashed), and $10^{-12}$~(dot-dot-dashed) of dark matter. To the left-hand side of the white gaps, we take dark matter to be half in antisexaquarks.}
    \label{fig:nu_superK}
\end{figure*}

Fig.~\ref{fig:nu_superK} presents the number of oxygen nuclei in the Super-Kamiokande which would be scattered by muon neutrinos produced from the annihilation of antisexaquarks with nucleons in the Earth, both for the Galactic dark matter wind and the accumulated antisexaquarks. The left panel's depletion at large scattering cross sections is due to the stopping power of the atmosphere and the crust, while the depletion at low scattering cross section and $B$ is from antisexaquarks which would fully cross the Earth without slowing down nor annihilating.

Assuming ideal conditions, i.e. perfect efficiency for tagging and reconstructing, as well as statistics in our favor, this process would give a single neutrino scattering in the Super-Kamiokande detector all the way down to $B = 10^{-20}$ (1000 events for $B= 10^{-17}$) for wind dark matter. For accumulated dark matter and scattering cross sections above $10^{-32}~\text{cm}^2$, we can get one event all the way down to $B = 10^{-35}$ (1000 events for $B = 10^{-32}$). For smaller scattering cross sections yet still larger than the Earth radius, this is instead ${B = 10^{-24}}$ ($B = 10^{-21}$). These numbers correspond to rates roughly seven orders of magnitude below their counterparts in Sec.~\ref{sec:SuperK}.

Another caveat in this situation is that it may be much harder to distinguish a small number of neutrinos coming from antisexaquarks annihilating in the Earth from other neutrino sources.

Both antisexaquark annihilation signatures (directly in Super-Kamiokande and indirectly via the emission of neutrinos which will then scatter in Super-Kamiokande) scale with $N_{n,0} \ev{\sigma v}_{\bar{S} n}^\text{ann} t_E$. The distinction is the second calculation additionally has a factor for the number of nucleons in the Earth ($10^{51}$) times the ratio of the neutrino-nucleus cross section over the geometric cross section of the Earth ($10^{-57}$), and finally a factor of $10^{-1}$ to go from the number of nucleons to the number of oxygen nuclei in the detector. Putting these numbers together, the signature of Eq.~\eqref{eq:nu_SK} is smaller than Eq.~\eqref{eq:NSuperK} by seven orders of magnitude.

\bibliography{biblio}

\end{document}